\documentclass[twocolumn,traditabstract]{aa} % for the abstract without structuration 
\usepackage{graphicx}
\usepackage{longtable,lscape}
\usepackage{txfonts}
\usepackage{natbib}
\bibpunct{(}{)}{;}{a}{}{,}

\begin{document}
   \title{Chemically tagging the Hyades Supercluster:}

   \subtitle{A homogeneous sample of F6-K4 kinematically selected  northern stars\thanks{
        Based on observations made with the Mercator Telescope, operated on the island of La Palma by the Flemish Community, at the Spanish Observatorio del Roque de los Muchachos of the Instituto de Astrof\'isica de Canarias.  }
}

   \author{H.M. Tabernero
          \inst{1},
          D. Montes\inst{1}, 
	\and
	J.I. Gonz\'alez Hern\'andez\inst{1,2,3}}

   \institute{Dpto. Astrof\'isica, Facultad de CC. F\'isicas, Universidad Complutense de Madrid, E-28040 Madrid, Spain.
              \email{htg@astrax.fis.ucm.es}
         \and
             Instituto de Astrof\'isica de Canarias, C\/ Via
	     Lactea s/n, E-38200 La Laguna, Tenerife, Spain
	\and
	Dept. Astrof\'isica, Universidad de La Laguna (ULL),
	E-38206 La Laguna, Tenerife, Spain
             %\email{...}
             %\thanks{}}
}
   \date{Received 17 June, 2011 ; accepted 17 April, 2012}

% \abstract{}{}{}{}{} 
% 5 {} token are mandatory
 
  \abstract
  % context heading (optional)
  % {} leave it empty if necessary  
{Stellar kinematic groups are kinematical coherent groups of stars that might have  a common origin. These
groups are dispersed throughout the Galaxy over time by the tidal effects of both Galactic rotation and disc heating, although their chemical  content remains unchanged. The aim of {\it chemical tagging} is to establish that the abundances of every element in the analysis  are  homogeneus among the members.
We study the case of the Hyades Supercluster to compile a reliable list of members (FGK stars) based on our {\it chemical tagging}  analyisis. 
For a total of 61 stars from the Hyades Supercluster, stellar atmospheric parameters ($T_{\rm eff}$, $\log{g}$, $\xi$, and [Fe/H]) are determined using our code called {\scshape StePar}, which is based on the sensitivity to the stellar atmospherics parameters of the  iron $EWs$ measured in the spectra.
 We derive the chemical abundances of 20 elements and find that their [X/Fe] ratios are consistent with Galactic abundance trends reported in previous studies.
The {\it chemical tagging} method is applied with a carefully developed differential abundance analysis of each candidate member of the Hyades Supercluster, using a well-known member of the Hyades cluster as  a reference (vB~153). We find that only 28 stars (26 dwarfs and 2 giants) are members, i.e.  that 46 \% of our candidates are members based on the differential abundance analysis. This result confirms that the Hyades Supercluster cannot originate solely from the Hyades cluster.}  
% aims heading (mandatory)
 {} 
 % methods heading (mandatory)
   { }
  % results heading (mandatory)
   {} 
  % conclusions heading (optional), leave it empty if necessary 
  {}


   \maketitle

\section{Introduction}

Stellar  kinematic groups (SKGs)
--superclusters (SCs) and moving groups (MGs)-- are kinematic
coherent groups of stars \citep{Eggen1994} that could have a common
origin. The youngest and most well-studied SKGs are: the Hyades
SC (600 Myr), the Ursa Major MG (Sirius SC, 300 Myr), 
the Local Association or Pleiades MG (20 to
150 Myr), the IC 2391 SC (35-55 Myr), and the Castor
MG (200 Myr) \citep{mon01a}.
Since Olin Eggen introduced the concept of a MG and the idea that
stars can maintain a kinematic signature over long periods of
time, their existence (mainly in the case of the old MGs)  has been disputed. 
There are two factors that can disrupt a MG: the Galactic
differential rotation  which tends to disperse the stars, and the disc
heating, which causes velocity the dispersion of the disc stars to increase with age.

The  overdensity of stars in some regions of the Galactic velocity 
UV-plane may also be the result of global
dynamical mechanisms  linked with the non-axisymmetry of the Galaxy \citep{fae05},
namely the presence of a rotating central bar 
\citep[e.g.][]{deh98, Fux01, Minchev10}
and spiral arms 
\citep[e.g.][]{QuillenMinchev05, ant09,ant11}, or
both 
\citep[see][]{Quillen03, MinchevFamaey10}.
However, several works have shown that different age subgroups are situated in the
same region of the Galactic velocity plane as the classical MGs
\citep{asi99} suggesting that both field-like
stars and young coeval ones can coexist  in MGs \citep{fae07, fae08, ant08, 
Klement08, sil08, FrancisAnderson09a, FrancisAnderson09b, Zhao09}. 

Using different age indicators such as the
lithium line at 6707.8 \AA, the chromospheric activity level, and the X-ray emission,
it is possible to quantify the contamination by younger or older field stars among late-type candidate members
of a SKG \citep[e.g.][]{mon01b, mar10, lop06, lop09, lop10, mal10}.
However, the detailed analysis of the chemical content (\textit{chemical tagging}) is 
another powerful method that provide clear  constraints on the membership to these structures \citep{fre02}. 
Studies of open clusters such as the Hyades and Collinder 261 \citep{pau03, sil06, sil07a, sil09}
have found high levels of chemical homogeneity, showing that chemical information is preserved within
the stars and  that the possible effects of any external sources of pollution
are negligible. This \textit{chemical tagging} method  has so far only been used in the three old SKGs:
The Hercules stream \citep{ben07} whose stars have different ages and chemistry 
(associated with dynamical resonances namely a bar or spiral structure), HR 1614 \citep{sil07b, sil09}, and Wolf 630 \citep{BubarKing10}  all of which appear to be true MGs
(debris of star-forming aggregates in the disc). In addition, \citet{SoderblomMayor93}, \citet{King03}, \citet{KingSchuler05}, and \citet{Ammler09} studied the Ursa Major MG and demonstrated, in terms of chemical abundances and spectroscopic age determinations, that some of the candidates are consistent with being members of a MG with a mean [Fe/H] =-0.085. 
The  \textit{chemical tagging} method has been applied to some smaller samples of possible Hyades SC members. 
\citet{pom11} studied a sample of 21 kinematically selected stars and identified two candidate members of the Hyades SC, whereas 
 \citet{sil11} analysed 26 southern giant candidates finding also four candidate members. These results show that  the Hyades Supercluster may not originate uniquely from the Hyades cluster.
In this paper,  we apply the \textit{chemical tagging} method to
 a homogeneous sample of 61 F6-K4 northern kinematically selected Hyades Supercluster (Hyades stream, or Hyades moving group) candidates most of whom are main sequence stars,  although some are giant stars.
In Sect.~2, we give details of our sample selection.  Our observations and data reduction methodology are described in Sect.~3.
Descriptions of our codes to determine the stellar parameters and chemical abundances  are provided in Sect.~4.
The results about the absolute and differential abundances are given in Sect.~5. Finally in Sect.~6, we summarize our
results for the membership of the Hyades Supercluster derived from the  \textit{chemical tagging} method.

\begin{figure}
%\centering 
\centerline{
\includegraphics[scale=0.50]{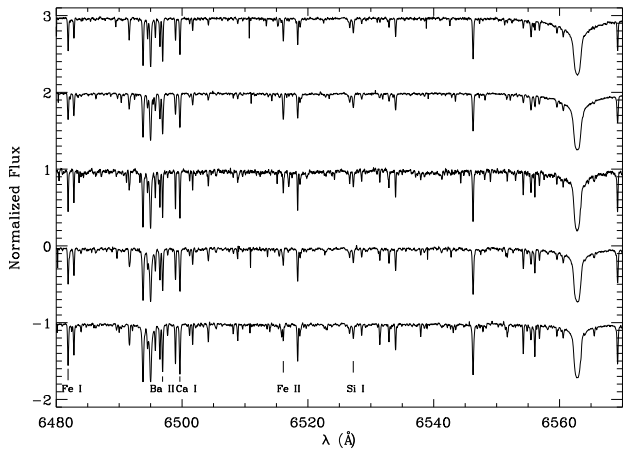}
}
%[bb= 40 83 500 657, clip=true, angle=90, scale=0.40]{tabernero_h_fig2.eps}
\caption{High resolution spectra for some representative stars from our sample (from top to bottom):
HD 27285 (G4 V), HD 53532 (G3 V), HD 98356 (K0 V), these three stars satisfy {\it chemical tagging} membership conditions, vB~153 (a K0 V reference star used in the differential abundances analysis) and BZ Cet (a K2 V, also member of the Hyades cluster). Lines used in the abundance analysis are highlighted on bottom. }
\label{spec}
\end{figure}

Identification of a significant number of late-type members of these young MGs
would be extremely important for a study of their chromospheric and
coronal activity and their age evolution, and could lead to a
clearer understanding of the star formation history in the solar neighborhood.
In addition, these stars can also be  selected as
targets for direct  imaging campaigns to detect of sub-stellar companions
(brown dwarfs and exoplanets).

\section{Sample selection}

The sample analyzed in this paper (see Table~\ref{tablavel}) was selected using kinematical criteria based on the $U$, $V$ and $W$ Galactic velocities of a chosen target being within  approximately 10 km s$^{-1}$ of the mean velocity of the group \citep{mon01a}. \\
We also selected additional candidates and spectroscopic information about some of these stars from \citet{lop10}, \citet{mar10}, and \citet{mal10}. Some exoplanet-host star candidates were also taken from \citet{mon10}.\\

After the first stage of selection based on kinematical criteria, we then eliminated stars that were unsuitable for our particular abundance analysis, namely stars cooler than K4 and hotter than F6, because for these stars we would have been unable to measure the spectral lines required for our particular abundance analysis. We  also discarded  stars with high rotational velocities,  as well as those known to be spectroscopical binaries to avoid contamination from the companion star during the analysis.\\

\begin{figure*}
\centering \includegraphics[scale=0.70,angle=270]{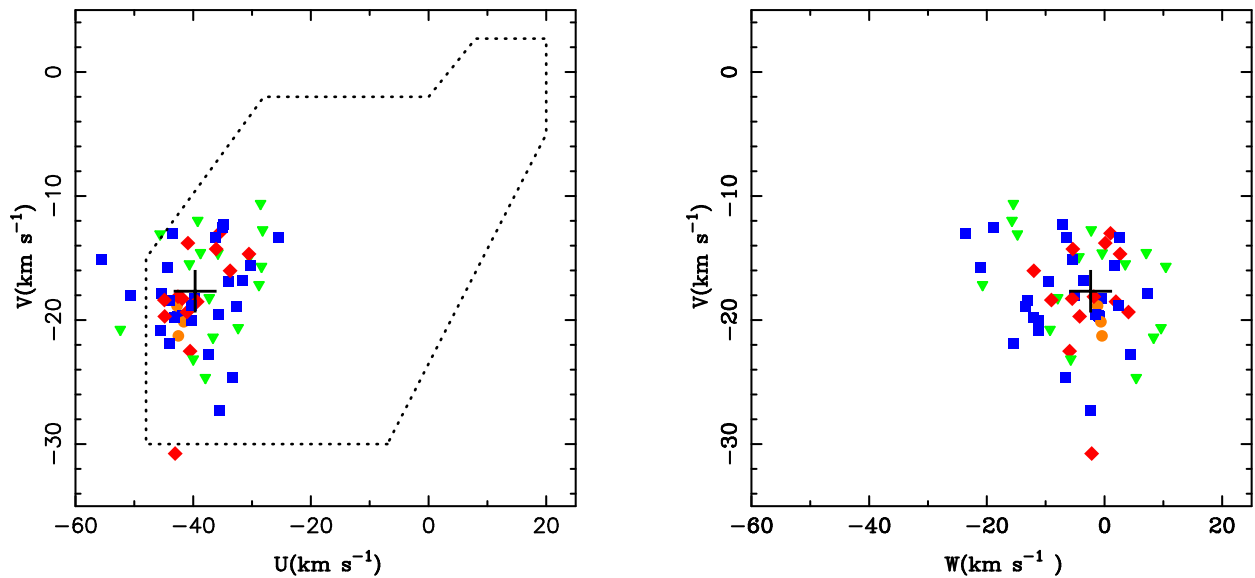}
\caption{$U$, $V$, and $W$ recalculated velocities for the possible members of the Hyades Supercluster. Blue  squares represent  stars selected as members by the {\it chemical tagging} approach, red diamonds represent  stars that have similar Fe abundances,  but not for all the elements. Orange circles  represent three Hyades cluster stars (BZ Cet, V683 Per, and $\epsilon$ Tau). Green triangles represent stars that do have similar Fe abundances (as well as similar values of other elements). The  big black cross indicate the $U$, $V$, and $W$ central location of the Hyades Supercluster \citep[see ][]{mon01a}. Dashed lines show the region where the majority of the young disk stars tend to be according to \citet{egg84,egg89}.}
\label{uvwhyad}
\end{figure*}

We recalculated the Galactic velocities of our selected targets by employing the radial velocities and uncertainties derived by the HERMES spectrograph automated pipeline \citep{ras11}. However, for some spectra  were taken when the automated radial velocity was not available, in these cases, we applied the cross-correlation method using the routine {\sc fxcor} in IRAF \footnote{IRAF is distributed by the National Optical Observatory, which is operated by the Association of Universities for Research in Astronomy, Inc., under contract with the National Science Foundation.}, by adopting a spectrum of the asteroid Vesta (as a solar reference) that had been previously corrected for Doppler shift with the Kurucz solar {\it ATLAS} \citep{kur84}. The radial velocities were derived after applying the heliocentric correction to the observed velocity. Radial velocity errors were computed by {\sc fxcor} based on the fitted peak height and the asymmetric noise, as described by \citet{td79}.
The obtained radial velocities and their associated errors are given in Table~\ref{tablavel}. Proper motions and  parallaxes were taken from the $Hipparcos$ and Tycho catalogues \citep{esa97}, the Tycho-2 catalogue  \citep{hog00}, and  
the latest reduction of the $Hipparcos$ catalogue \citep{lee07}.\\
The method employed in the determination of the $U$, $V$, and $W$ velocities is the same  as that  used in \citet{mon01a}. The Galactic velocities are in a right-handed coordinated system (positive in the directions of the Galactic center, Galactic rotation, and the North Galactic Pole, respectively). \citet{mon01a}  modified the procedures of
 \citet{jo87} to perform the velocity calculation and associated errors. This modified program uses  coordinates adapted to the epoch J2000 in the International Celestial Reference System (ICRS). The recalculted velocities are given in Table~\ref{tablavel} and  plotted in the $U$, $V$, and $W$ diagram in Fig.~\ref{uvwhyad}.

\section{Observations}

Spectroscopic observations (see Fig.~\ref{spec}) were obtained at the 1.2 m Mercator Telescope at the \emph{ Observatorio del Roque de los Muchachos}  (La Palma, Spain) in January, May, and November 2010 with HERMES \citep[High Efficiency and Resolution Mercator Echelle Spectrograph, ][]{ras11}.
Using the high resolution mode (HRF), the spectral resolution is 85000 and the wavelength range is from $\lambda$3800~{\AA} to $\lambda$8750~{\AA} approximately. Our signal-to-noise ratio ($S/N$) ranges from 70 to 300 (160 on average) in the $V$ band. A total of 92 stars were observed. We analyzed single main-sequence and giant stars (from F6 to K4), including 61 candidates in total \citep[including the Hyades cluster members BZ Cet, V683 Per, and  $\epsilon$ Tau, see][]{per98} and the reference used in the differential abundance analysis (vB~153).
All the obtained \emph{echelle} spectra were reduced with the automatic pipeline \citep{ras11} for HERMES. We later  used several IRAF tasks to normalize the spectra order  by order, using a low-order polynomial fit to the observed continuum, merging the orders into a unique  one-dimensional spectrum and applying the Doppler correction required for its observed radial velocity. When several exposures had been performed for the same star we combined all of the individual spectra to obtain a unique spectrum with higher $S/N$.

\section{Spectroscopic analysis}

\subsection{Stellar parameters}

Stellar atmospheric parameters and abundances were computed using the 2002 version of the MOOG code \citep{sne73} and a grid of Kurucz ATLAS9 plane-parallel model atmospheres \citep{kur93}. As damping prescription, we used the Uns$\ddot{{\rm o}}$ld approximation multiplied by a factor recommended by the Blackwell group (option 2 within MOOG). The atmospheric parameters were inferred from 263 $\ion{Fe}{i}$ and 36 $\ion{Fe}{ii}$ lines \citep[the iron lines and their atomic parameters were obtained from ][]{sou08} iterating until the slopes of $\chi$ versus (vs.) $\log{\epsilon(\textrm{\ion{Fe}{i}})}$ and $\log{(EW / \lambda)}$ vs. $\log{\epsilon(\textrm{\ion{Fe}{i}})}$ were zero (excitation equilibrium) and imposing ionization equilibrium, such that $\log{\epsilon(\textrm{\ion{Fe}{i}})} = \log{\epsilon(\textrm{\ion{Fe}{ii}})}$. However, we imposed that the [Fe/H] obtained from the iron lines matched the model metallicity. To simplify the iterative procedure, we built an automatic code called {\scshape StePar} that employs a Downhill Simplex Method \citep{pre92}. The function to minimize is a quadratic form composed of the excitation and ionization equilibrium conditions. The code performs a new simplex optimization until the model's metallicity and the iron abundance are the same. The {\scshape StePar} code finds the best solution in the paramater space within minutes. The obtained solution for a given star is independent of the initial set of parameters employed, hence we used the canonical solar values as initial input values ($T_{\rm eff}$=5777 K, $\log{g}$=4.44 dex, $\xi$=1 $km$ $s^{-1}$).\\

The uncertainties in the stellar parameters were determined as follows: for the microturbulence, we changed $\xi$ until the slope of $\log{\epsilon(\textrm{\ion{Fe}{i}})}$ vs. $\log{(EW / \lambda)}$ varied with its error (divided by the square root of the number of $\ion{Fe}{i}$ lines). We varied the effective temperature until the slope $\log{\epsilon(\textrm{\ion{Fe}{i}})}$ vs. $\chi$ increased up to its own error (divided by the square root of the number of $\ion{Fe}{i}$ lines). By increasing $\xi$ on its error, we recomputed the effective temperature, these two sources of error are added in quadrature.

We varied the surface gravity until the $\ion{Fe}{ii}$ abundance increases by a quantity equal to the  standard deviation divided by the square root of the number of $\ion{Fe}{ii}$ lines. We also took into account the previous errors in $\xi$ and $T_{\rm eff}$ by varying these quantities separately, thus recomputing the gravity. These differences  were later added in quadrature.
Finally, for the Fe abundance, we varied the stellar atmospheric paramaters in their respective errors, thus adding all the variations and the standard deviation in quadrature.

\begin{figure}
\centering
\includegraphics[scale=0.53]{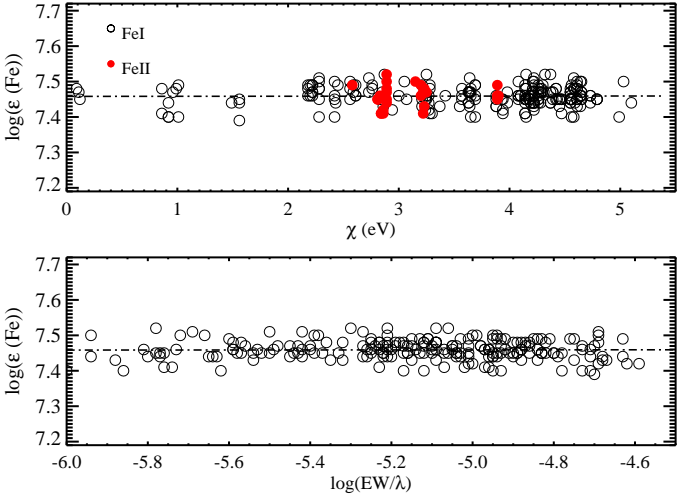} 
\caption{ $\log{\epsilon(\ion{Fe}{i,ii})}$ versus the excitation potential, $\chi$, and the reduced equivalent width, $\log{(EW / \lambda)}$, for the Sun (spectrum of the asteroid Vesta).The dashed lines represent the least squares fit to the data points, which are very close to a constant fit as expected from the iterative procedure of the stellar parameter determination. }
\label{grafpardet}
\end{figure}

The $EW$ determination of the Fe lines was carried out with the ARES\footnote{The ARES code can be downloaded at http://www.astro.up.pt/} code \citep{sou07}. We followed the approach of \citet{sou08} to adjust the $rejt$ parameter of ARES according to the $S/N$ of each spectrum. The other ARES parameters we  employed were $smoother$ = 4, $space$ = 3, $lineresol$ = 0.07, and $miniline$ = 2.

In addition, we performed a 2-$\sigma$ rejection of the $\ion{Fe}{i}$ and $\ion{Fe}{ii}$ lines after a first determination of the stellar parameters, therefore we re-run the {\scshape StePar} program again without the rejected lines.\\
As a test, we  performed the parameter determination in the case of the Sun. Employing a HERMES spectrum of the asteroid Vesta with the same instrumental configuration as the other stars. In this case, we obtained: $T_{\rm eff}=$ 5775 $\pm$~15 (K), $\log{g}=$4.48 $\pm$~0.04 (dex), $\xi=$0.965 $\pm$~0.020 ($km$ $s^{-1}$), and $\log{\epsilon(\textrm{\ion{Fe}{i}})}=$7.46 $\pm$  0.01~(dex), which are very close to the canonical solar values of the atmospheric parameters.
We ended up with 220 $\ion{Fe}{i}$ and 27 $\ion{Fe}{ii}$ spectral lines for this solar reference. To illustrate this iterative procedure we present in Fig.~\ref{grafpardet} a representation of both $\log{\epsilon(\ion{Fe}{i,ii})}$ vs. $\chi$ and  $\log{\epsilon(\ion{Fe}{i})}$ vs. $\log{(EW / \lambda)}$. In both panels of Fig.~\ref{grafpardet} the null slope is indistinguisable from the mean. In the case of $\ion{Fe}{ii}$  we are only interested in its mean abundance value, which  it is the same as the mean abundance value obtained from the $\ion{Fe}{i}$ lines. \\
The derived stellar parameters for our solar reference are used as a zero-point. The determined abundances are presented with respect to our solar values in  a self-consistent manner.

\begin{figure}[hbt]
\centering
\includegraphics[scale=0.5]{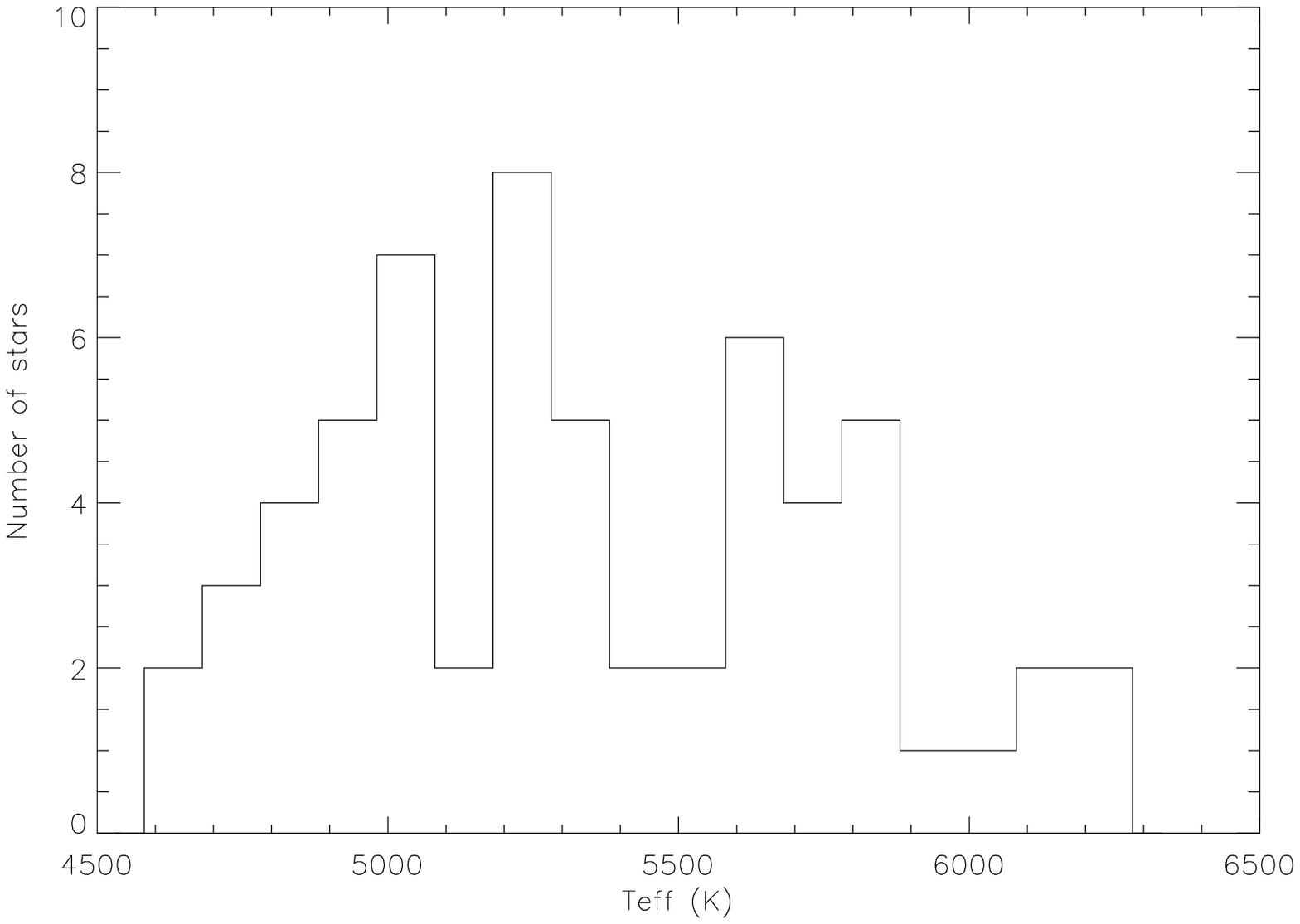}
\includegraphics[scale=0.5]{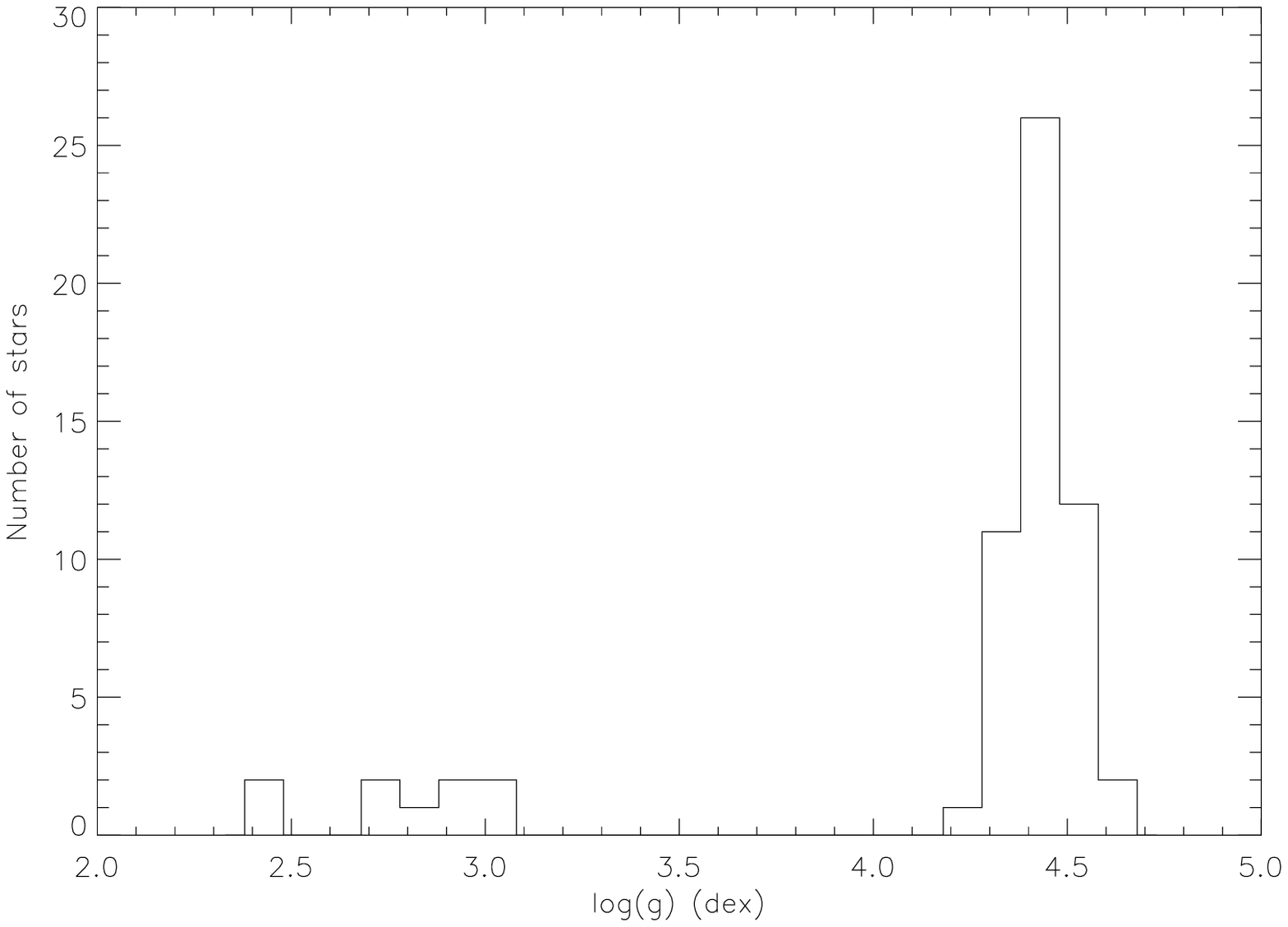}
\caption{ Histograms for the determined $T_{\rm eff}$ and $\log{g}$ of the candidate stars.}
\label{histpar}
\end{figure}

The obtained stellar parameters $T_{\rm eff}$, $\log{g}$, $\xi$, $\log{\epsilon(\ion{Fe}{i})}$, $\log{\epsilon(\ion{Fe}{ii})}$, and [Fe/H] (using our solar reference) are given in Table~\ref{tablapar}, together with  the internal uncertainties in the stellar parameters. In Fig.~\ref{histpar}, we show the histrogram distribution of $T_{\rm eff}$ and $\log{g}$ which is also available in tabular form in the online version. The effective temperature ranges approximately from 4500 K to 6300 K. The surface gravities of most stars in the sample are those typical main sequence stars (52 out of 61),  and the rest are low gravity stars. 

\begin{figure}
\centering
\includegraphics[scale=0.5]{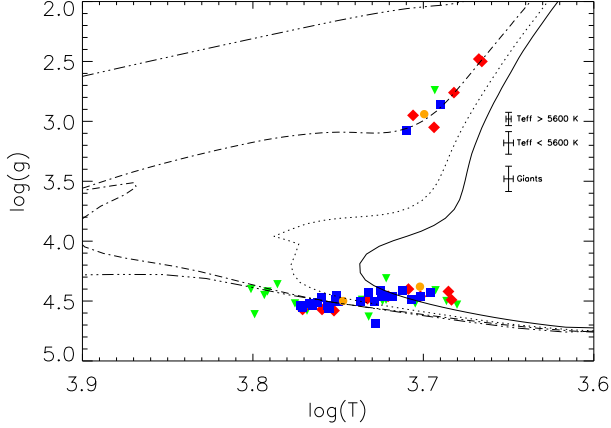}
\caption{Spectroscopic $\log{T_{\rm eff}}$ vs. $\log{g}$ for the candidate stars. We have employed the Yonsei-Yale ischrones \citep{dem04} for Z=0.025 and 0.1, 0.7, 4,
and 13 Gyr (from left to right). Mean error bars are represented at the middle right. Blue squares represent stars selected as members by the {\it chemical tagging} approach, red diamonds represent those stars that have similar Fe abundances, but different values of other elements. Orange circles represent three Hyades cluster stars (BZ Cet, V683 Per and $\epsilon$ Tau). Green triangles represent those stars that do not have similar Fe abundances (as well as dissimilar values of other elements).}
\label{hyadesiso}
\end{figure}

As a complementary stellar parameter test, we compiled a $\log{g}-\log{T_{\rm eff}}$ diagram by employing the determined stellar parameters in Fig.~\ref{hyadesiso}. Most of the giants and the solar-like stars tend to fit within the depicted isochrones, being consisent with the isochrone  for 0.7 Gyr  \citep[the accepted Hyades age, see][]{per98}. In constrast, some cooler stars tend to deviate 0.1 dex from the ischrones, particulary at the lowest temperatures. However, the mean gravity error for these cooler stars is about 0.1 dex, being compatible with the Hyades isochrone within their error bars. On the other hand, the error may be systematic, the $EW$s of the $\ion{Fe}{ii}$ lines having larger systematic errors, since they get weaker as $T_{\rm eff}$ drops. This may in turn lead to an underestimation of the gravity derived assuming excitation equilibrium. The effect on the $\ion{Fe}{ii}$ lines may propagate as an increasing understimation of $\log{g}$.
\subsection{Chemical abundances}

The selection of the chemical elements that we considered in this study includes those in the line list of  \citet{nev09}, which also provides the atomic parameters for each line. In addition we considered  some lines from \citet{gon10} and \citet{pom11} of some  neutron-capture elements (their atomic parameters are given in Table~\ref{linerstab}). 

Chemical abundances were calculated using the equivalent width ($EW$) method. The $EW$s were determined with the ARES code \citep{sou07} following the approach described  in Sect.~4.1.\\
Once the $EW$s had been measured, the analysis was carried out with the LTE MOOG code \citep{sne73} using the ATLAS model corresponding to the derived atmospheric parameters. We determined the elemental abundances (see Tables~\ref{galtab2} and \ref{galtab3}) relative to solar values using the spectrum of the asteroid Vesta (with the same instrumental configuration) as the solar reference. These were determined by computing the mean of the line-by-line differences of each chemical element and candidate star with respect to our solar reference (see Table~\ref{tablasol} for the solar reference elemental abundances).\\  
\begin{table*}
\scriptsize
\caption{Elemental abundances for our solar reference (spectrum of the asteroid Vesta). The uncertainties are the line-to-line scatter divided by the square root of the number of spectral lines measured. }
\label{tablasol}
\centering
\begin{tabular}{c c c c c c c c c c c c c c c c c c c c c}
\hline\hline
Element & Fe  & Na & Mg & Al & Si & Ca & Sc & Ti & V & Cr & Mn & Co & Ni &  Cu  & Zn  & Y  & Zr & Ba & Ce  & Nd \\
\hline
$\log{\epsilon(X)}$ & 7.46  & 6.37 & 7.64  & 6.44  & 7.55  & 6.34  & 3.19  & 4.99  & 4.00 & 5.66  & 5.41  & 4.91  & 6.26  & 4.33  &  4.54 &  2.17 & 2.61  &  2.35  &  1.61 &  1.47  \\  

$\sigma \log{\epsilon(X)}$ & 0.01 & 0.01 & 0.06 & 0.01 & 0.01 & 0.03 & 0.04  & 0.01  & 0.04 & 0.01  &  0.02 &  0.01  &  0.01 & 0.18  &  0.06 & 0.04  & 0.10  & 0.08 &  0.03 &  0.08   \\

\hline
\end{tabular}
\end{table*}

A total of 20 elements were analyzed: Fe, the $\alpha$-elements (Mg, Si, Ca, and Ti), the Fe-peak elements (Cr, Mn, Co, and Ni), the odd-Z elements (Na, Al, Sc, and V) and the s-process elements (Cu, Zn, Y, Zr, Ba, Ce and Nd), see Tables~\ref{galtab2} and \ref{galtab3}. The spectral lines used and their atomic parameters were taken from \citet{nev09}. However, to avoid incorrect $EW$ measurements (e.g. caused by an incorrect continuum placement), we  rejected  lines that were separated by more than a factor two of the standard deviation ($\sigma$) from the median  differential abundance derived for each line.\\      
The differential abundances (see Table~\ref{diftable2} and \ref{diftable3}) were also determined to establish mermbership of each candidate using a Hyades cluster member. For this purpose, we have obtained a HERMES spectrum of a well-known member of the Hyades cluster (vB~153), which is the same reference star employed in \citet{pau03} and \citet{sil06} in their differential analysis of the Hyades cluster. This differential treatment minimizes errors due to the uncertainities in the oscillator strengths ($\log{gf}$) of each line treated in the analysis. We also computed the abundance sensitivities to changes in the stellar atmospheric parameters (see Table~\ref{tablasense} and \ref{tablasense2}). These sensitivities due to the  internal uncertainties in the stellar parameters are small for all the elements (about a few hundredths of a dex) except for titanium, vanadium, barium, and zirconium, whose sensitivities to variations in the parameters is quite high (tenths of a dex).\\ 

\section{Discussion}

We will compare our derived element abundances with those of thin disc stars \citep{gon10} to determine whether our values follow Galactic trends. We also verify  the chemical homogeneity of the Hyades Supercluster and  whether some of the stars indeed have homogeneus values of  all the considered elements.

\subsection{Element abundances}

\begin{figure*}
\centering
\centerline{ \includegraphics[scale=0.55]{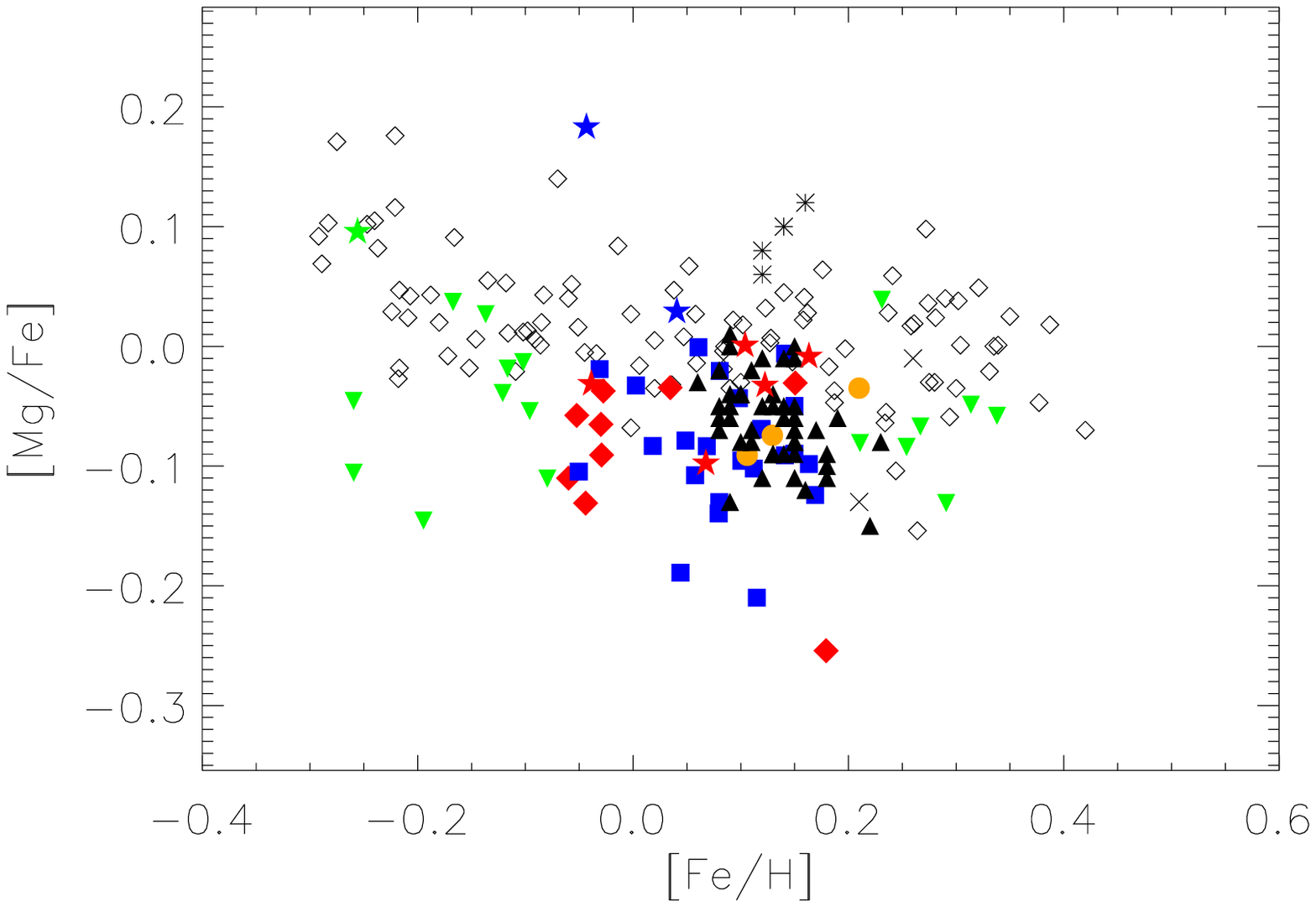}
 \includegraphics[scale=0.55]{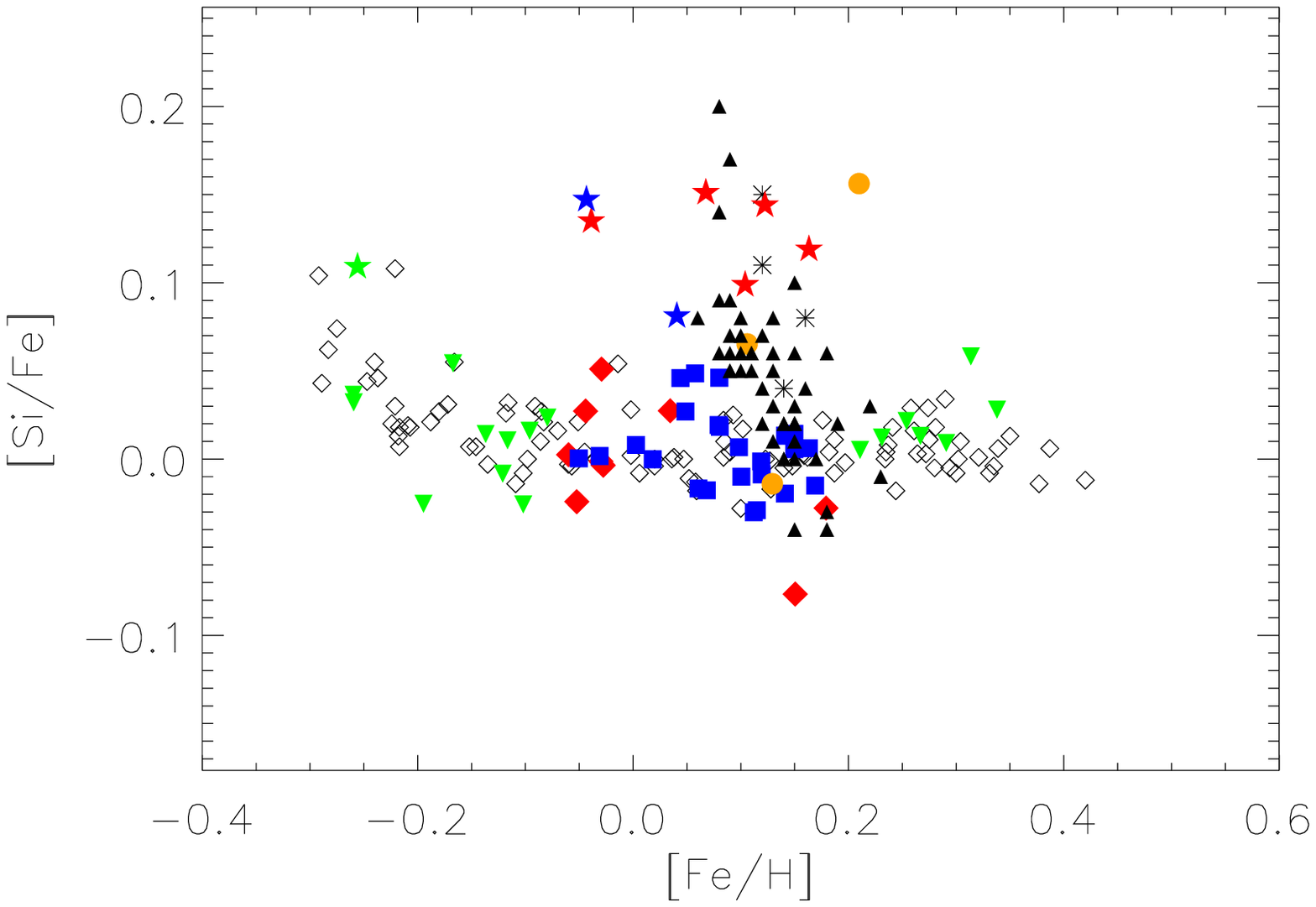}}
\centerline{ \includegraphics[scale=0.55]{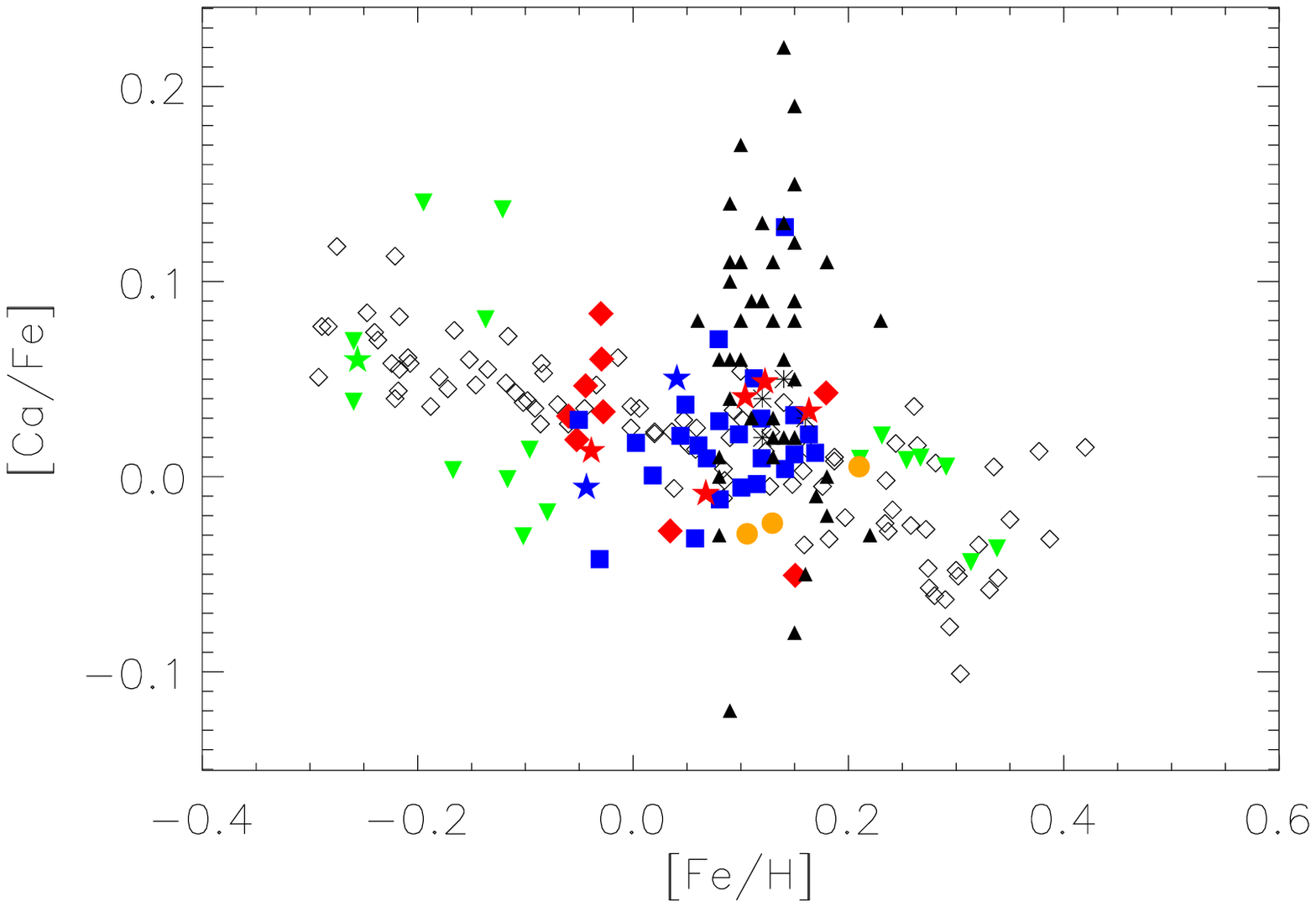}
 \includegraphics[scale=0.55]{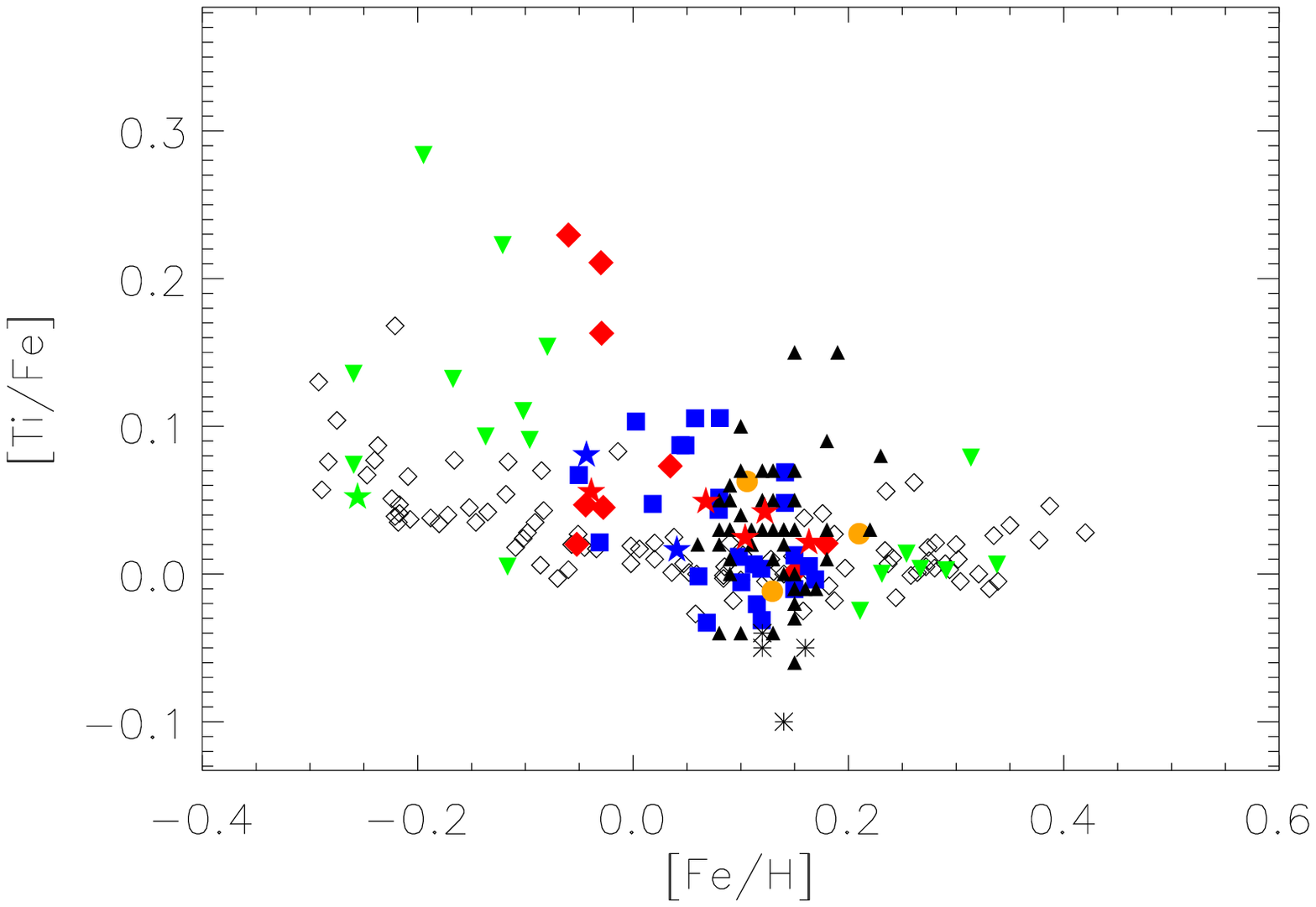}}
\caption{[X/Fe] vs. [Fe/H] for the $\alpha$-elements Mg, Si, Ca, and Ti: open diamonds represent the thin disc data \citep{gon10},  black upward-pointing filled triangles represent Hyades cluster data \citep{pau03}, red  diamonds are our stars compatible to within 1-rms with  the Fe abundance but not for all elements, blue squares and blue starred symbols are the candidates selected to become members of the Hyades Supercluster (see Figs.~\ref{hyadif1}, \ref{hyadif3}, \ref{hyadif4}, and \ref{hyadif5}). Green downward-pointing triangles show no compatible stars. BZ Cet, V683 Per, and $\epsilon$ Tau Hyades cluster members stars are marked with orange circles.  Starred points represent the giant stars. Black asterisks are the candidates selected by \citet{sil11} and black crosses represent the members selected by \citet{pom11}.}
\label{hyagal1}
\end{figure*}

\begin{figure*}
\centering
\centerline{\includegraphics[scale=0.55]{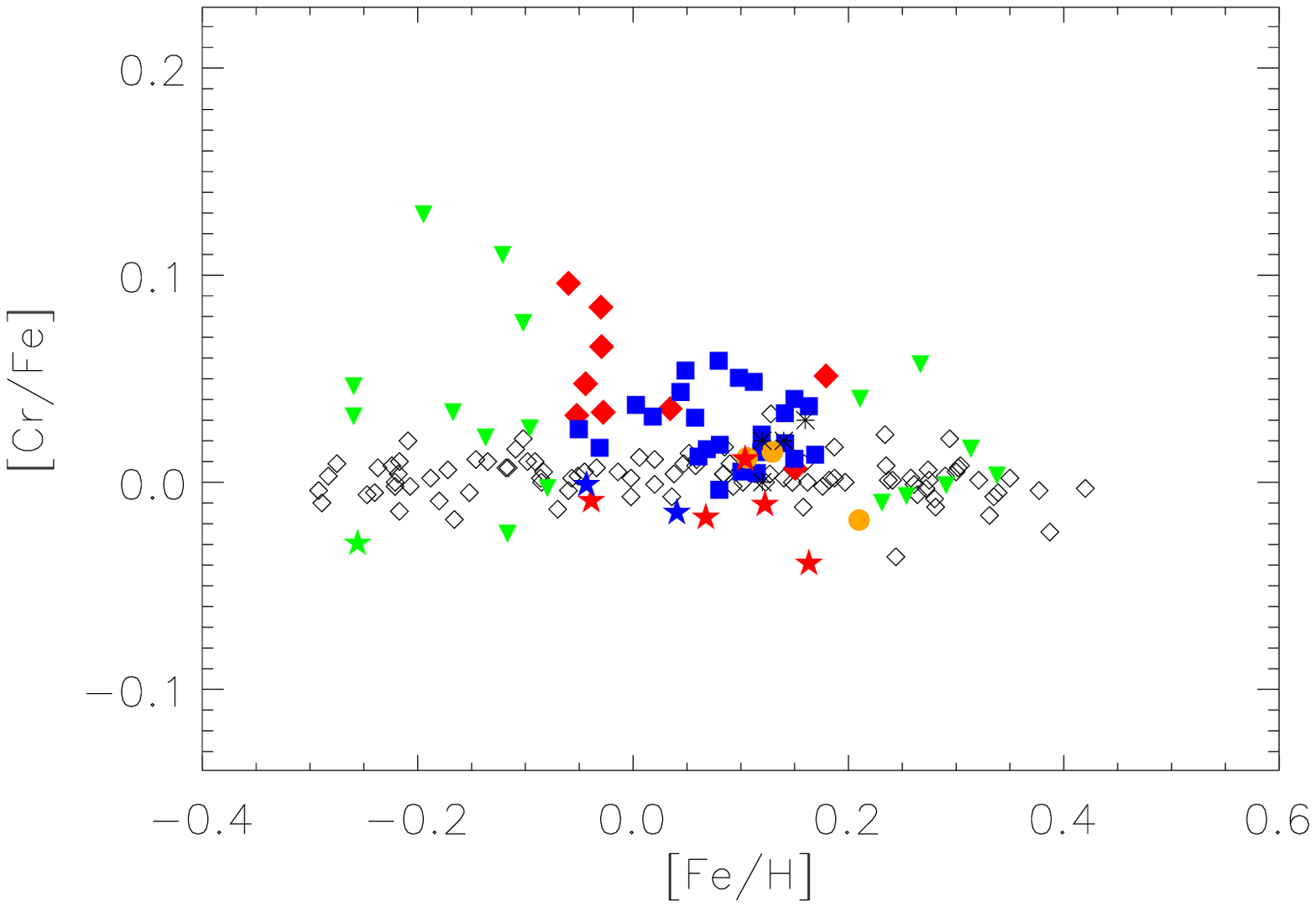}
\includegraphics[scale=0.55]{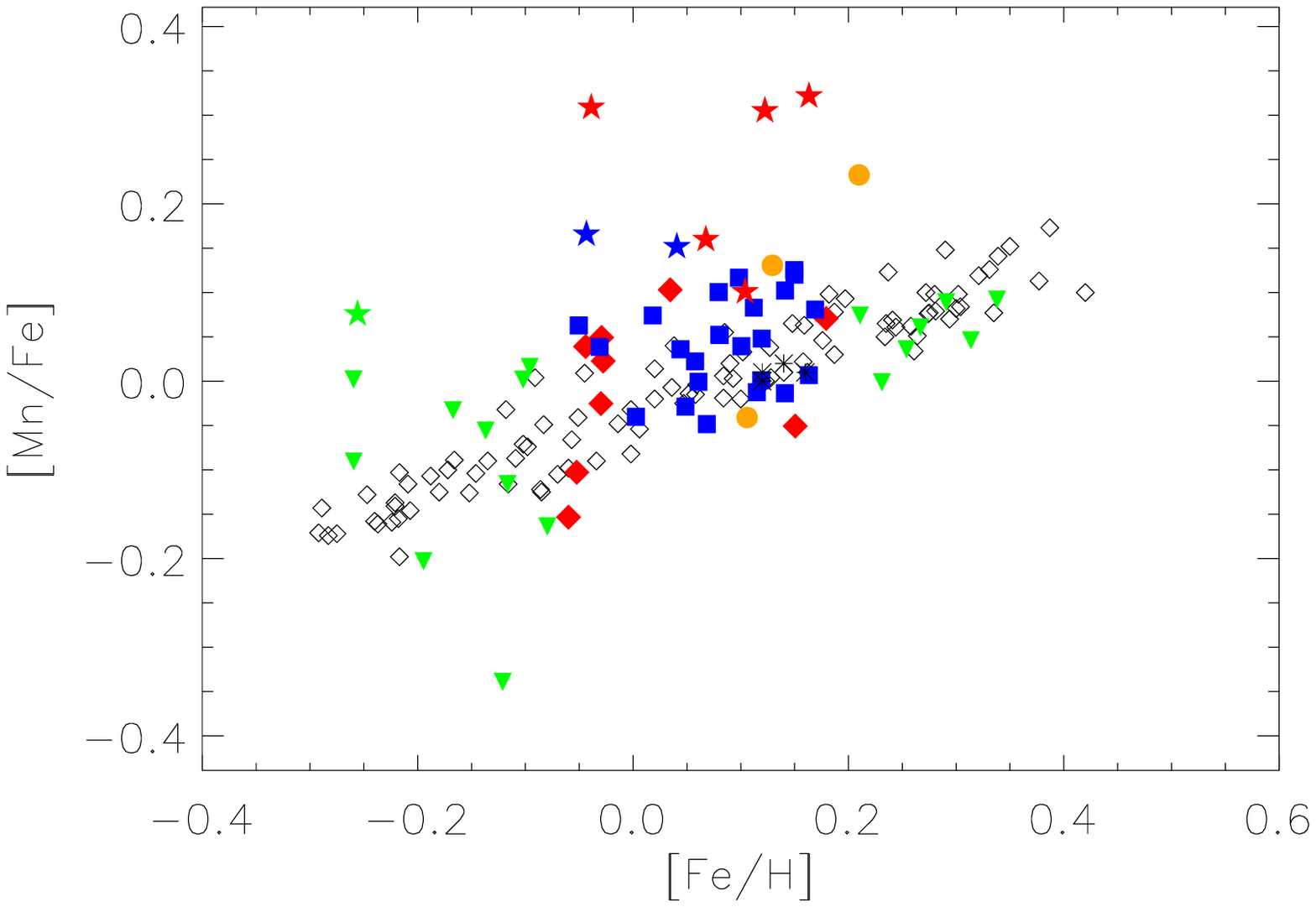}}
\centerline{\includegraphics[scale=0.55]{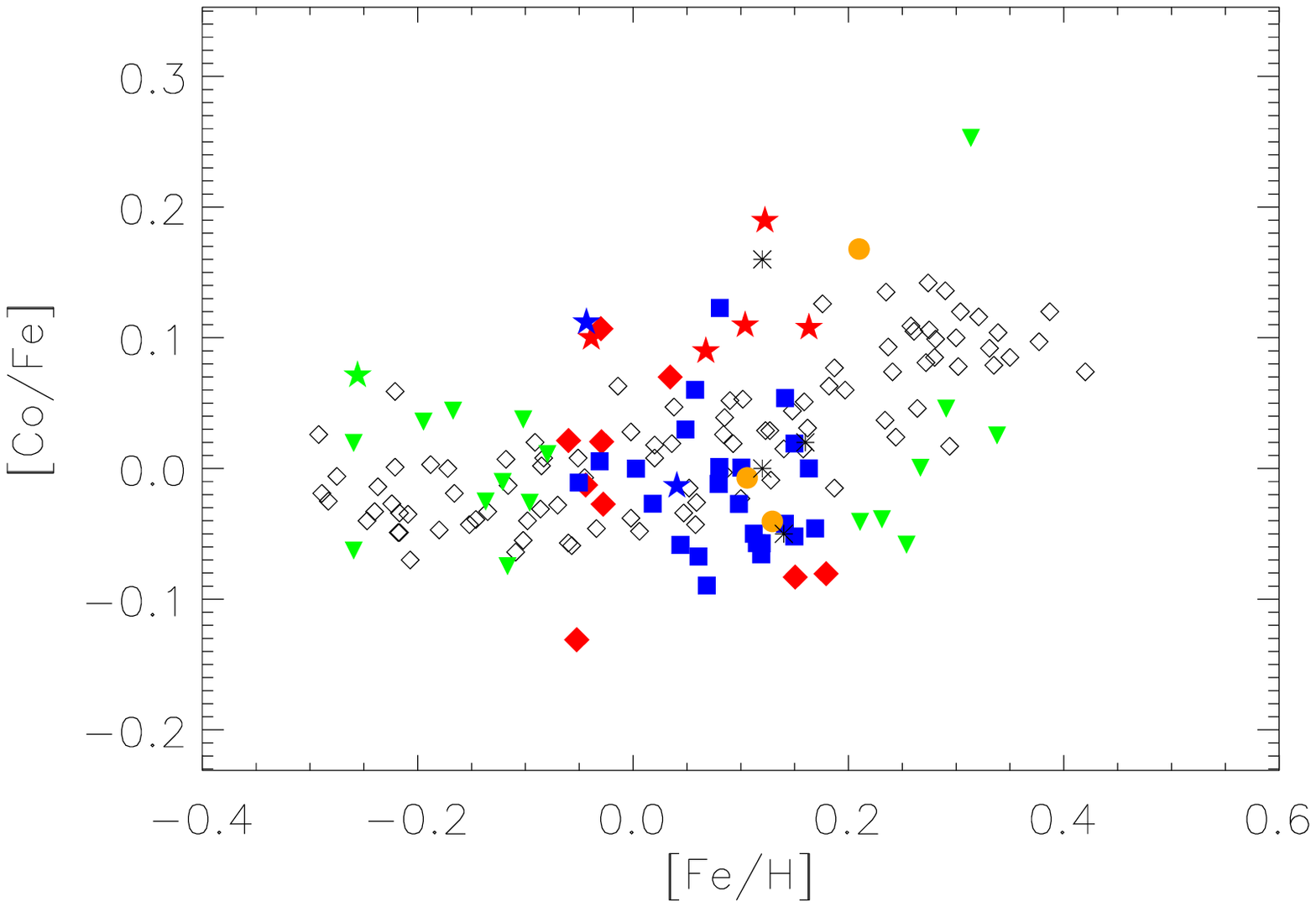}
\includegraphics[scale=0.55]{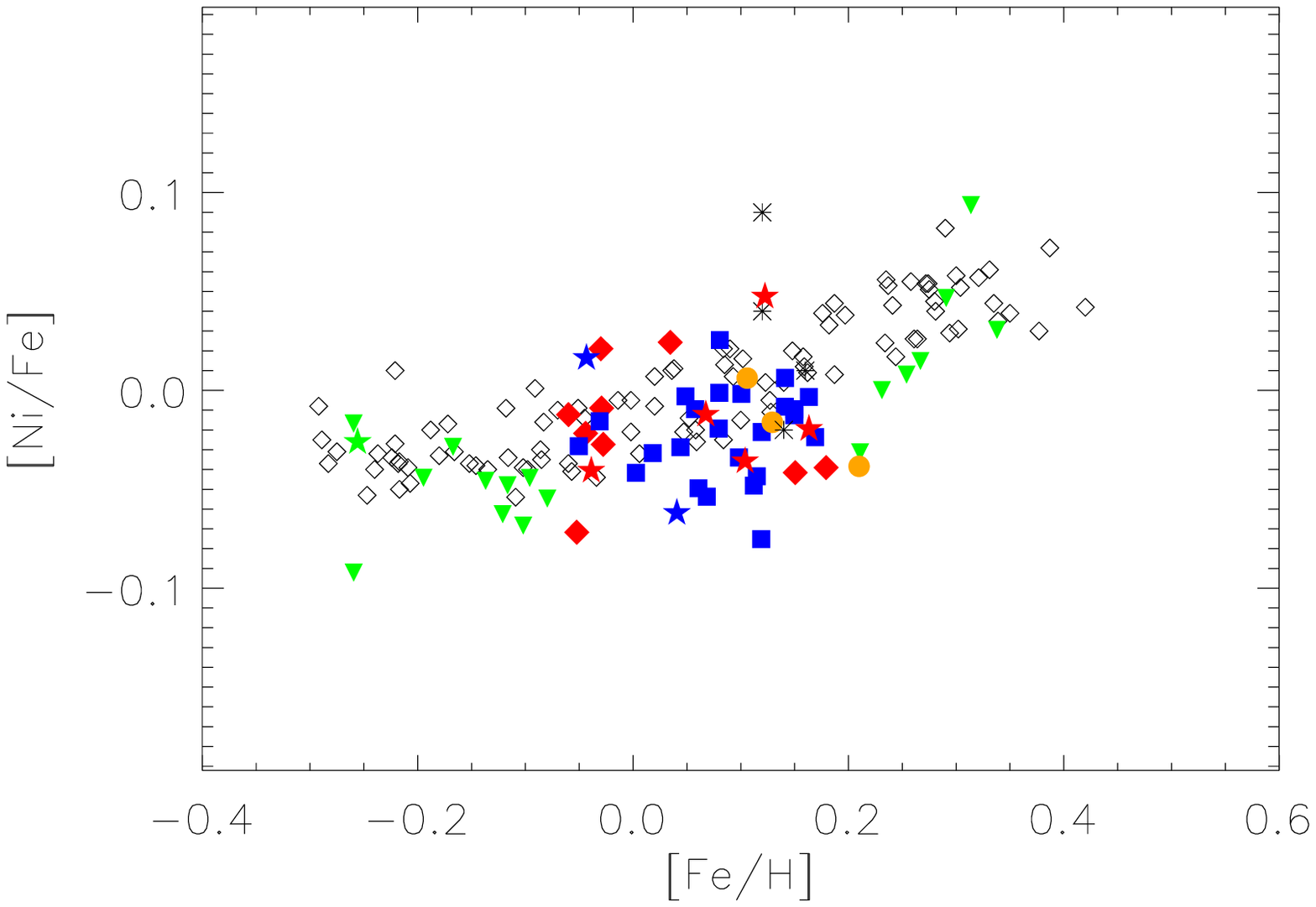}}
\caption{Same as Fig.~\ref{hyagal1} but for the Fe-peak elements Cr, Mn, Co, and Ni.}
\label{hyagal2}
\end{figure*}

The element abundances were determined in a fully differential way by comparing them with those derived for a solar spectrum (as stated in Section~4.1). The choice of elements is taken from \citet{gon10}, which we also compared (see Figs.~\ref{hyagal1}, \ref{hyagal2}, \ref{hyagal3}, \ref{hyagal4}, and \ref{hyagal5}) with the data from \citet{pau03} for the elements in common (Na, Mg, Si, Ca, Ti, and Zn).  Data for those elements in common with \citet{sil11} (Na, Mg, Al, Si, Ca, Sc, Ti, Cr, Mn, Co ,Ni, Zn, Ba, and Ce) and \citet{pom11} (Na, Mg, Zr, Ba, Ce, and Nd) were also added to these figures.\\

The $\alpha$-elements (Mg, Si, Ca, and Ti) seem  to follow the Galactic trends \citep[see][]{ben05,red06,gon10}, although we note that giant stars deviate in the case of Si. There is a noticable scatter in Ti and Mg, but the Ti scatter tends to increase as the [Fe/H] decreases.\\
For the Iron-peak elements (Cr, Mn, Co, and Ni) we find a small scatter in Ni and Cr, although the  scatter in Cr increases at the lowest metallicities in this relatively narrow metallicity range as observed for Ti. For Cr we note that most of the stars lie above the Galactic trend, and that  Co has a larger scatter, and the giants tending to deviate more than the main sequence stars. In the case of Mn, there is a smaller scatter for the main sequence stars but the giants tend to deviate away from the Galactic trend. Ti may be affected by NLTE effects but as reported in \citep[][and references therein]{ben03} the deviations from LTE might be small if the atomic parameters are adecuate. \\
For the odd-Z elements (Na, Al, Sc, and V), the giant stars deviate from the Galactic trend, except for Sc. A high dispersion towards low metallicity is observed for Sc, as well as for Ti and Cr. We confirm a large  dispersion for V, which some authors interpret as a NLTE effect \citep[e.g.][]{bod03,gil06,nev09}. Vanadium lines are indeed difficult to measure and may require  high signal-to-noise data. \\ 

The neutron capture  elements (Cu, Zn, Ba, Ce, Y, Zr, and Nd) follow similar trends to those seen in solar analogs \citep{gon10}. We find some enhancement for Ba above the solar level as observed in the Hyades cluster \citep{sil06}, although the scatter in the data is relatively large. This enhancement is mostly observed around Hyades-like metallicity ([Fe/H]) values. Stars below solar metallicity are also enhanced, however the Galactic abundance distribution has a larger scatter below solar metalliticy. Ce and Y match the Galactic pattern but at lower metallicities the stars of our sample tend to have a larger dispersion. We note here that the dispersion seen in our data is reasonable according to the quality of the data. The Zr abundance values are consistent with the abundance Galactic pattern, but those of the giants are found to deviate from the main Galactic trend. This slight deviation seen for the giants  is also observed  for Cu and Zn,  although the dwarf stars match the Galactic trend,  with the exception of Nd, which is enhanced for almost all the sample stars.

When we compare the element abundance data from \citet{pau03} with our derived [X/Fe] abundances, we find that for the case of Na, Mg, Ti, and Zn some of the sample
stars are consistent with the Hyades cluster abundances. However, there is an small difference for Si and a larger scatter than for Ca. These differences might be caused by the use of different line lists for these elements and the solar reference they used. Candidates obtained by \citet{pom11} are displaced about 0.10 dex in metallicity from the Hyades cluster mean value. This offset is probably caused by their different line lists and their $\log{gf}$ choices.
  
\begin{figure*}
\centering
\centerline{\includegraphics[scale=0.55]{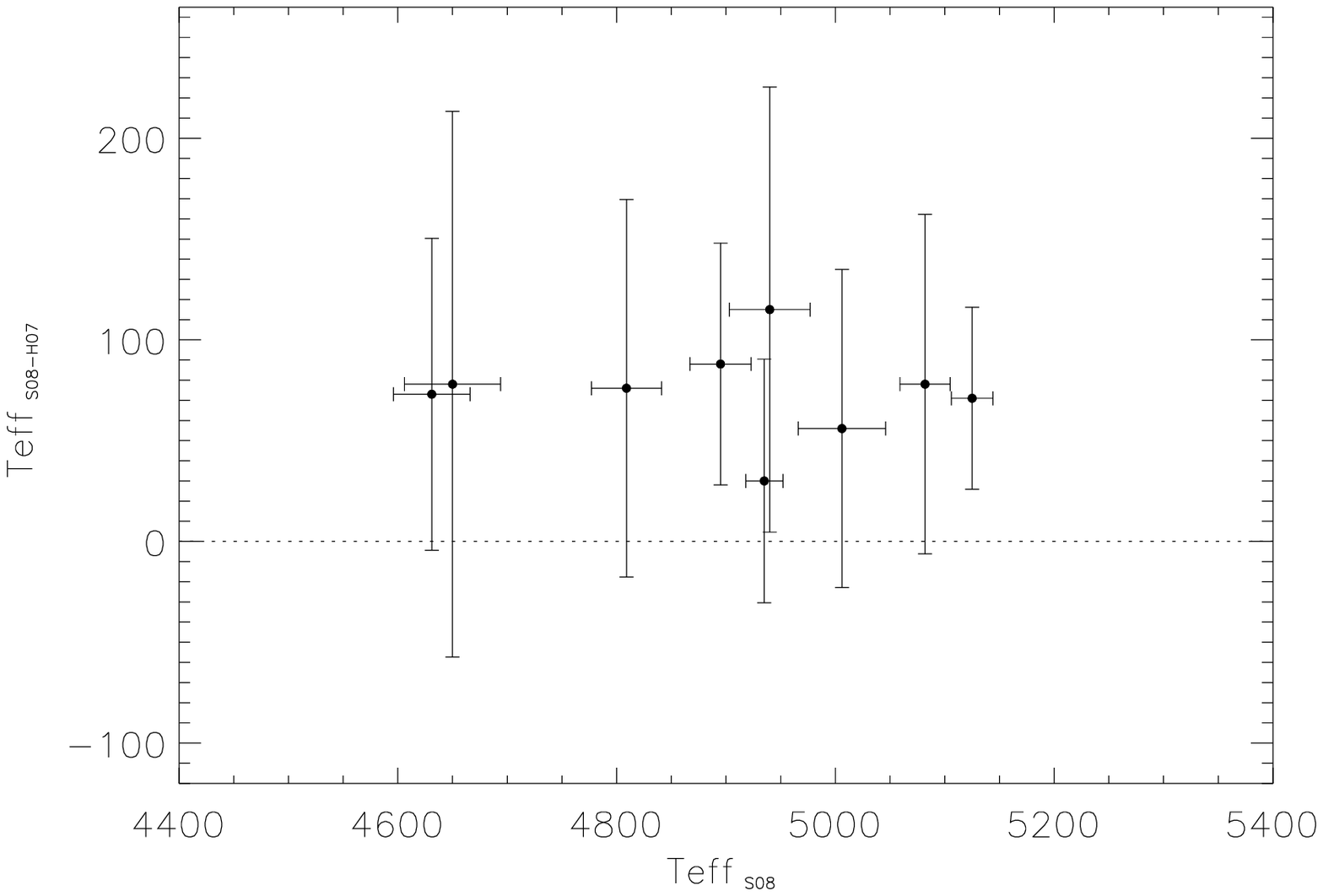}
\includegraphics[scale=0.55]{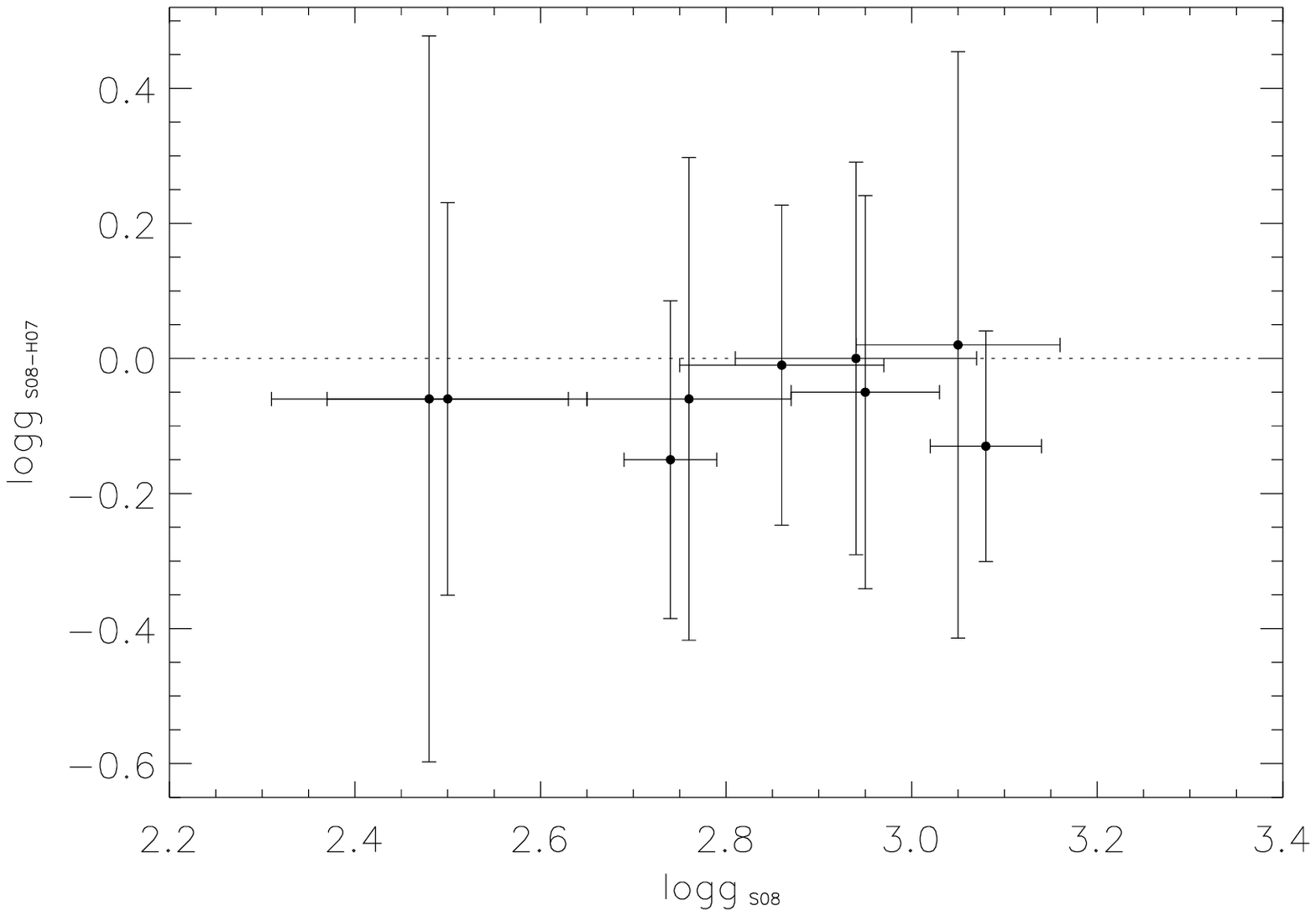}}
\centerline{\includegraphics[scale=0.55]{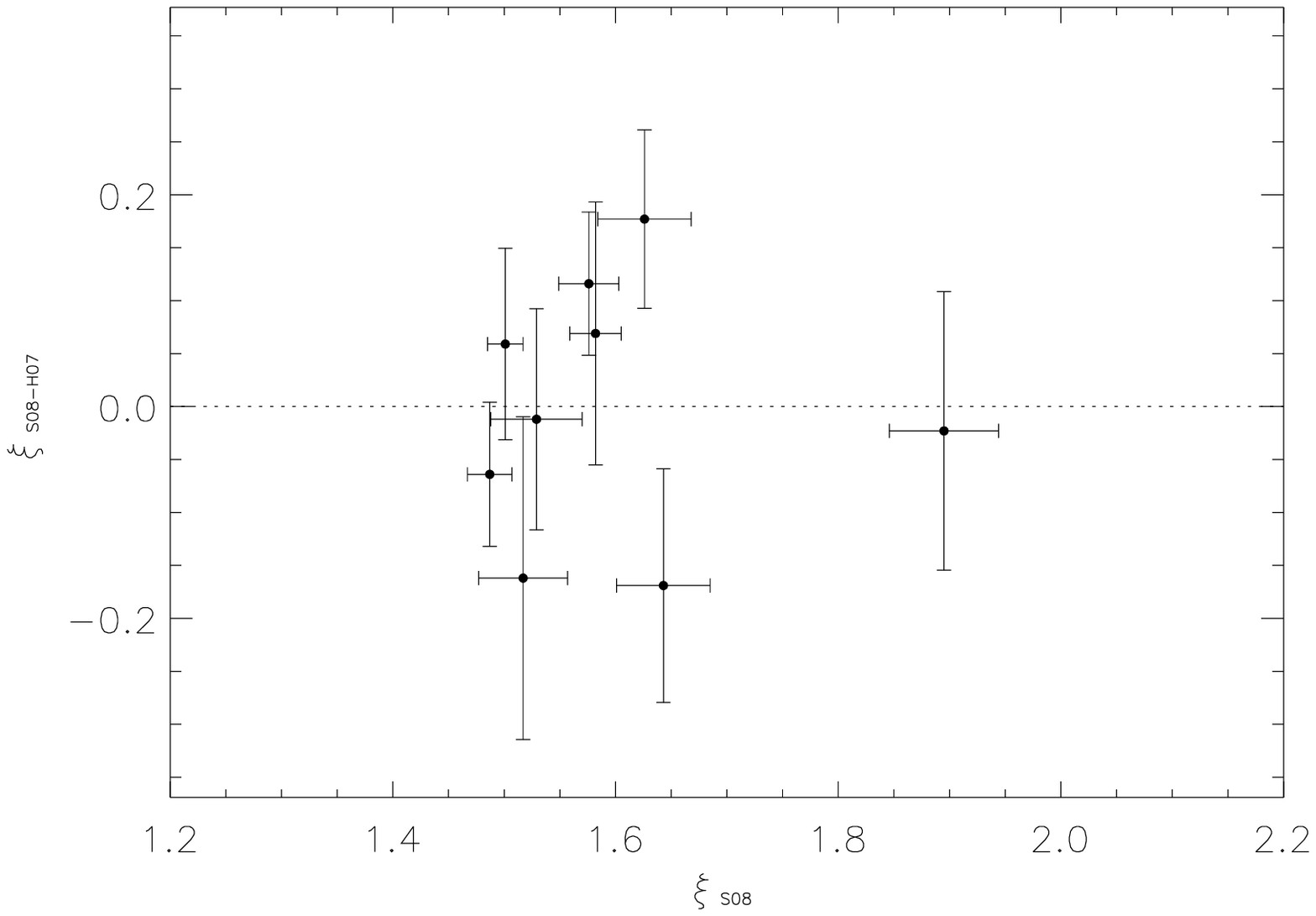}
\includegraphics[scale=0.55]{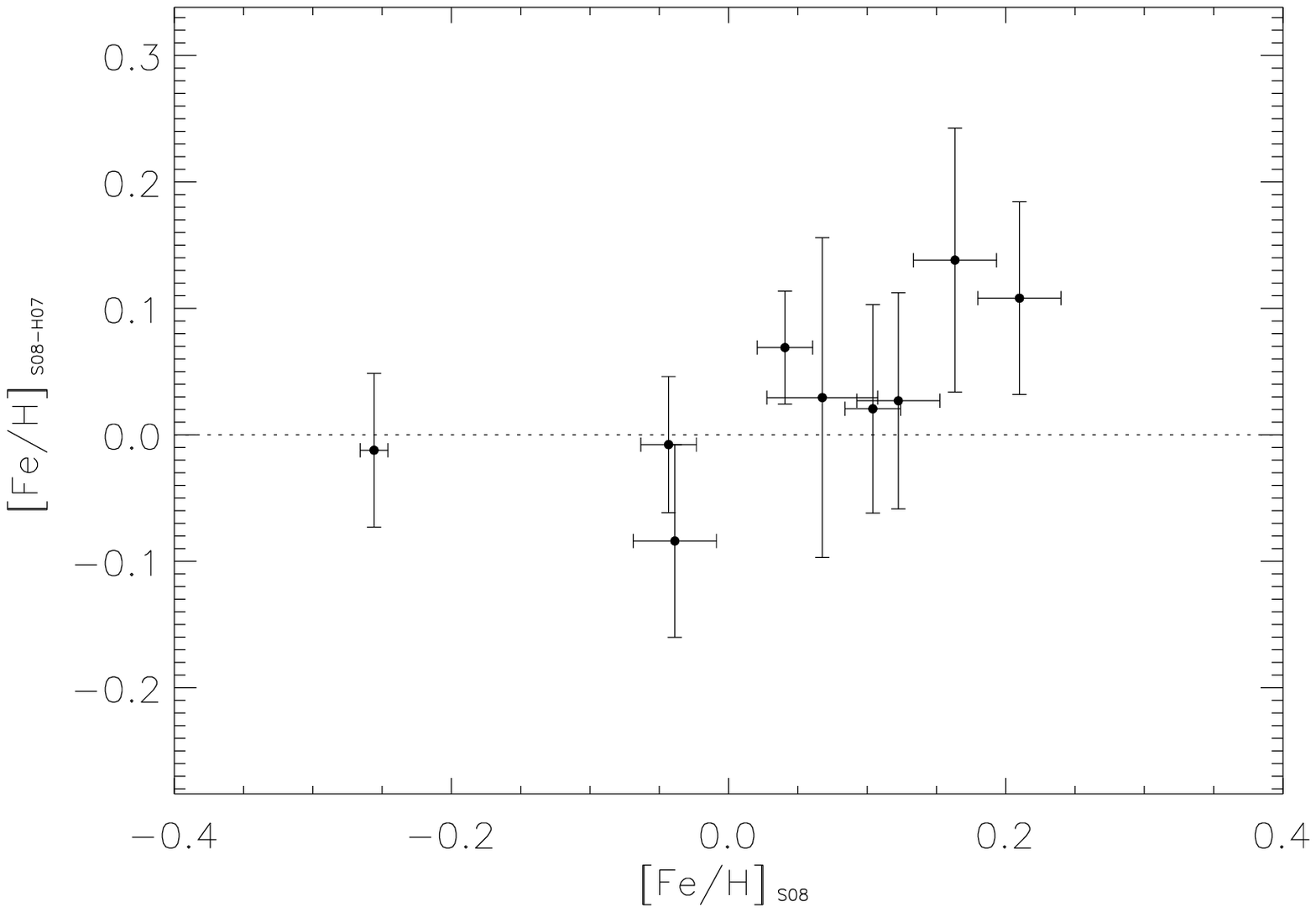}}

\caption{ Comparison of the stellar parameters determined with the \citet{sou08} $\ion{Fe}{i}$-$\ion{Fe}{ii}$ line list versus the list from \citet{hek07}.  }
\label{shpar}
\end{figure*}

\begin{figure*}
\centering
\centerline{\includegraphics[scale=0.55]{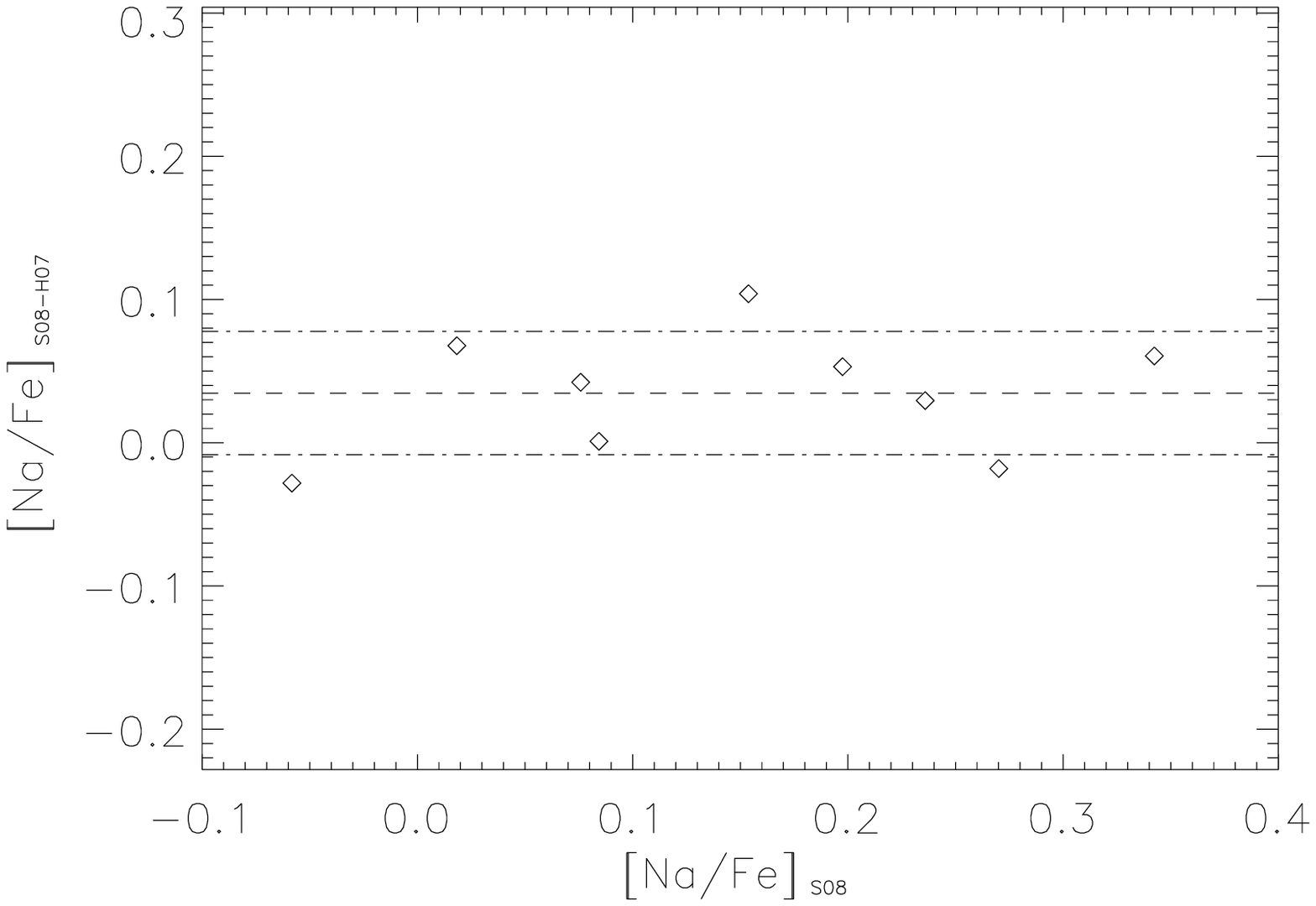}
\includegraphics[scale=0.55]{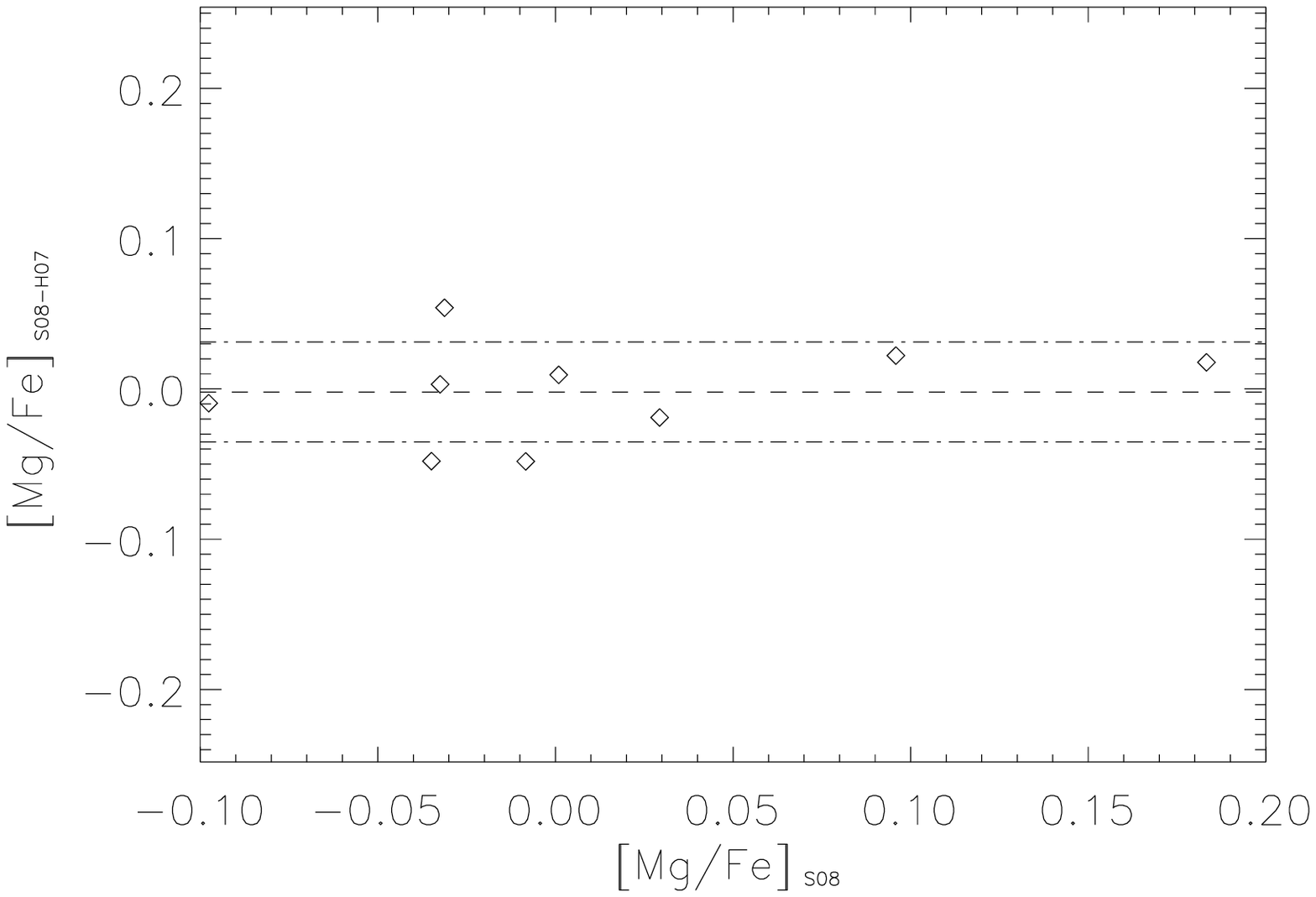}}
\centerline{\includegraphics[scale=0.55]{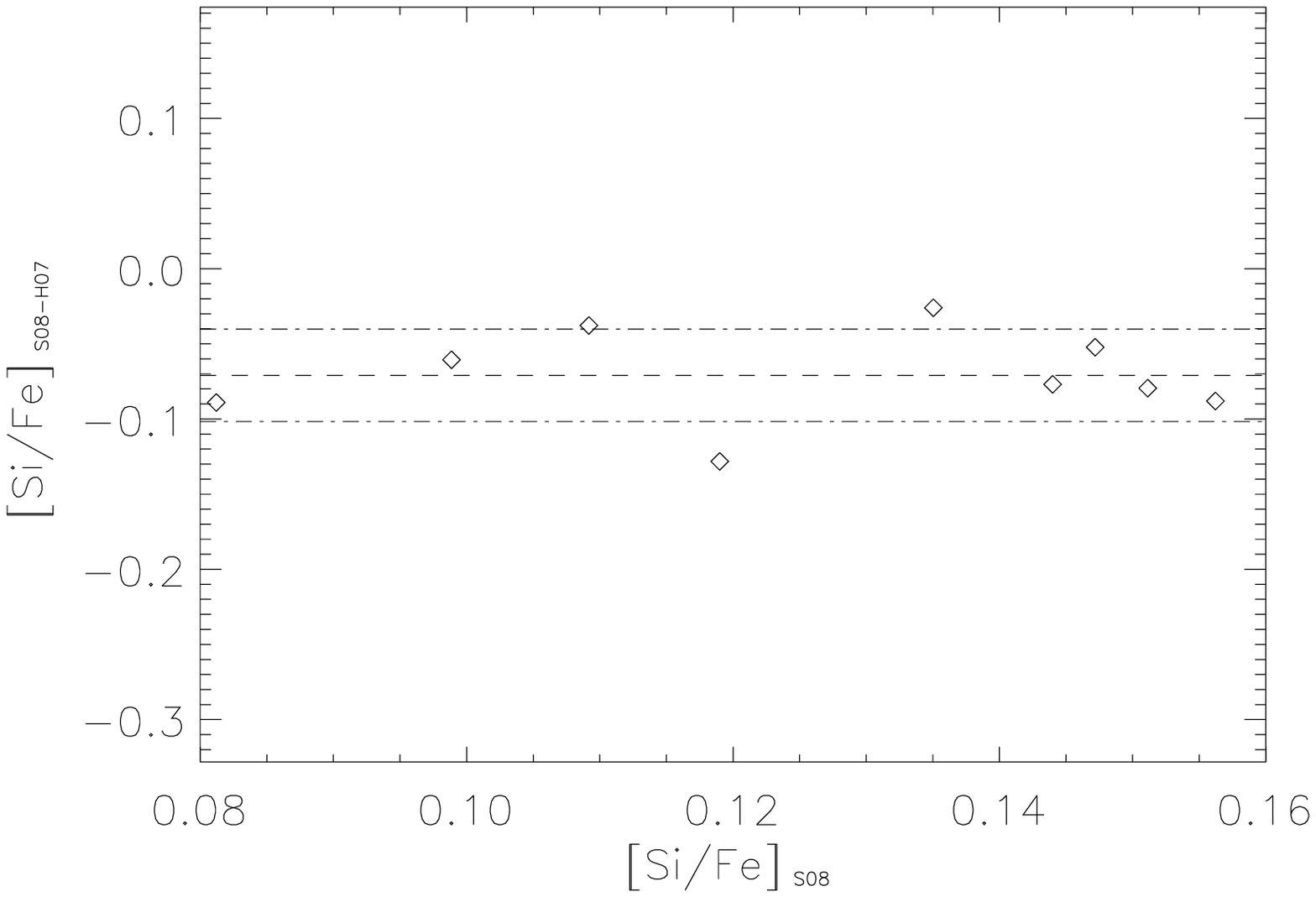}
\includegraphics[scale=0.55]{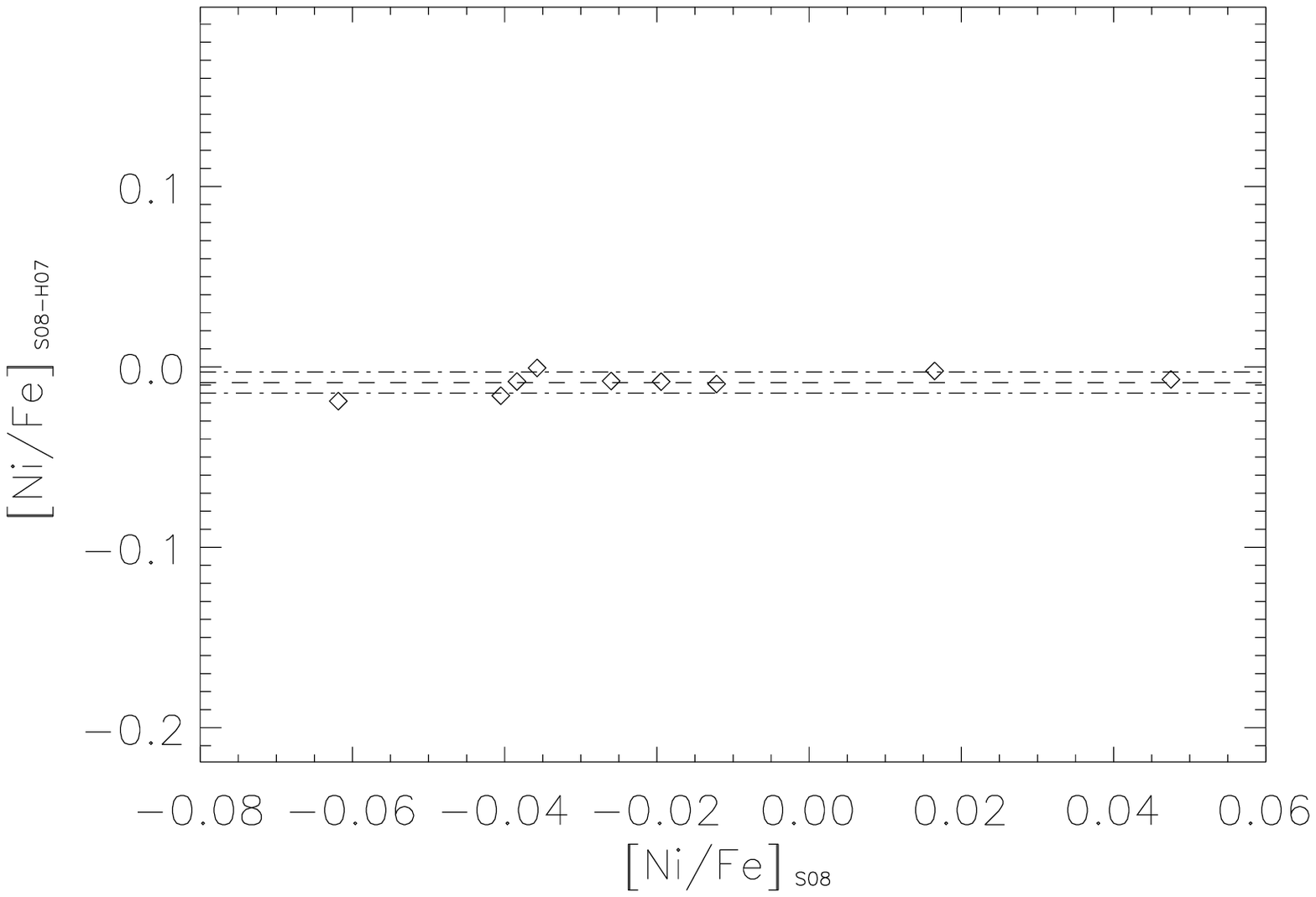}}
\caption{ Comparison of [X/Fe] derived with the stellar paramaters determined with the \citet{sou08}  $\ion{Fe}{i}$-$\ion{Fe}{ii}$  line list versus the list from \citet{hek07}. Dashed line represents the mean value for the differences. Dashed-dotted lines represent one standard deviation over the mean value.}
\label{xfedifs}
\end{figure*}

Giant stars tend to deviate from the Galactic trend in certain elements. The method presented in the present study does not seem to work for giants as well as for main sequence stars. For consistency purposes, we verified our derived stellar parameters for the giant stars. In the first place, we recomputed the stellar parameters  described in Section~4  by using instead the iron line list of \citet{hek07} to investigate the origin of any possible differences. This comparison can be seen in Fig.~\ref{shpar}, where we find an offset of $\approx$ 70 K compared with the results from the list of \citet{sou08}. The next step was to check  whether the effect on the parameter determination was sufficient to explain this deviation. For this purpose, we rederived the [X/Fe] abundances making use of the set of parameters derived with the \citet{hek07} line list and compared them with the original [X/Fe] (see Fig.~\ref{xfedifs}). As one would expect, we find systematic differences but in most cases these differences are at the level of hundreths of dex. The worst case is that of Si where the differences show an even higher [Si/Fe] abundance when derived using the new set of stellar parameters. In other cases such as Mg and Ni, we find no systematic differences with a dispersion of 0.03 dex and 0.01 dex, respectively. Comparison stars from \citet{sil11} (all of them are giants) also deviate from the Galactic trends in some cases and for some elements. In that article, the other metallicities of the stars considered also deviate from the Galactic trends \citep[when compared with][]{ben05}. \citet{sil11} argue that Na is higher than the Galactic trend when comparing with dwarfs from the Galactic disk, perhaps owing to the internal mixing in some of the giant stars. Systematically higher Na abundances were also found for giants belonging to the Hyades cluster by \citet{smi12} and  attributed to the internal processes operating in giant stars. This enhancement in [Na/Fe] ratio is similar to the one we found for some supercluster candidates (of about 0.30 dex) as shown in Fig.~\ref{hyagal3}. Since this effect has been observed by other authors and given the consistency tests, we can assume that the deviations from the Galactic trend for some elements may be real. 
\subsection{Differential abundances}

We determined differential abundances $\Delta$[X/H]  by comparing our measured abundances with those of a reference star known to be a member of the Hyades cluster (vB~153) on a line-by-line basis. There are additional dwarf Hyades cluster members in our sample, V683 Per, and BZ Cet. Although we chose an identical reference star  to  the one used in \citet{pau03} and \citet{sil06}, these studies showed that the Hyades cluster have a high degree of homogeneity. We also expect to find certain degree of pollution due to field stars that have to be identified and discarded as members. As stated in Section~5.1, giants might have systematically higher abundances than dwarfs. For this purpose, we chose a giant star as a reference for the giants instead of a dwarf star (vB~153). The chosen star was $\epsilon$ Tau \citep[also discussed in ][]{smi12} since it is known to belong to the original cluster. We find no significant differences in membership numbers when analyzing differential abundances using this second reference star. We therefore prefer to be consistent and only use as a reference the dwarf star vB~153.\\
 
\begin{figure*}
\centering
\centerline{ \includegraphics[scale=0.55]{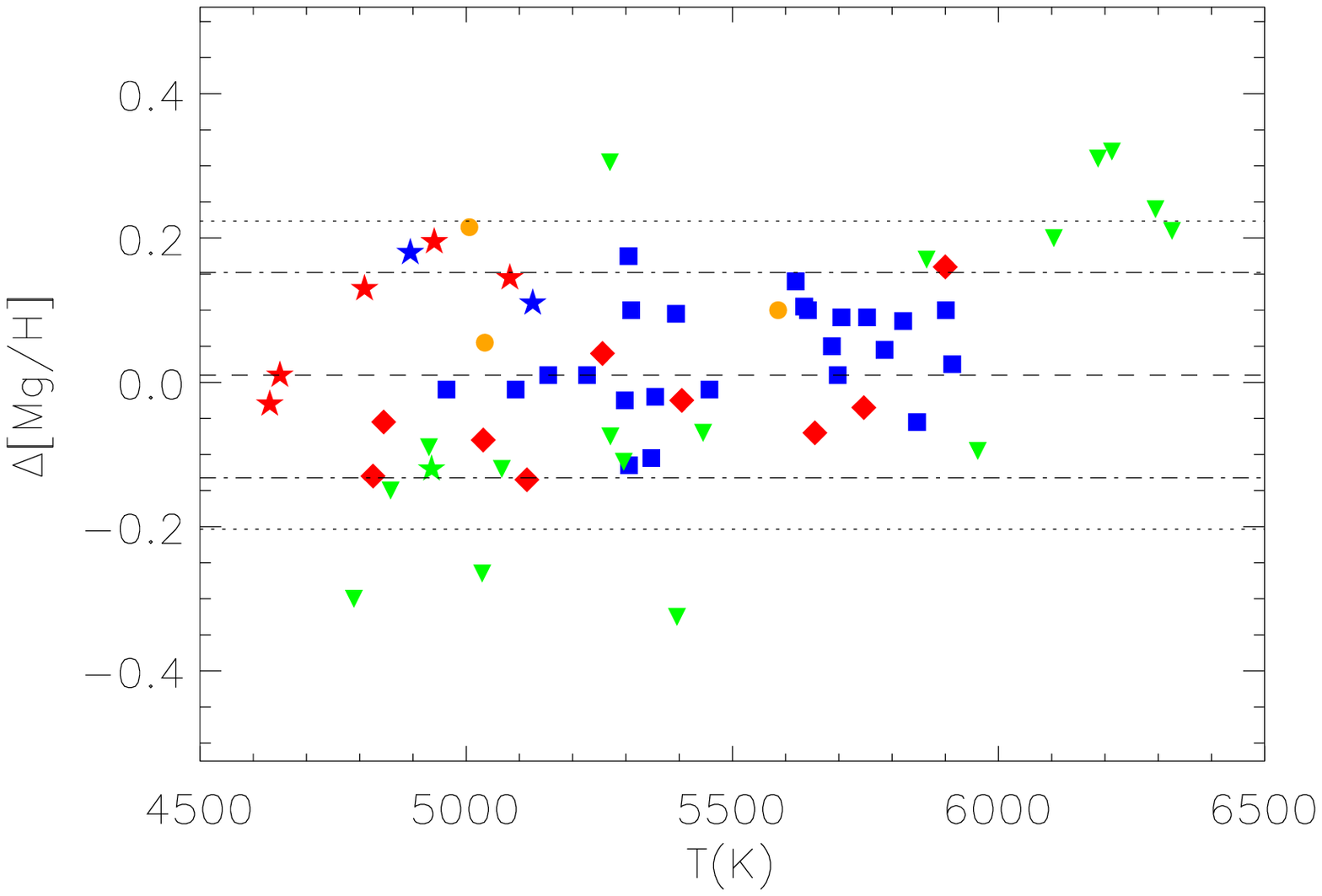}
\includegraphics[scale=0.55]{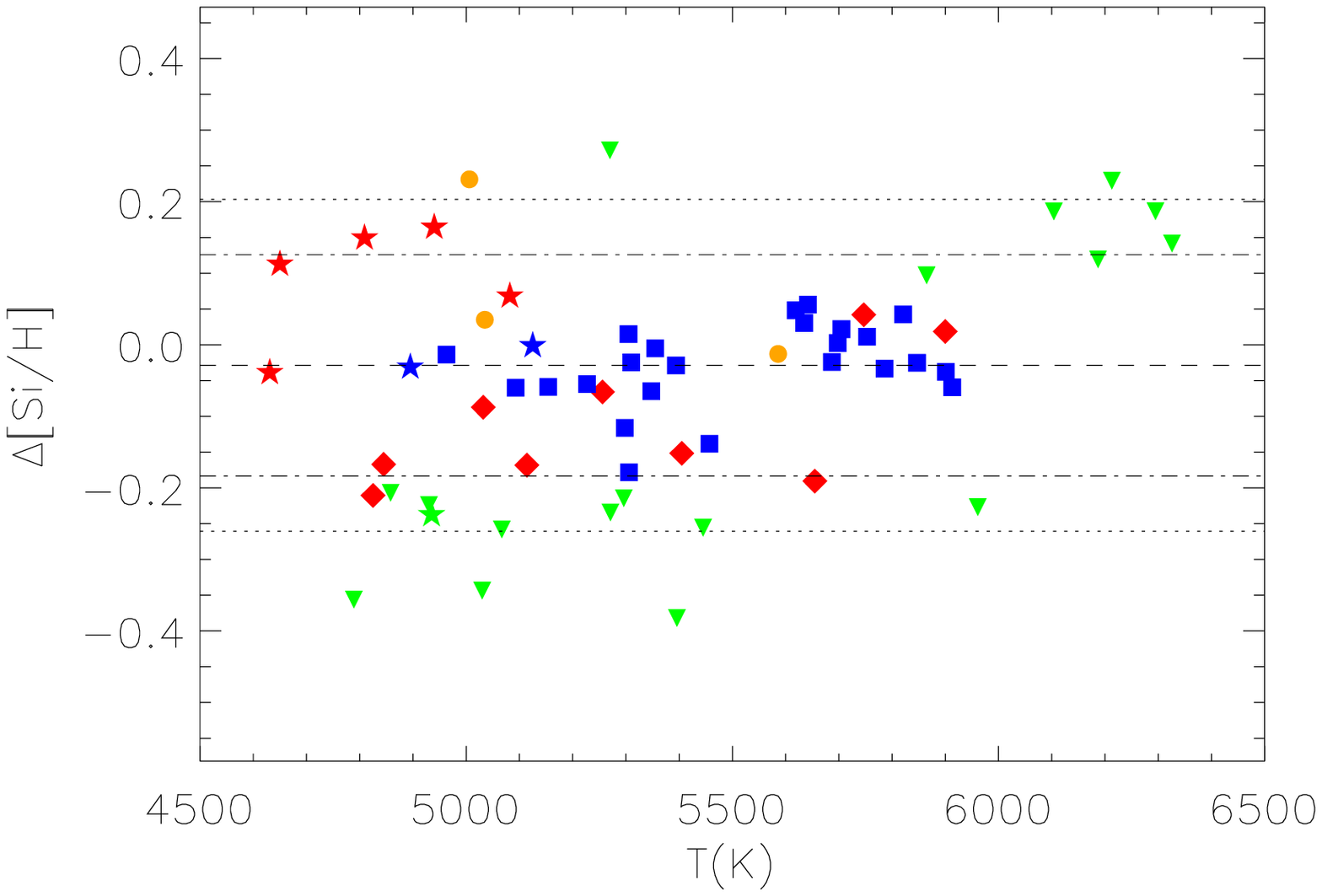}}
\centerline{\includegraphics[scale=0.55]{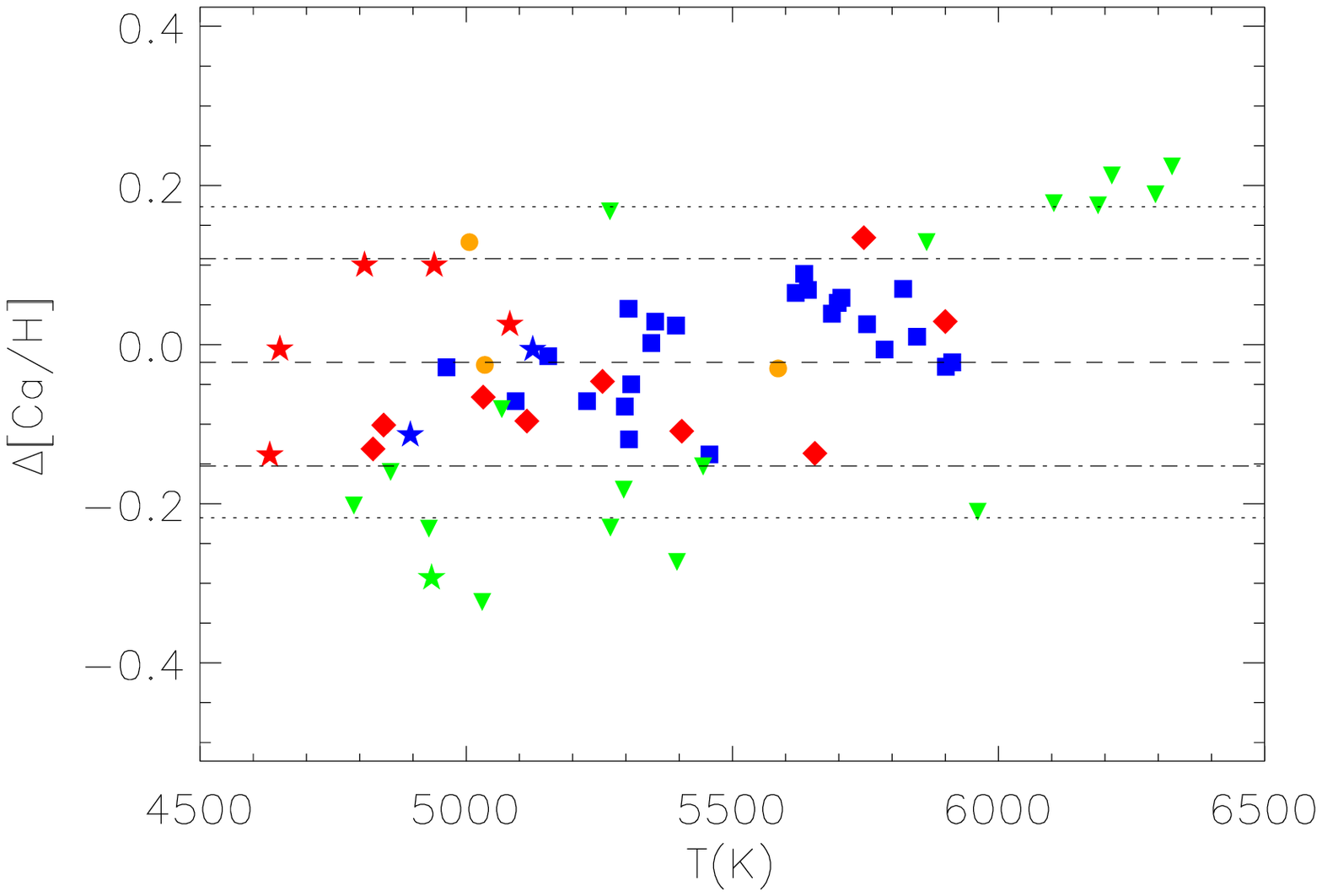}
\includegraphics[scale=0.55]{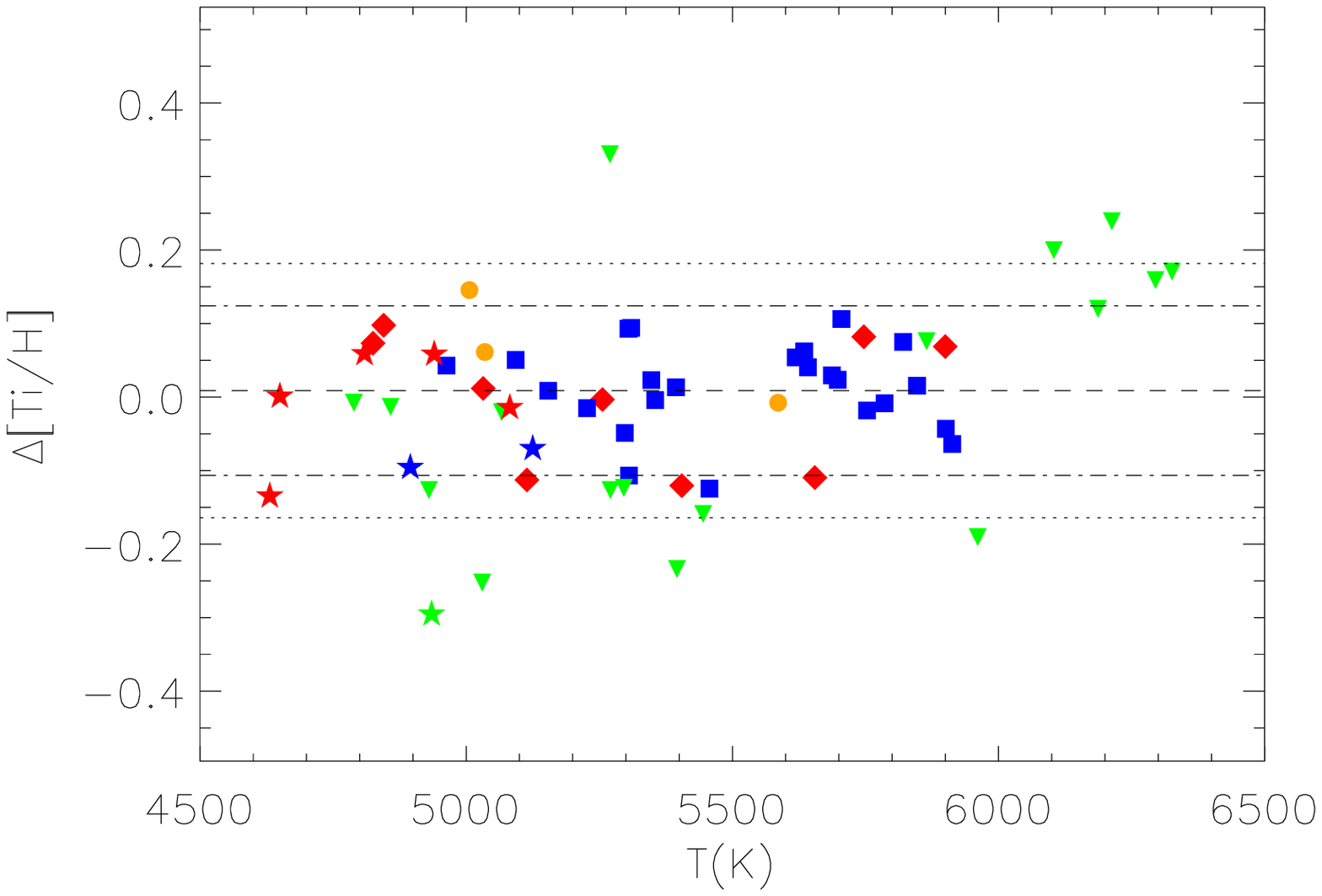}}
\caption{$\Delta$[X/H] differential abundances (for the $\alpha$-elements Mg, Si, Ca, and Ti vs. $T_{\rm eff}$). Dashed-dotted lines represent 1-rms over and below the median for our sample,  whereas dotted lines represent the 1.5-rms level. Dashed lines represent the mean differential abundance. Red diamonds are stars compatible within 1-rms with the Fe abundance but not for all elements, blue squares are the candidates selected to become members of the Hyades Supercluster, while green triangles are rejected candidates and starred points are the giants. Blue squares  and starred points are the final selected candidates to become members of the Hyades Supercluster. BZ Cet, V683 Per, and $\epsilon$  Tau  are indicated by orange circles.}
\label{hyadif1}
\end{figure*}

\begin{figure*}
\centering
\centerline{\includegraphics[scale=0.55]{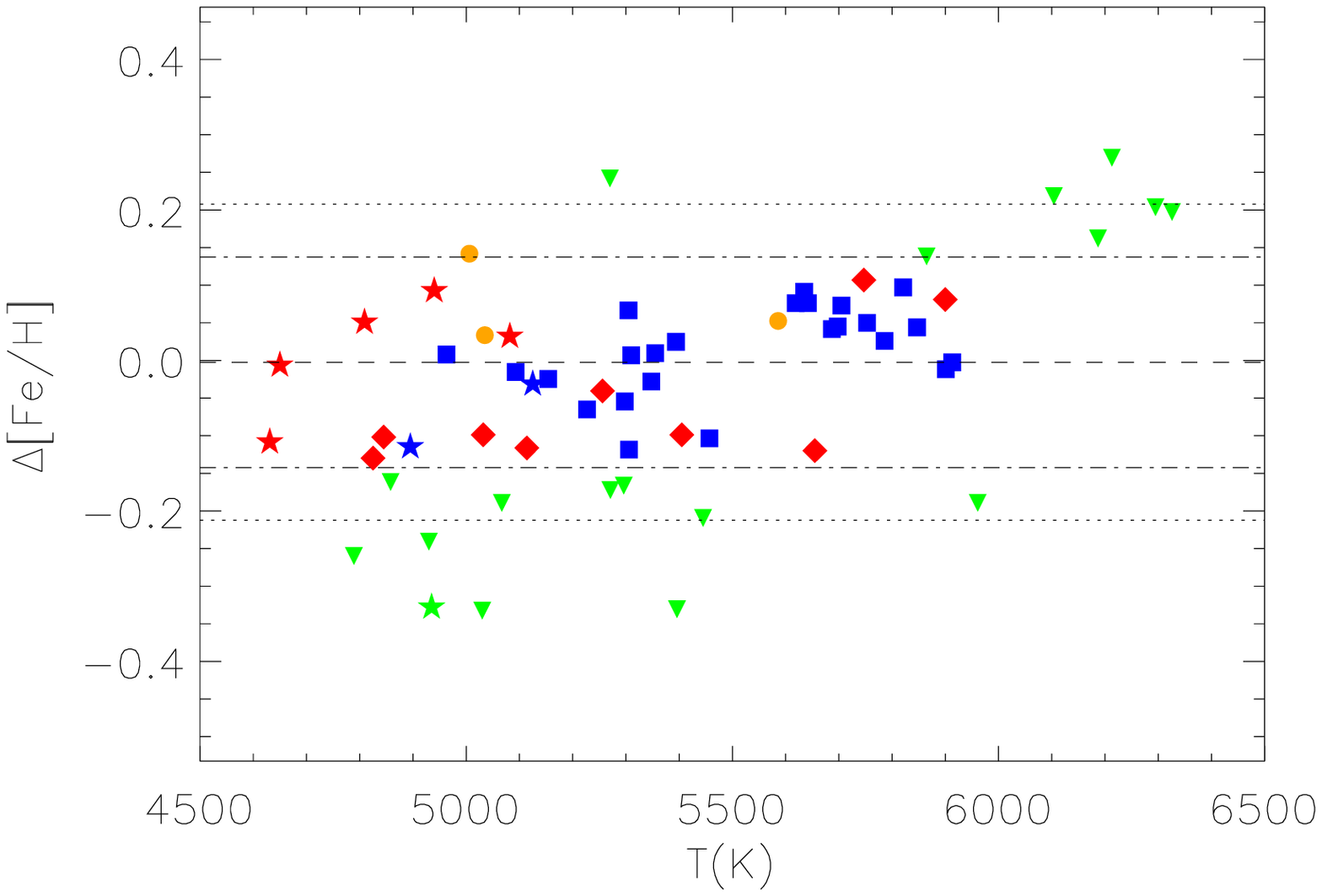}
\includegraphics[scale=0.55]{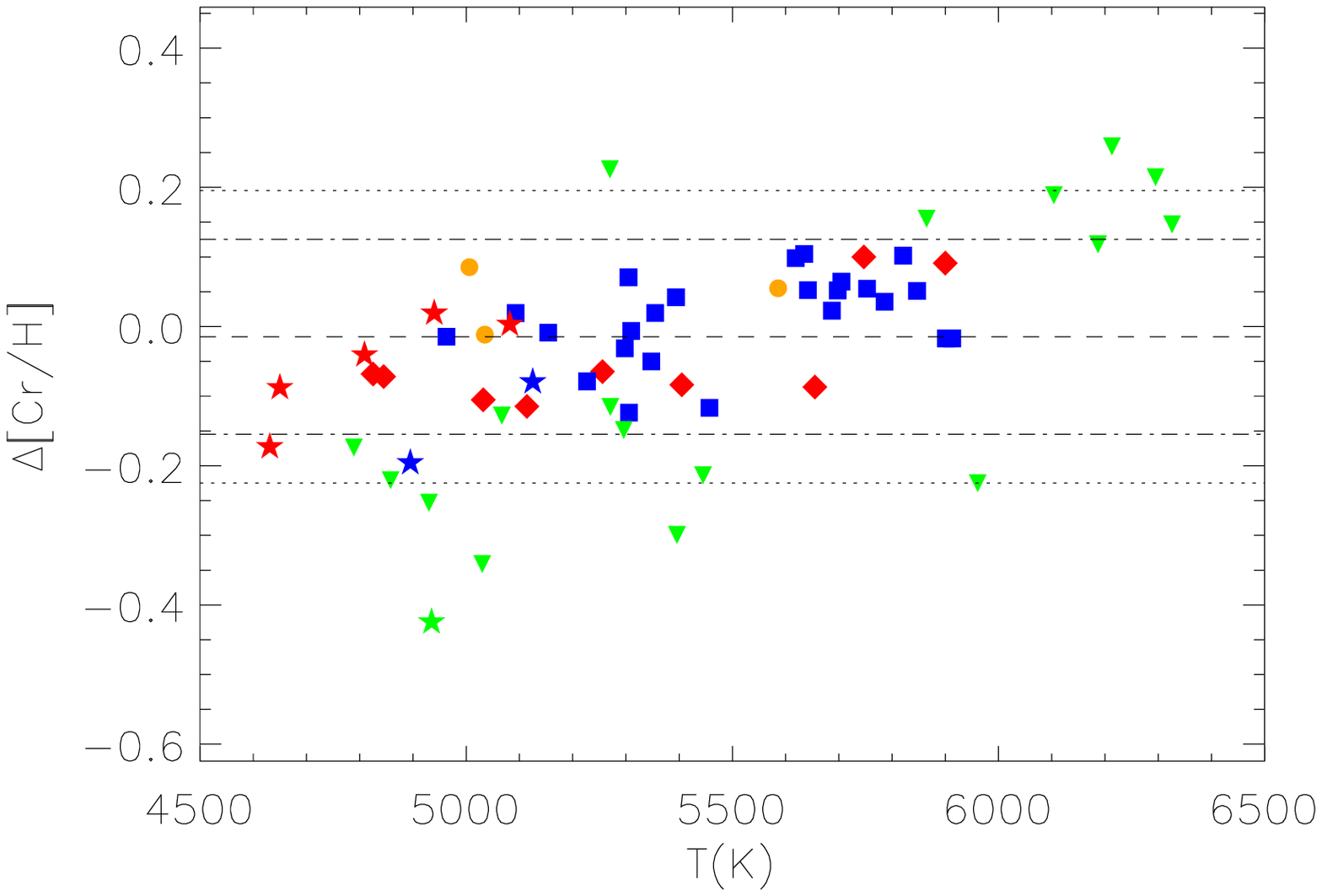}}
\centerline{\includegraphics[scale=0.55]{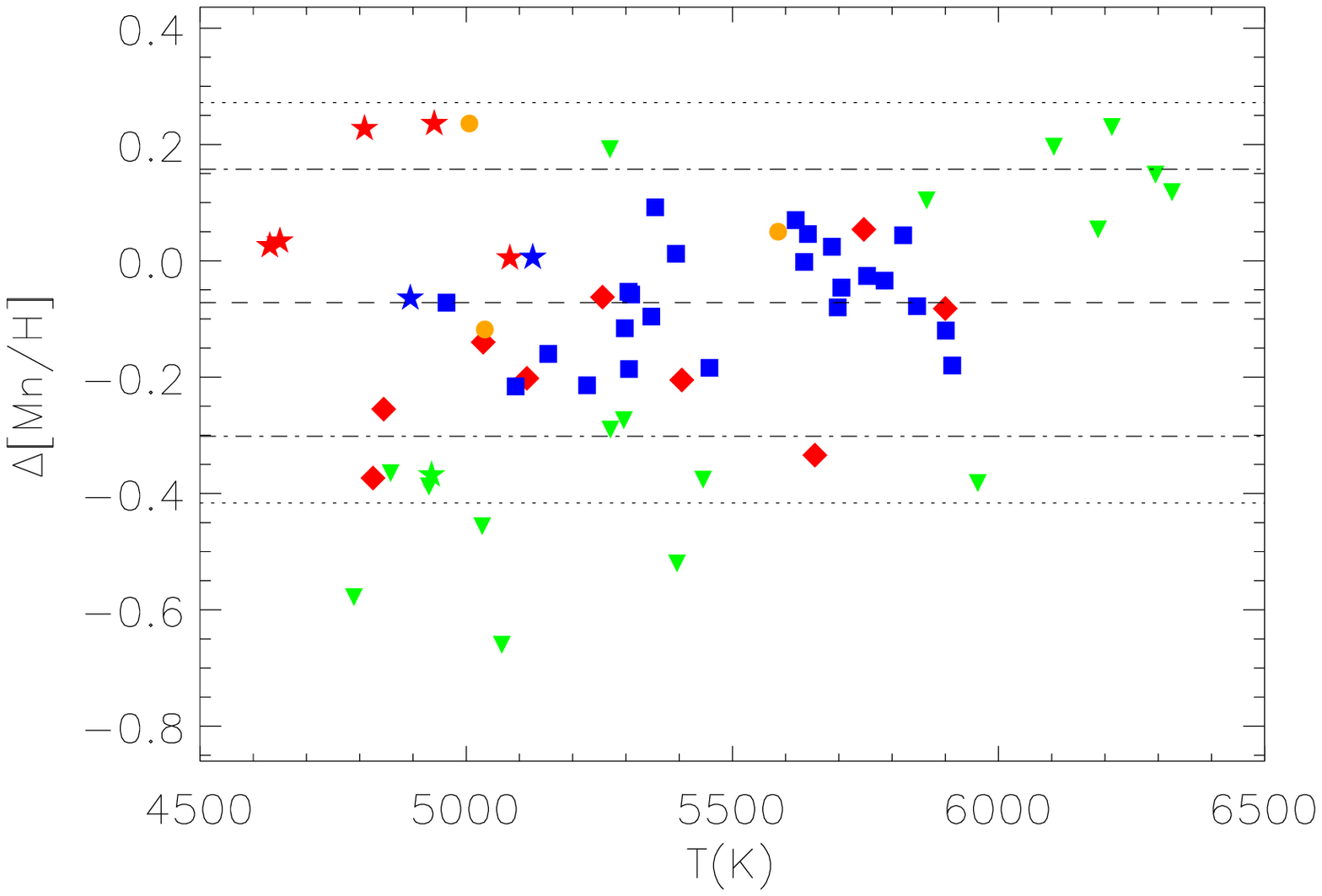}
\includegraphics[scale=0.55]{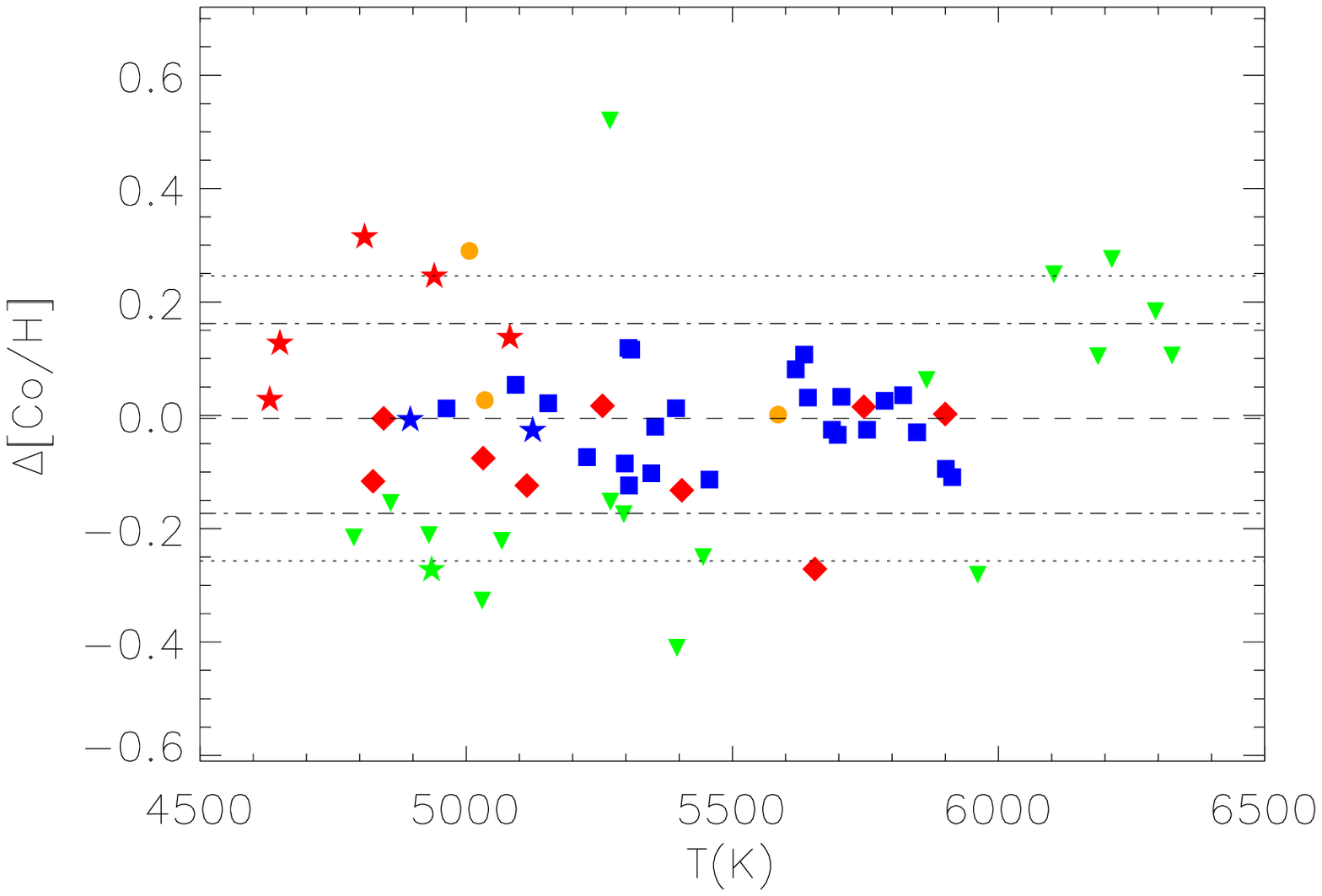}}
\centering 
\includegraphics[scale=0.55]{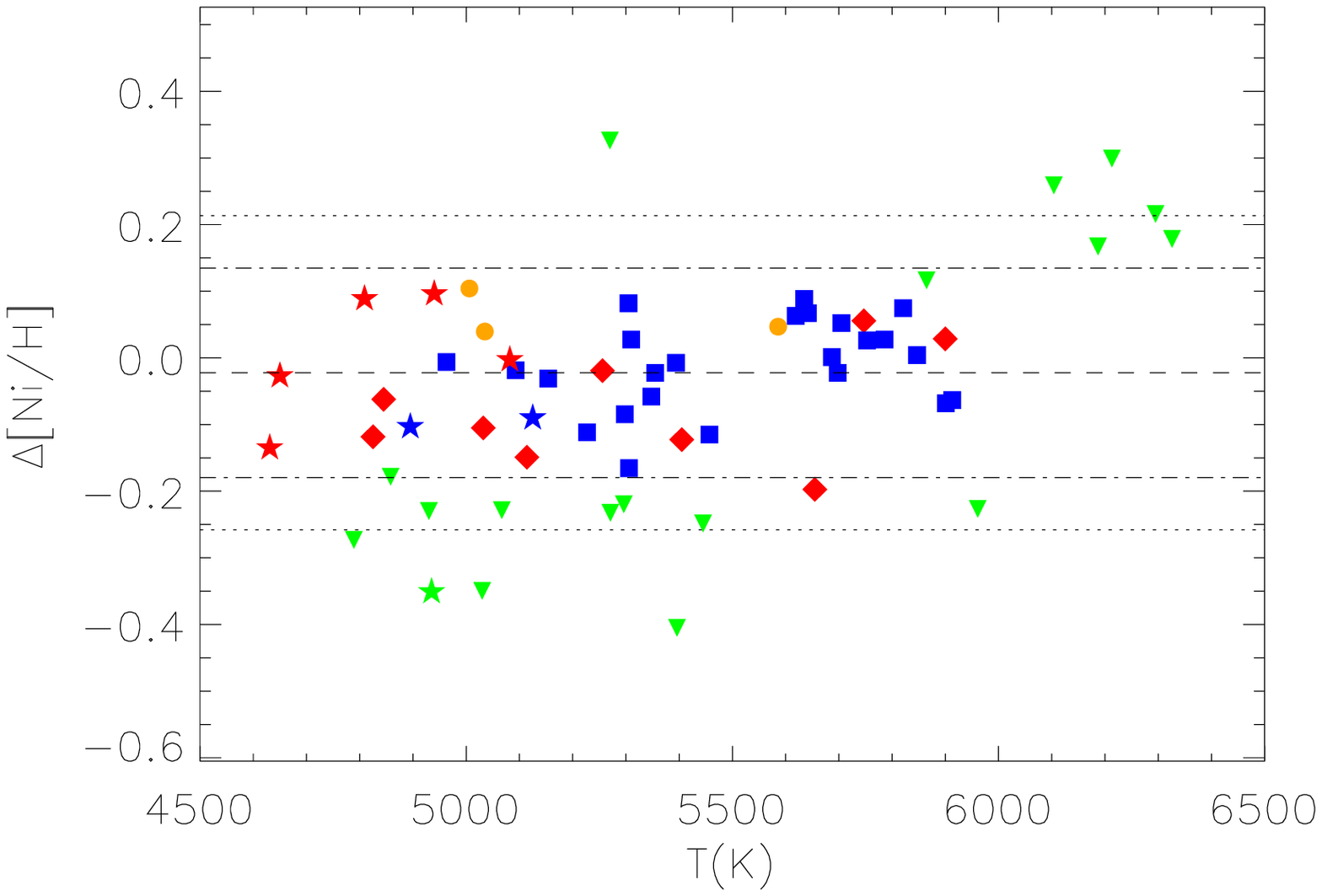}
\caption{Same as Fig.~\ref{hyadif1} but for  the Fe-peak elements Cr, Mn, Co,  and Ni.}
\label{hyadif2}
\end{figure*}

  The candidate selection within the sample was determined by applying a one root-mean-squared (rms, thereafter) rejection over the median for almost every chemical element treated in the analysis. The rejection process considers the rms in the abundances of the sample for each element. At first, we rejected every star that deviated by more than 1-rms from  the median abundance denoted by the dashed-dotted lines in Figs.~\ref{hyadif1}, \ref{hyadif2}, \ref{hyadif3}, \ref{hyadif4}, and \ref{hyadif5}.  The initial rms values considered during the candidate selection are given in Table~\ref{tablarms}. The initial 1-rms rejections led to the identification of 15 candidate members. We subsequently applied the more flexible criteria allowing stars to become members when their abundances were within the 1-rms region for  90 \% of the elements considered and the remaining 10 \% within the 1.5-rms region (18 elements and 2 elements respectively). The final rms is the one applied to the final selected candidates being members of the Hyades Supercluster. The error analysis considers only the standard deviation in the line-by-line differences. We also made a previous selection of candidates more likely to contain members based on their differential Fe abundances (see Table~\ref{tablapar}), since these selected stars tend to maintain their abundance coherence between elements, as shown in Figs.~\ref{hyadif1}, \ref{hyadif2}, \ref{hyadif3},  \ref{hyadif4}, and \ref{hyadif5}.\\

This more flexible rms-based analysis was made in order to see to the degree which the sample is homogeneous, and to take care of the more likely contamination of the sample by field stars. Therefore, to assess this degree of homogeneity one must take into account the number of stars that lie within 1-rms, 1.5-rms, 2-rms, and 3-rms regions (see Table~\ref{tablatotal}). In a pure 1-rms rejection, we find that 15 stars (25 \%) as possible members. Hence, allowing two elements (maximum) to satisfy the 1.5-rms criteria results in 28 candidate members (46 \%). The last three columns of Table~\ref{tablapar} give information about membership based on the differential abundances of Fe and the other elements following these criteria. The preliminary study of \citet{tab10} found a 64 \% of the candidates are members, although the sample in this study is larger than that previous one, hence our results here are more reliable. Adopting these criteria, we conclude  that the membership of the Hyades Supercluster ranges from 25 to 46 \%.
\begin{table}
\caption{Percentage analysis based on the rejection level of the differential abundances.}
\label{tablatotal}
\centering
\begin{tabular}{c c c}      
\hline\hline
                       
rms & \# stars & \% stars \\
\hline 
   1.0 & 15 &  25 \\
   1.5 & 35 &  57 \\
   2.0 & 47 &  77 \\ 
   2.5 & 57 &  93 \\
   3.0 & 59 &  97 \\
\hline 
\end{tabular}

\end{table}

In this differential abundance discussion, we have only taken into account the neutral elements when there are lines available to measure. We also determined the abundances of $\ion{Cr}{ii}$, $\ion{Ti}{ii}$, $\ion{Sc}{ii}$, and $\ion{Zr}{ii}$, which are typically  consistent with those abundances obtained for the neutral species. However, we made this choice because the Galactic abundances for the neutral elements have a smaller dispersion than for the ionized one. Vanadium  is considered as a reliable element in the differential analysis, despite the NLTE effects \citep[e.g.][]{bod03,gil06,nev09}: when vanadium is compared with the neutron-capture elements, the differential treatment seems to work well given the large dispersion and sensitivities intrinsic to these other elements.\\

As a first consistency test, we scrutinized at the differential abundances in Figs.~\ref{hyadif1}, \ref{hyadif2}, \ref{hyadif3}, \ref{hyadif4}, and \ref{hyadif5}. The dwarf stars known to belong to the Hyades cluster within our sample (V683 Per and BZ Cet) occupy the 1-rms region for all the considered abundances, which also agrees with the conclusions for them of \citet{pau03} and \citet{sil06}. These stars were also assumed to be cluster members by \citet{pom11}. \\

\begin{table}
\caption{Median abundance, and both initial and final rms values for all considered elements. }
\label{tablarms}
\centering
\begin{tabular}{lrrr}     
\hline\hline
Element  & $\Delta$ [X/H] & rms$_o$ & rms$_f$ \\
\hline
 Fe   &  0.03  & 0.15  & 0.06 \\
 Na   &  0.09  & 0.21  & 0.11 \\
 Mg   &  0.06  & 0.14  & 0.09 \\
 Al   &  0.06  & 0.15  & 0.08 \\
 Si   & $-0.02$  & 0.16  &  0.06 \\
 Ca   & $-0.06$ & 0.13  & 0.06 \\
 Ti   &  0.02  &  0.12 & 0.06 \\
 V    & $-0.10$  & 0.17 & 0.14 \\
 Sc   & $-0.03$  & 0.16 & 0.10 \\
 Cr   &  0.02  &  0.14  & 0.07 \\
 Mn   & $-0.06$ &  0.23    &  0.11 \\
 Co   & 0.00  & 0.17   & 0.07 \\
 Ni   & $-0.01$  & 0.16 &  0.07 \\
 Cu  & 0.03  & 0.21 &  0.09  \\
 Zn   & $-0.20$ & 0.24 & 0.20  \\
 Y    & 0.01 & 0.13 & 0.07 \\
 Zr   & $-0.23$ & 0.29 & 0.24 \\
 Ba   & 0.11 & 0.16 & 0.13 \\
 Ce   & 0.05 & 0.10 & 0.08 \\
 Nd   & 0.01 & 0.11 & 0.07 \\
\hline
\end{tabular}
\end{table}

The determined rms for the selected stars is shown on Table~\ref{tablarms}: we found that Na and Mn have larger dispersions than the other elements used in the differential analysis. In addition Sc, Al, Mg, Y, and Ce show an intermediate values of dispersion that are not as larger as those for Na, Mn,  and Ba. Zn and Zr have an even larger values of rms (of about 0.20 dex). The rest of the elements have an rms equal to or larger than 0.06 dex. The element abundances, relative to those also studied by \citet{sil11} also have large dispersion levels, such as those for Na and Sc, although our rms is smaller in the case of Co.

From the {\it chemical tagging} analysis, we found that 15-28 of the 61 stars analyzed have homogeneous abundances for all the elements we considered (a 25-46\% of the studied stars). This membership percentage implies that our selected sample contains a  mixture of field stars and stars that have been evaporated from the Hyades cluster. Our sample is not as homogeneous as the moving group HR 1614 \citep{sil07b}, but not similar to the Arcturus moving group \citep{wil09}, which does not correspond to an evaporated cluster. In contrast, \citet{fae07} argued that only  15 \% of the Hyades Supercluster candidates originate from the cluster itself. However, \citet{fae08} reported a cluster membership percentage of 52 \% (as an upper limit).  \citet{pom11} and \citet{sil11} found  10 \% and  15 \% membership fractions, respectively, their samples being composed of 21 and 26 stars. However, we analyzed a larger sample (61 stars) than these two previous studies, and confirm that the Hyades Supercluster is not composed entirely of Hyades cluster evaporated stars. In \citet{pom11}, the sample selection is based on the Geneva-Copenhagen survey \citep{nor04} and the mean velocity of  the Hyades overdensity studied by \citet{hol09}. In contrast, \citet{sil11} considered some stars with velocities in the expected direction of dispersion from the Hyades cluster.\\

Our analysis does not recover the same number of members as the two previous studies of \citet{pom11} and \citet{sil11}. The abundance selection in the case of \citep{pom11} is based on a statistical constrast based on a $\chi^2$ test that employs a few Hyades Cluster stars. This $\chi^2$ approach treats all the abundances as whole, thus does not concentrate on the individual elements one by one. On the other hand, \citet{sil11} discard the stars for which the Fe abundances are lower than the solar value. They identified a few stars that might have originated from the original cluster based on this Fe preselection. Later on, they analyzed the trends of other elements to determine whether thse stars have the same abundance as the Hyades Cluster. All these samples are incomplete but no robust assesment of the contamination levels of the Hyades Supercluster can be made. The method used in the present study \citep[as in][]{pom11,sil11} can only ascertain the origin of the Hyades Supercluster but cannot measure  the contamination levels by field stars. The common conclusion of this work and \citet{pom11},\citet{sil11} is that the Hyades Supercluster cannot originate entirely from the Hyades Cluster. However, we can still identify candidates that once belonged to the Hyades cluster.
\section{Conclusions}
We have computed the stellar parameters and their uncertainties for 61 Hyades Supercluster candidate stars, after which we have obtained their chemical abundances for  20 elements (Fe, Na, Mg, Al, Si, Ca, Ti, V, Cr, Mn, Co, Fe, Ni, Cu, Zn, Y, Zr, Ba, Ce, and Nd). All the abundances are obtained in a fully differential way employing a solar reference and a well-known member of the Hyades cluster.\\*

We have derived the Galactic space-velocity components for each star and used them to check the original selection based on Galactic velocities \citep{mon01a,lop10}, which were then improved using the radial velocities derived from  our data. We have employed  the new $Hipparcos$ proper motions and parallaxes \citep{hog00,lee07} employing the procedures described in \citet{mon01a}. To perform a preliminary consistency test, we have analyzed the $U$, $V$, and $W$ Galactic velocities of the final selected stars tend to diminish their dispersion in $V$ to values closer to those expected for a spread cluster \citep[see][]{Eggen1994,sku97}.\\*

As a  second test of the stellar parameters, we have compiled a $\log{g}$ vs. $\log{T_{\rm eff}}$ diagram to verify the consistency of the method employed to determine the stellar parameters. This diagram shows that most of the stars fall on the isochrone for the Hyades age (0.7 Gyr). This is an important but insufficient condition to ascertain that they have a common origin. Therefore, the abundance analysis must also be used to help stablish which stars share a common origin with those in the Hyades cluster. The abundance analysis shows that the final 28 selected stars  are compatible with the Hyades isochrone, as expected if they have evaporated from the Hyades cluster. The membership percentage that we find in this work ( 46 \%) compared with those of other authors demonstrates the importance of the sample selection and a detailed chemical analysis.

A yet more detailed analysis of different age indicators and  chemical homogeneity is in progress and will be presented in future publications. This analysis will lead to a more consistent means of confirming a list of candidate members from the abundance analysis.\\*
\begin{acknowledgements}

This work was supported by the Universidad Complutense de Madrid (UCM), the Spanish Ministerio de Ciencia e Innovaci\'on (MCINN) under grants BES-2009-012182, AYA2008-000695, and AYA2011-30147-C03-02, and The Comunidad de Madrid under PRICIT project S2009/ESP-1496  (AstroMadrid). We would like to thank the anonymous referee for helpful comments and corrections.

\end{acknowledgements}


%\begin{figure*}
%\centering
%\centerline{\includegraphics[scale=0.55]{hyagal_Na.ps}
%\includegraphics[scale=0.55]{hyagal_Al.ps}}
%\centerline{\includegraphics[scale=0.55]{hyagal_Sc.ps}
%\includegraphics[scale=0.55]{hyagal_V.ps}}
%\caption{Same as Fig.~\ref{hyagal1} but for the odd-Z elements.}
%\label{hyagal3}
%\end{figure*}

\clearpage

\begin{appendix}
\section{On-line material}

\begin{figure*}
\centering
\centerline{\includegraphics[scale=0.55]{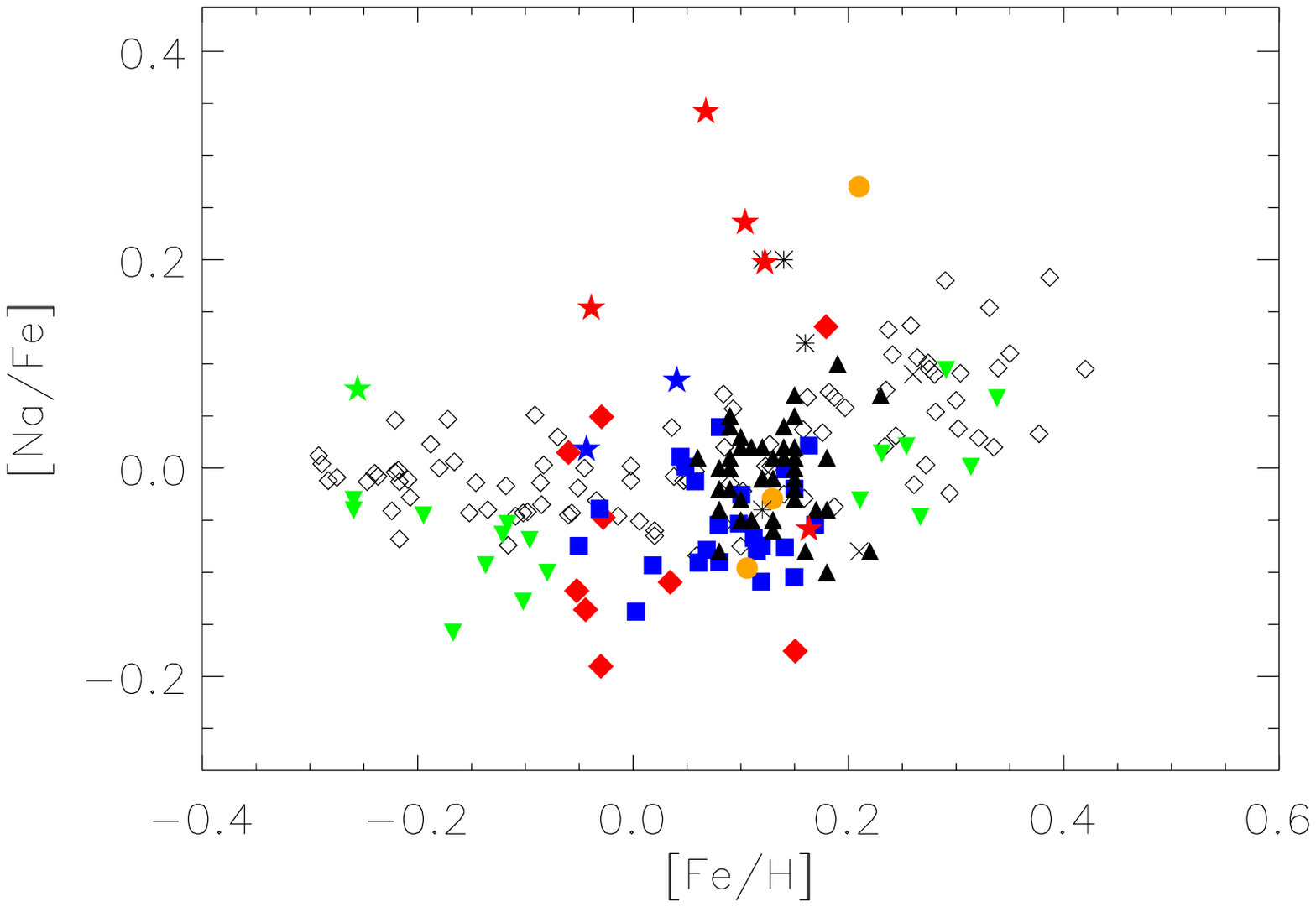}
\includegraphics[scale=0.55]{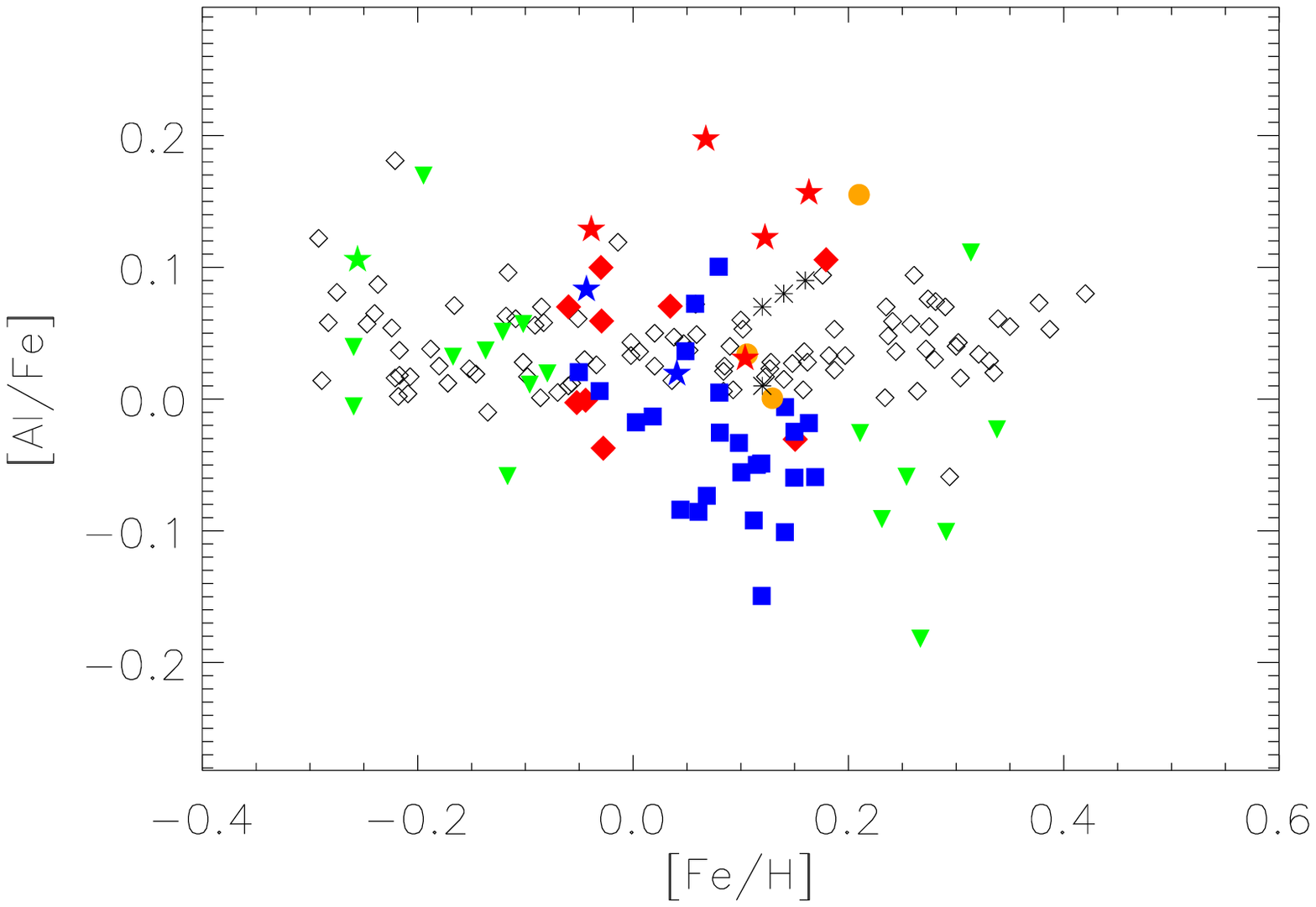}}
\centerline{\includegraphics[scale=0.55]{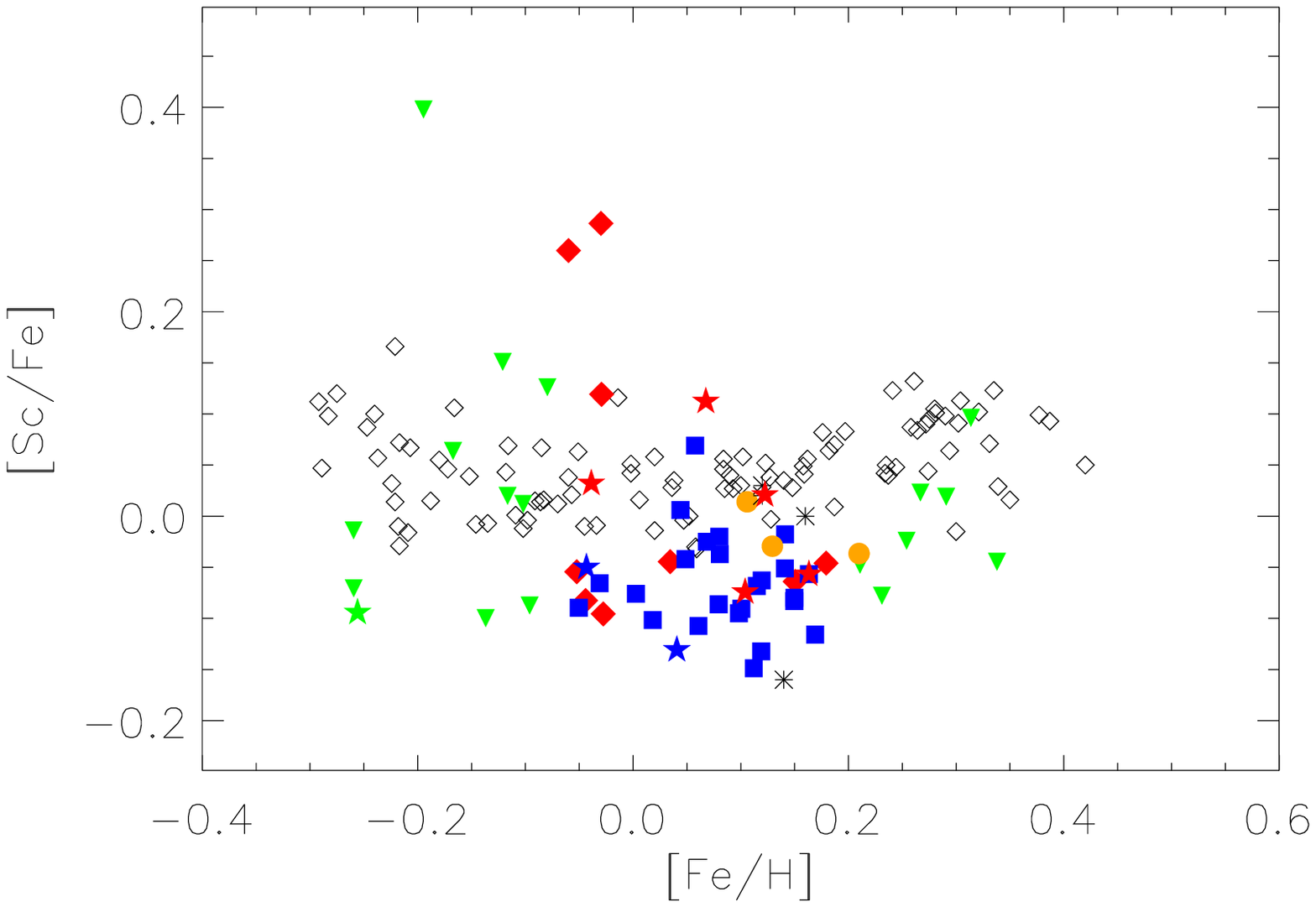}
\includegraphics[scale=0.55]{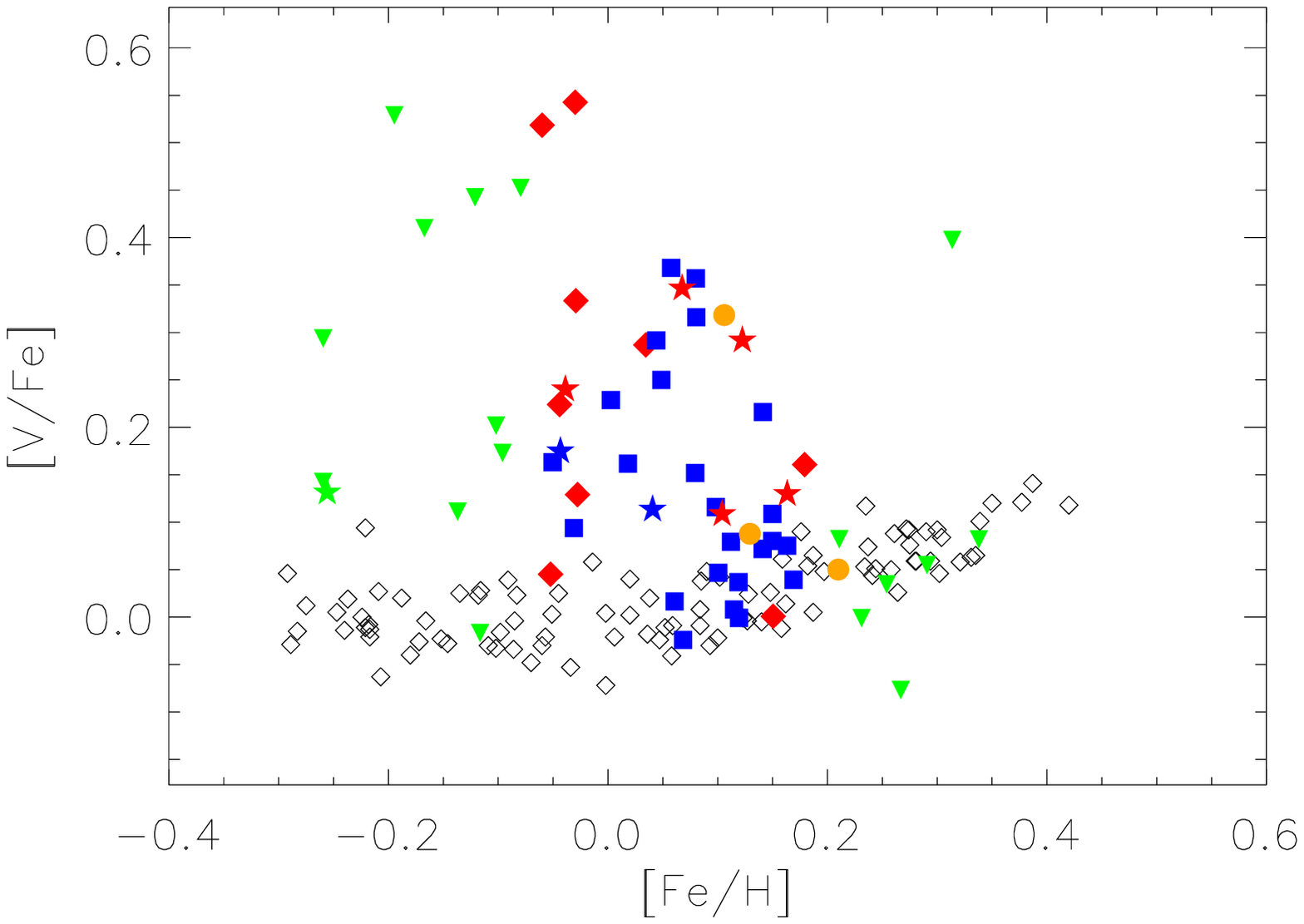}}
\caption{Same as Fig.~\ref{hyagal1} but for the odd-Z elements Na, Al, Sc, and V.}
\label{hyagal3}
\end{figure*}

\begin{figure*}
\centering
\centerline{\includegraphics[scale=0.55]{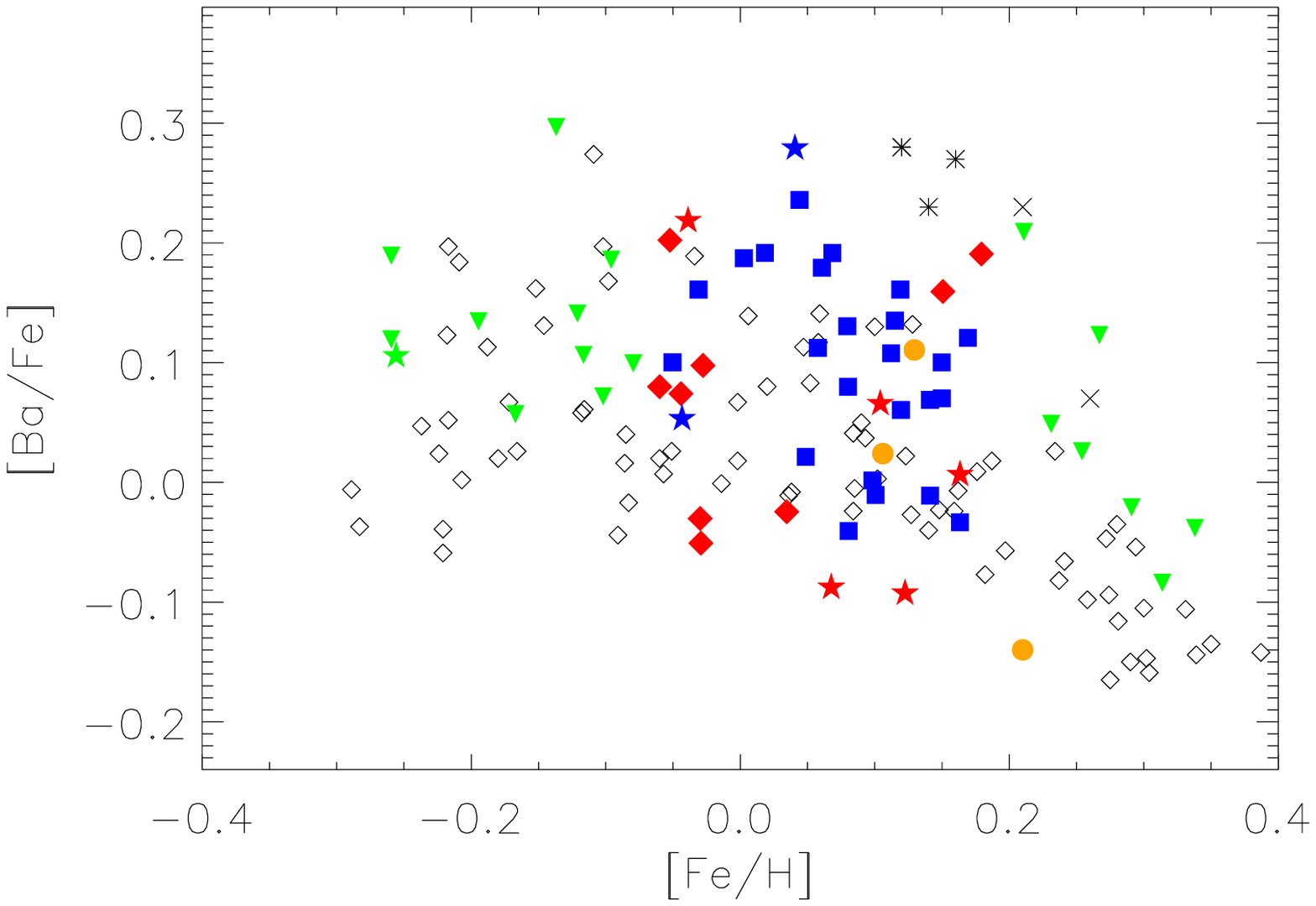}
\includegraphics[scale=0.55]{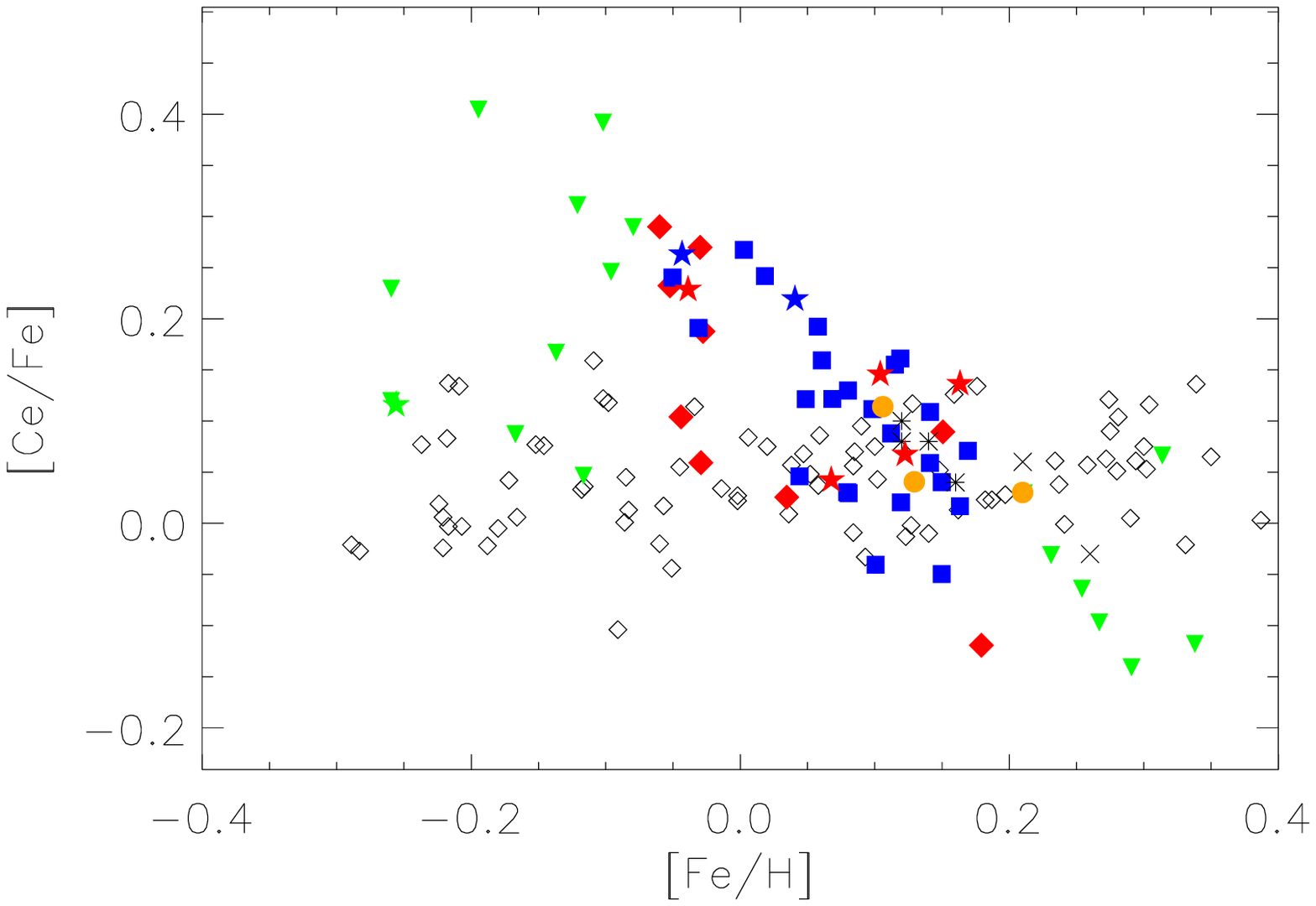}}
\centerline{\includegraphics[scale=0.55]{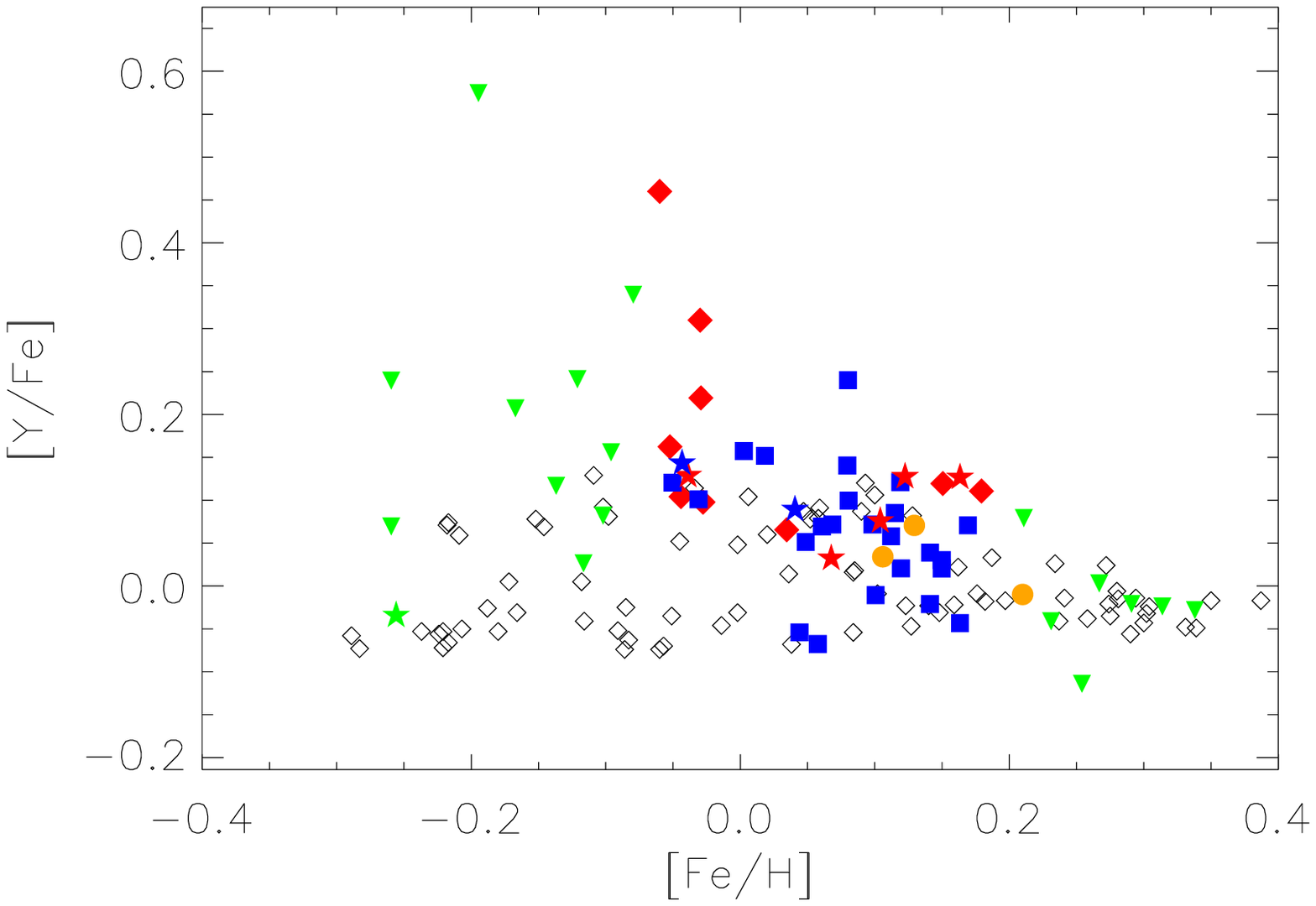}
\includegraphics[scale=0.55]{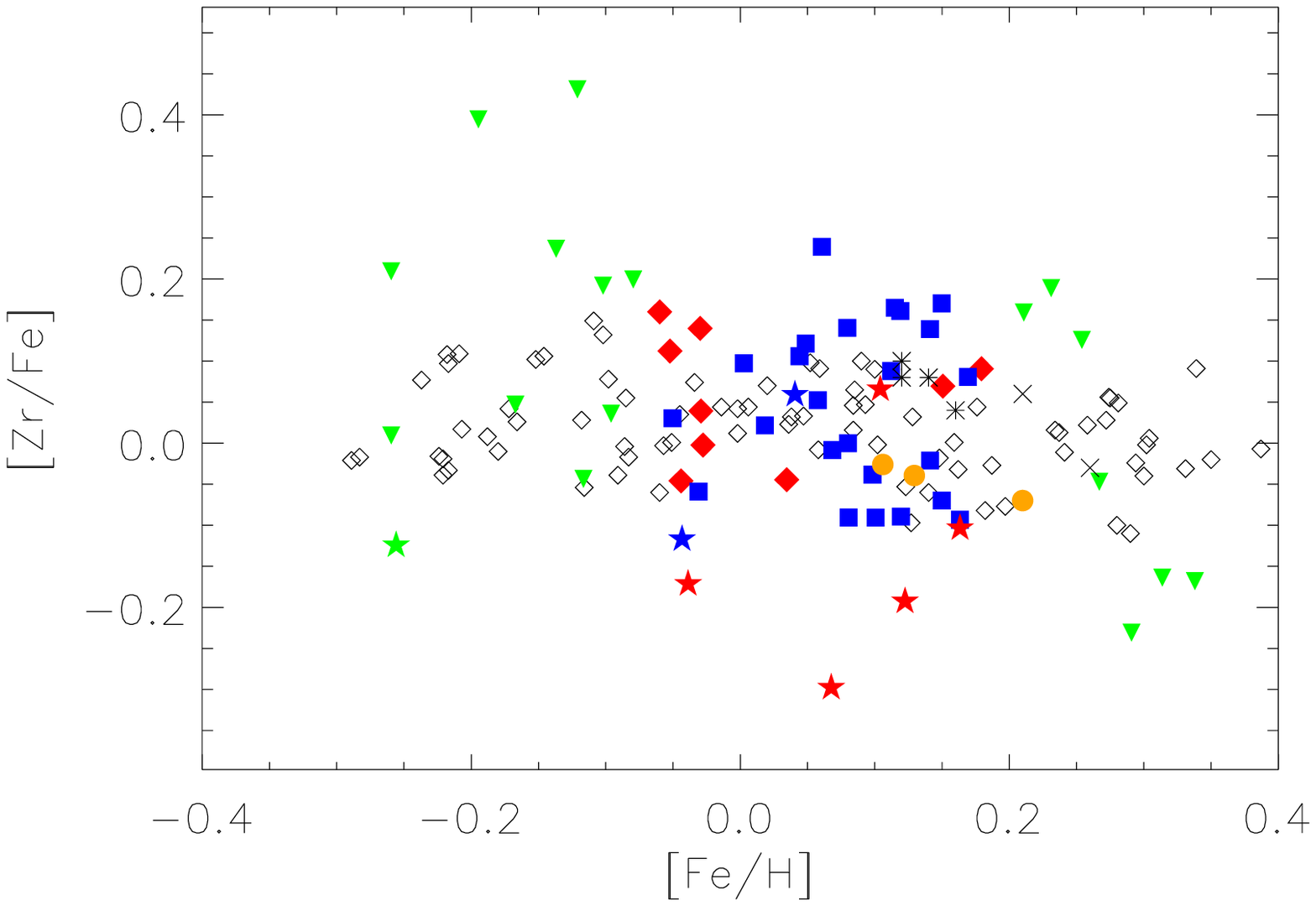}}
\caption{Same as Fig.~\ref{hyagal1} but for the s-process elements Ba, Ce, Y, and Zr.}
\label{hyagal4}
\end{figure*}

\begin{figure*}
\centering
\centerline{\includegraphics[scale=0.55]{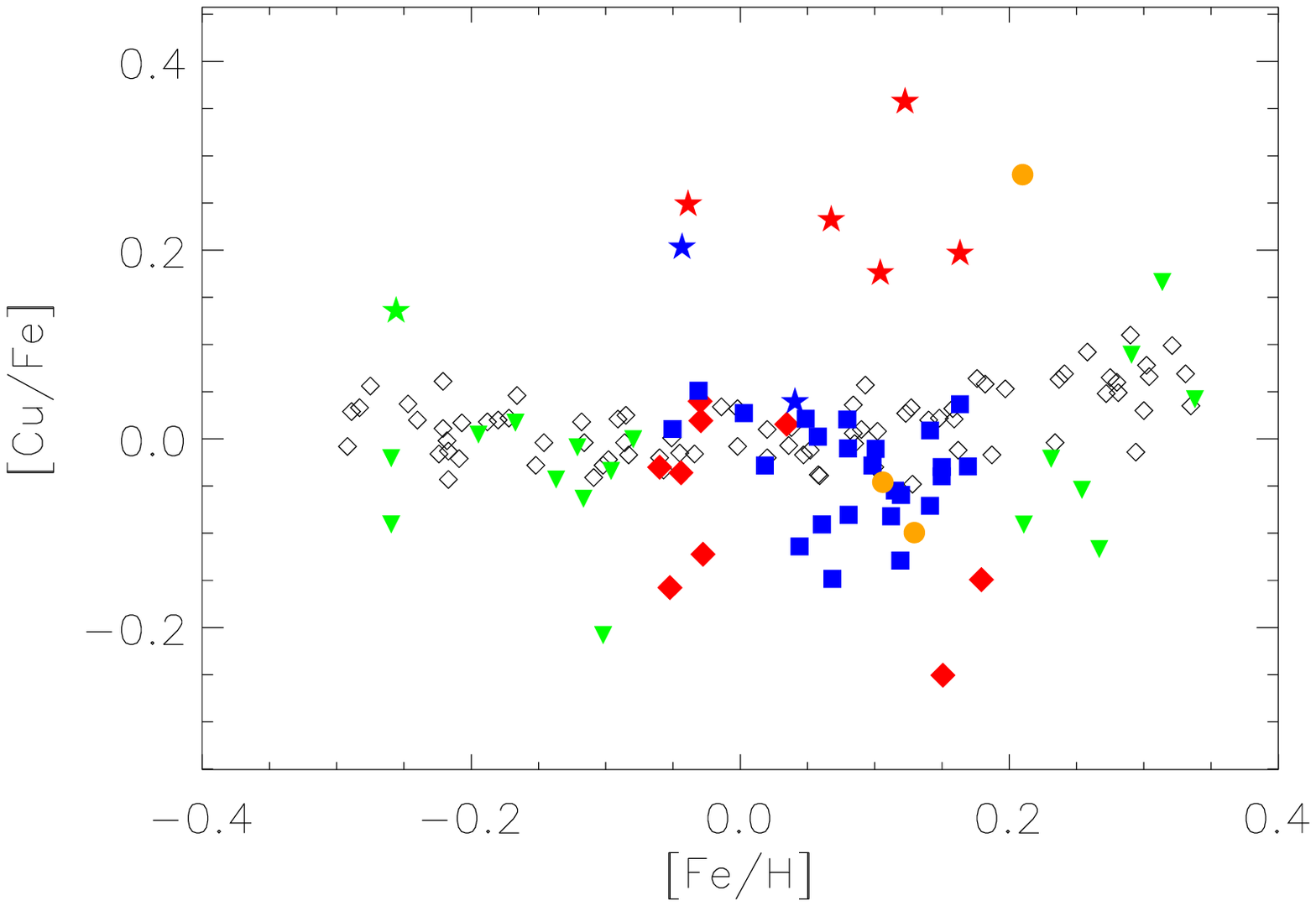}
\includegraphics[scale=0.55]{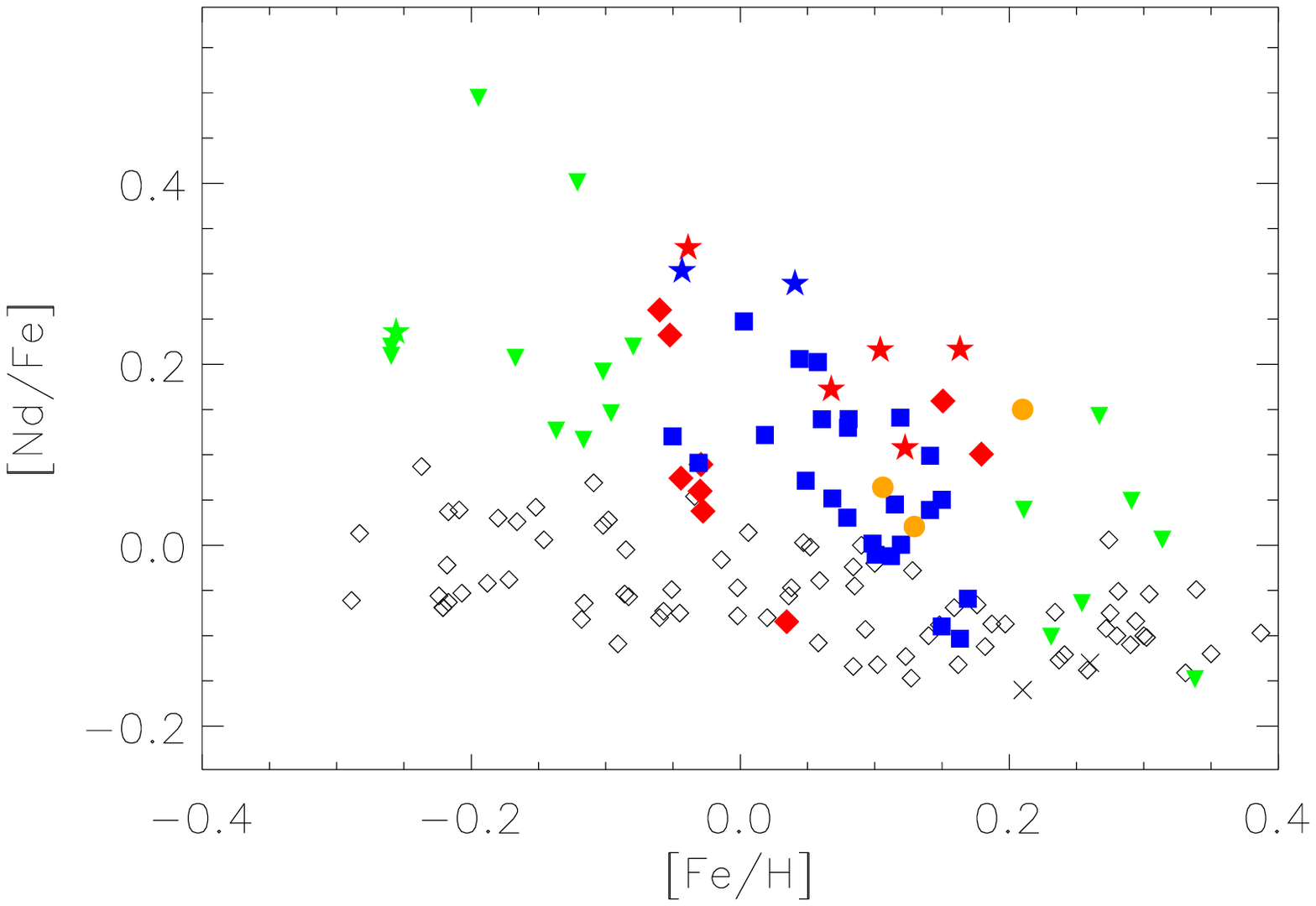}}
\includegraphics[scale=0.55]{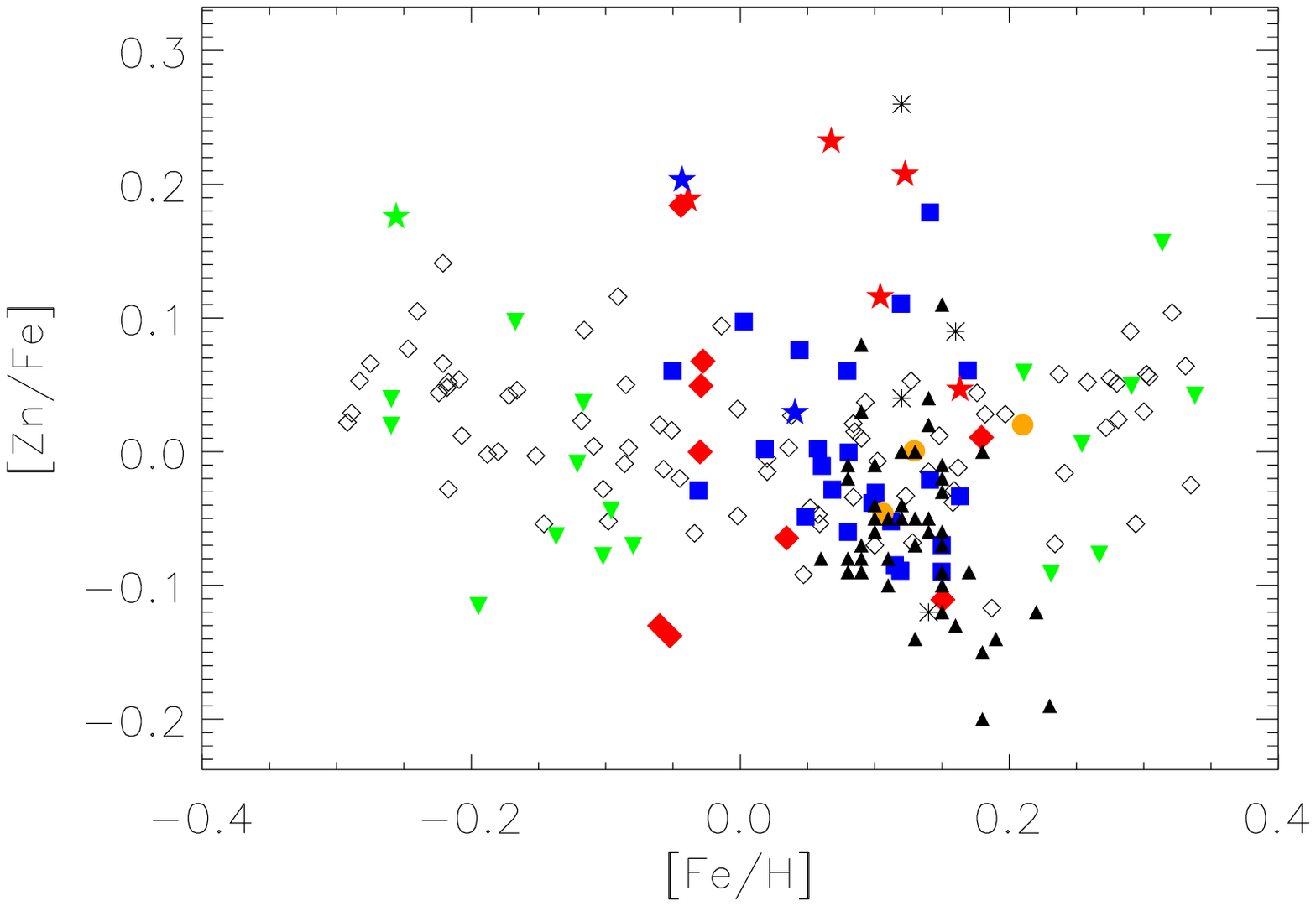}
\caption{Same as Fig.~\ref{hyagal1} but for the s-process elements Cu, Nd, and Zn.}
\label{hyagal5}
\end{figure*}

\begin{figure*}
\centering
\centerline{ \includegraphics[scale=0.55]{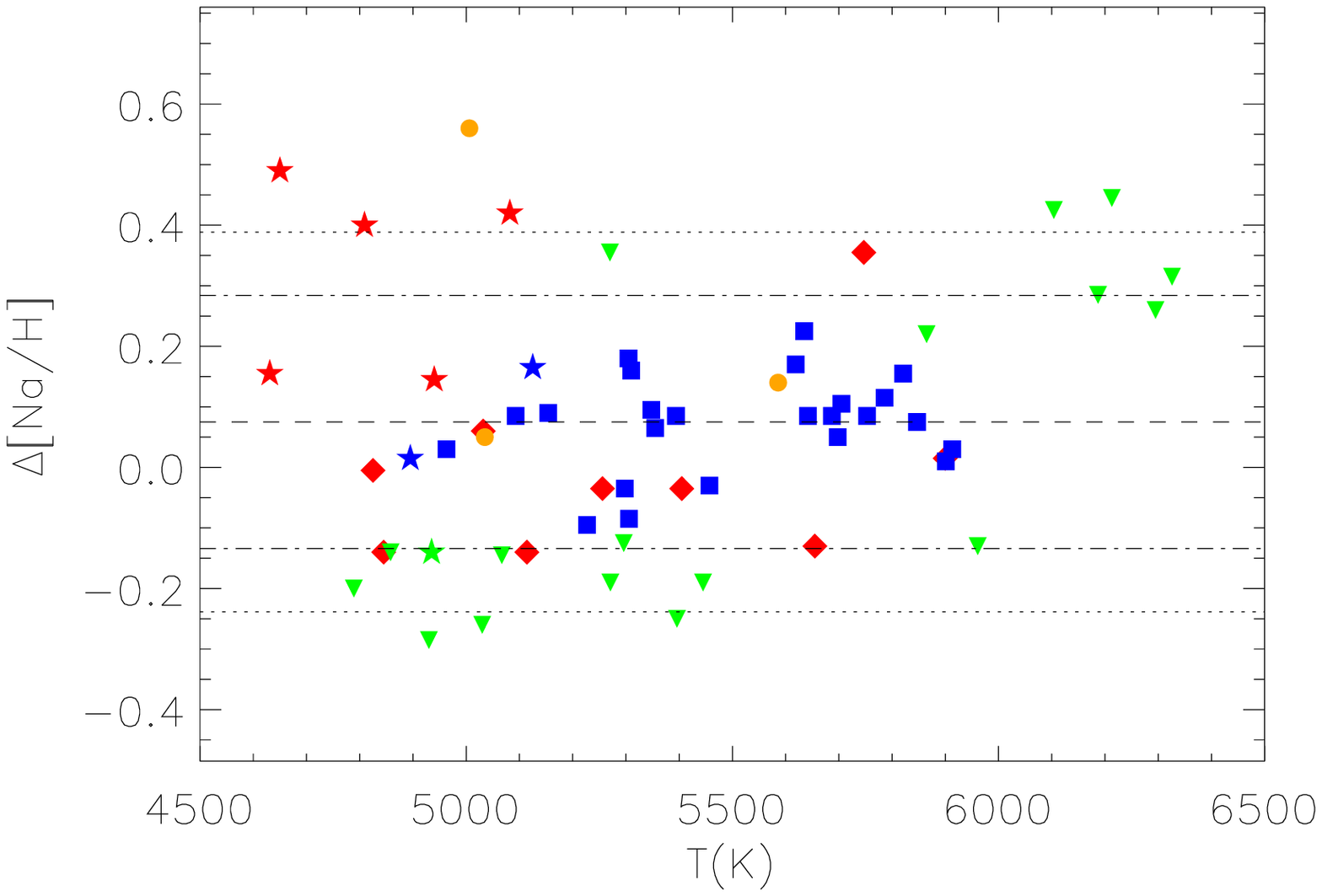}
\includegraphics[scale=0.55]{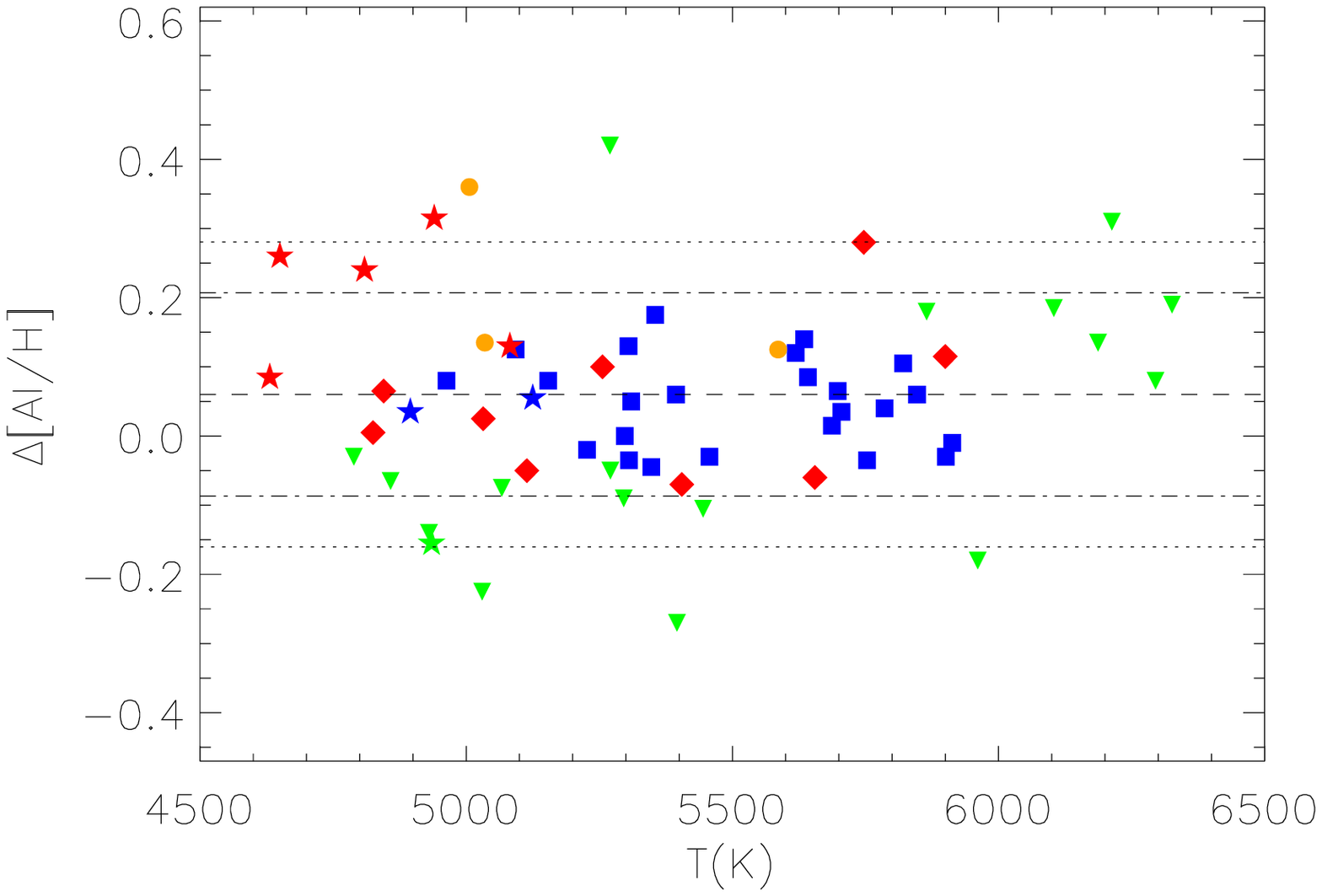}}
\centerline{ \includegraphics[scale=0.55]{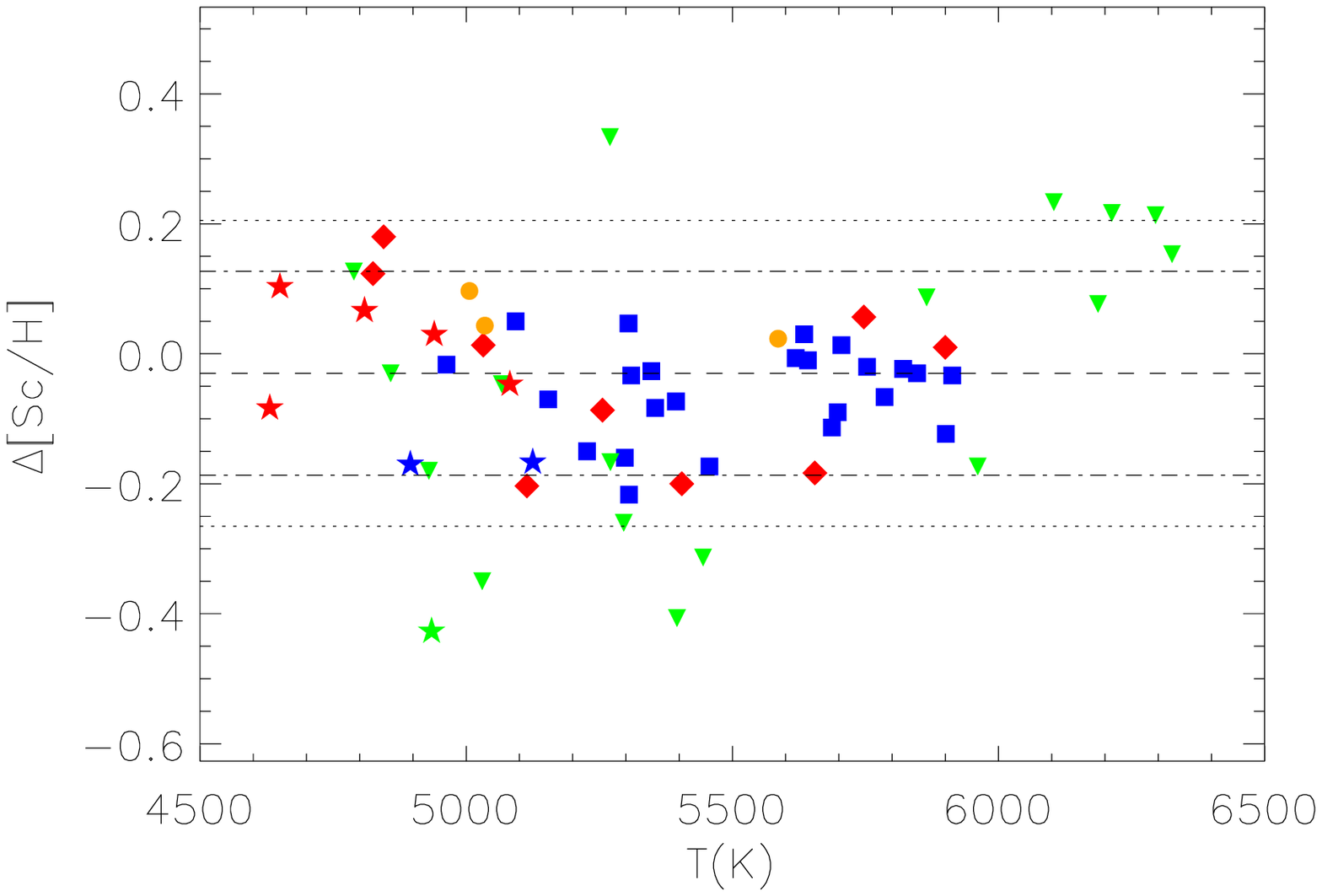}
 \includegraphics[scale=0.55]{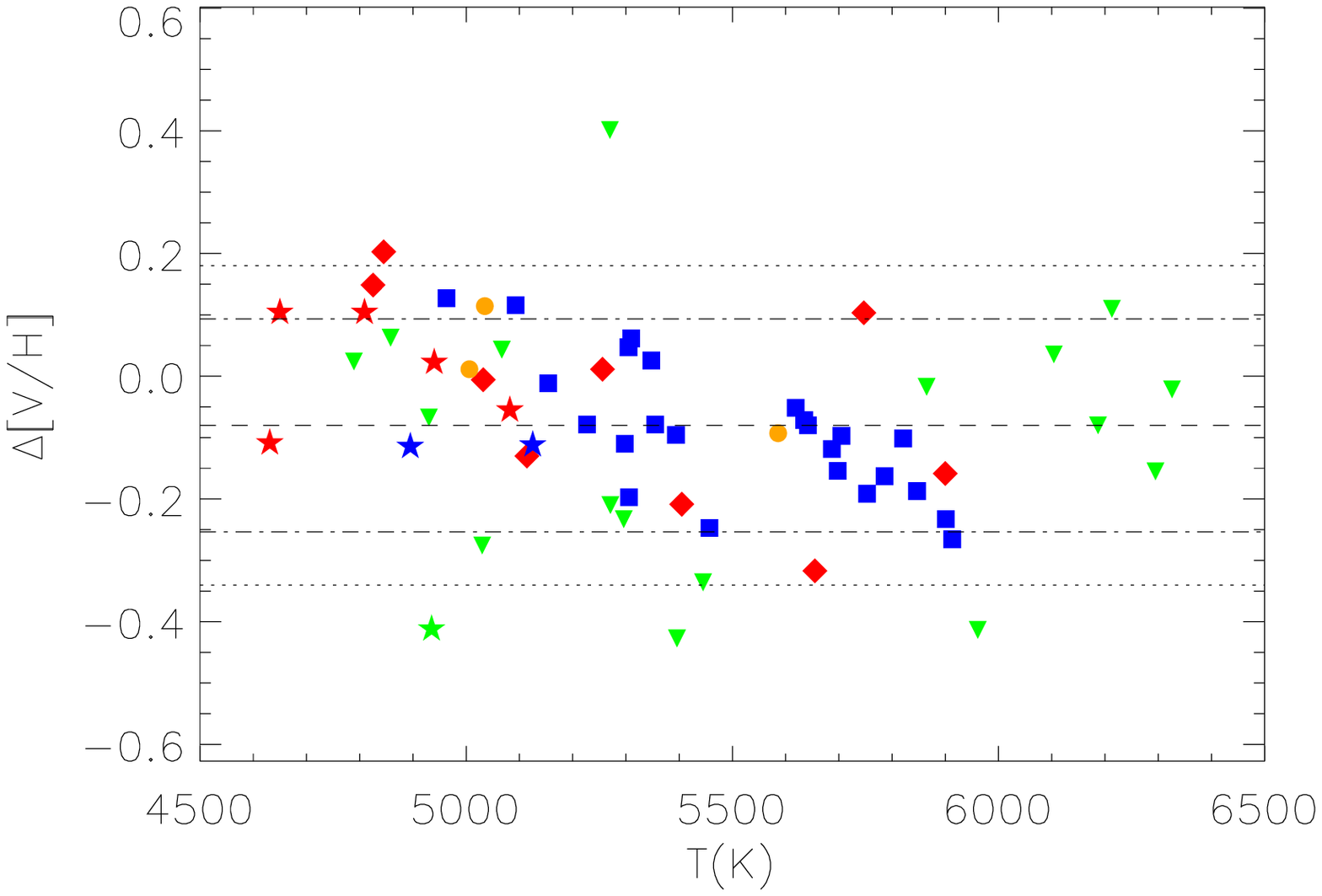}}
\caption{Same as Fig.~\ref{hyadif1} but for the odd-Z elements  Na, Al, Sc, and V.}
\label{hyadif3}
\end{figure*}

\begin{figure*}
\centering
\centerline{ \includegraphics[scale=0.55]{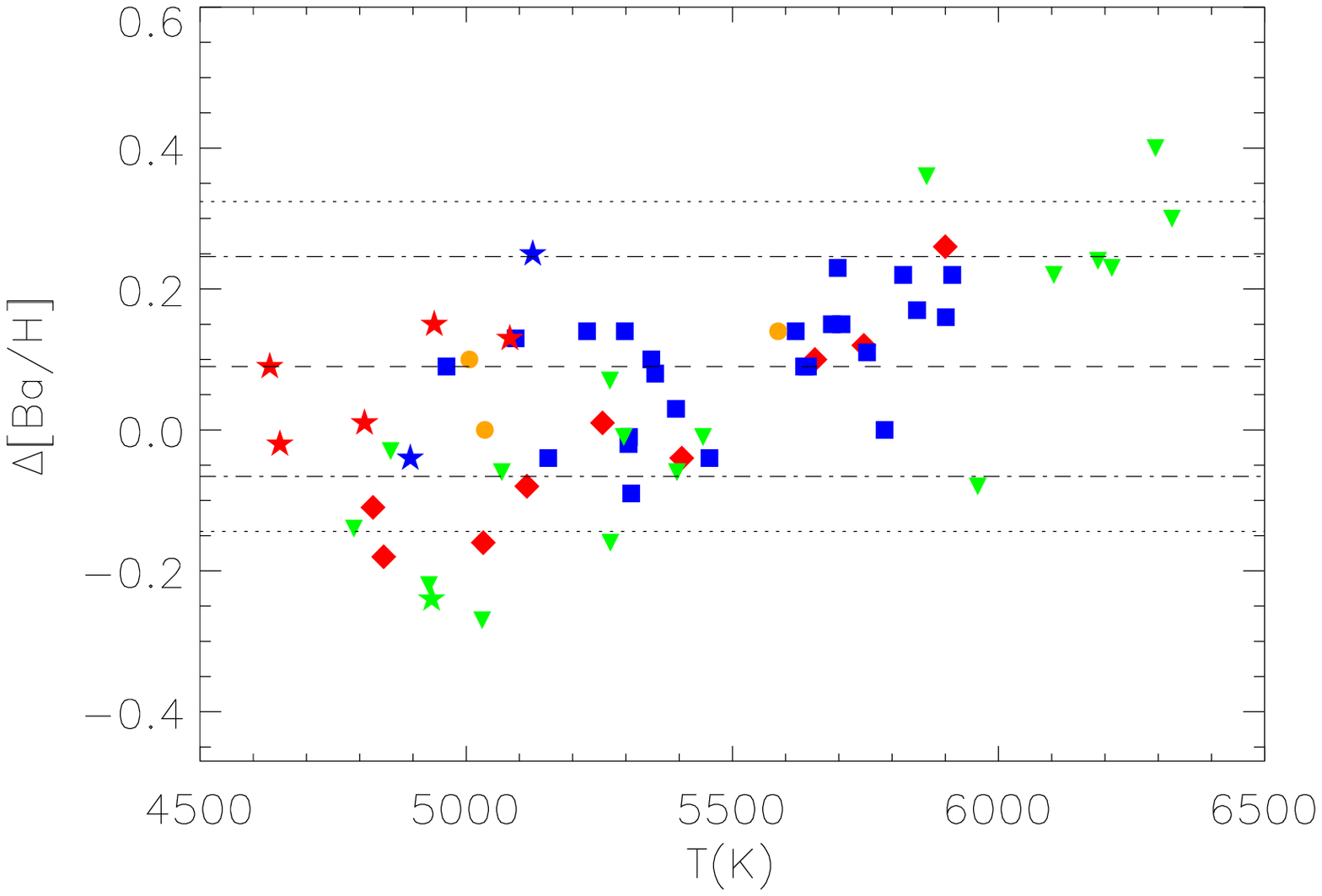}
\includegraphics[scale=0.55]{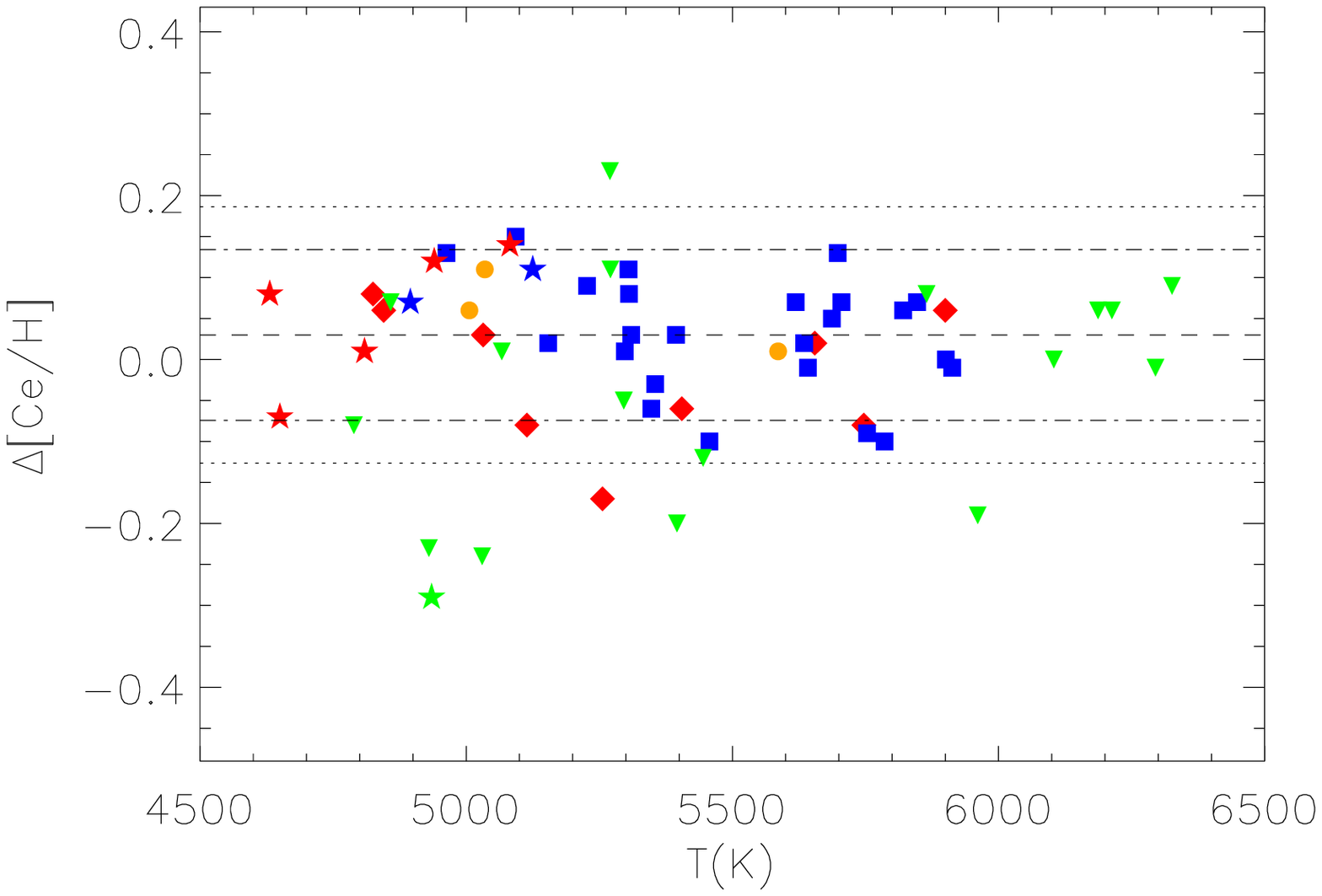}}
\centerline{ \includegraphics[scale=0.55]{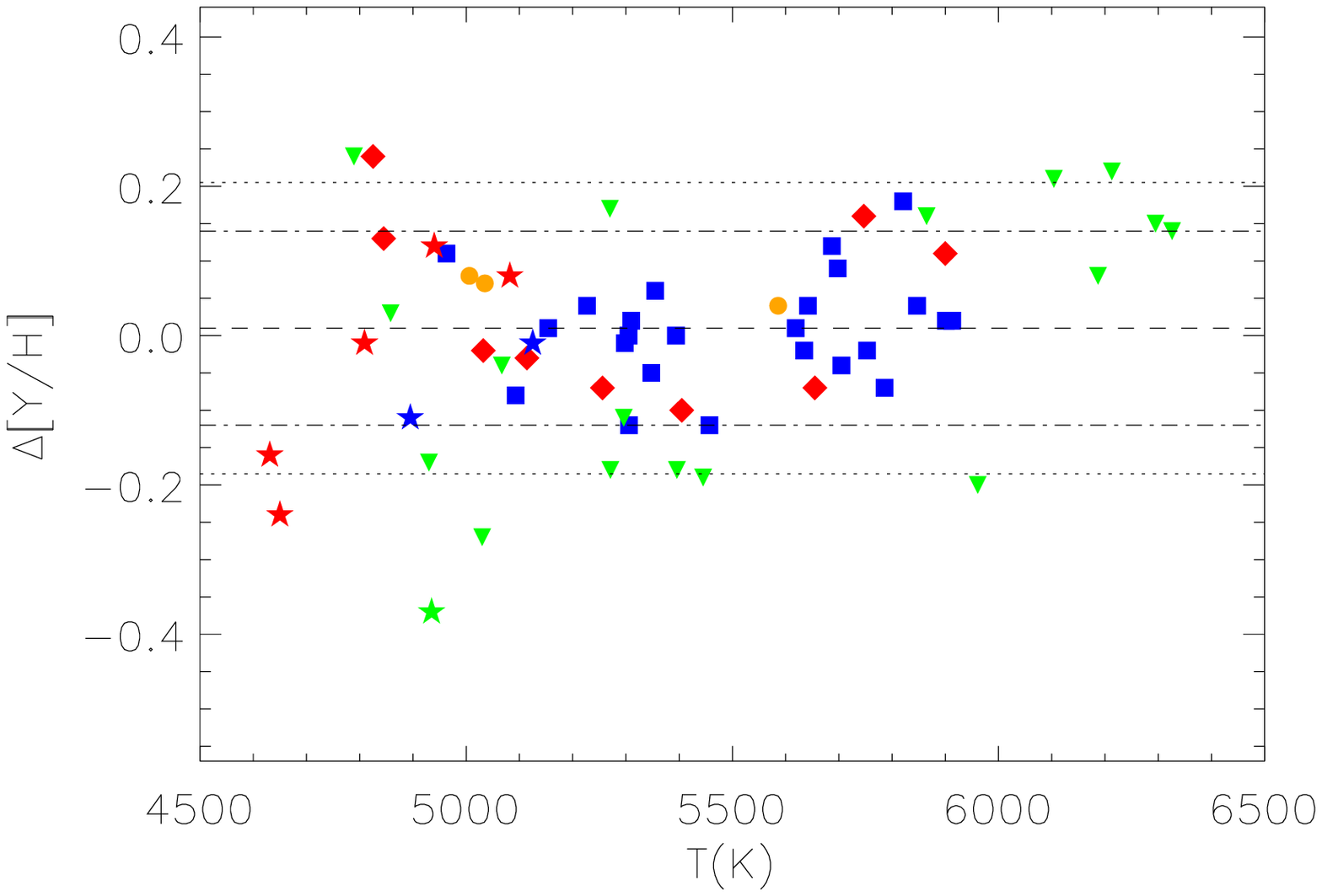}
 \includegraphics[scale=0.55]{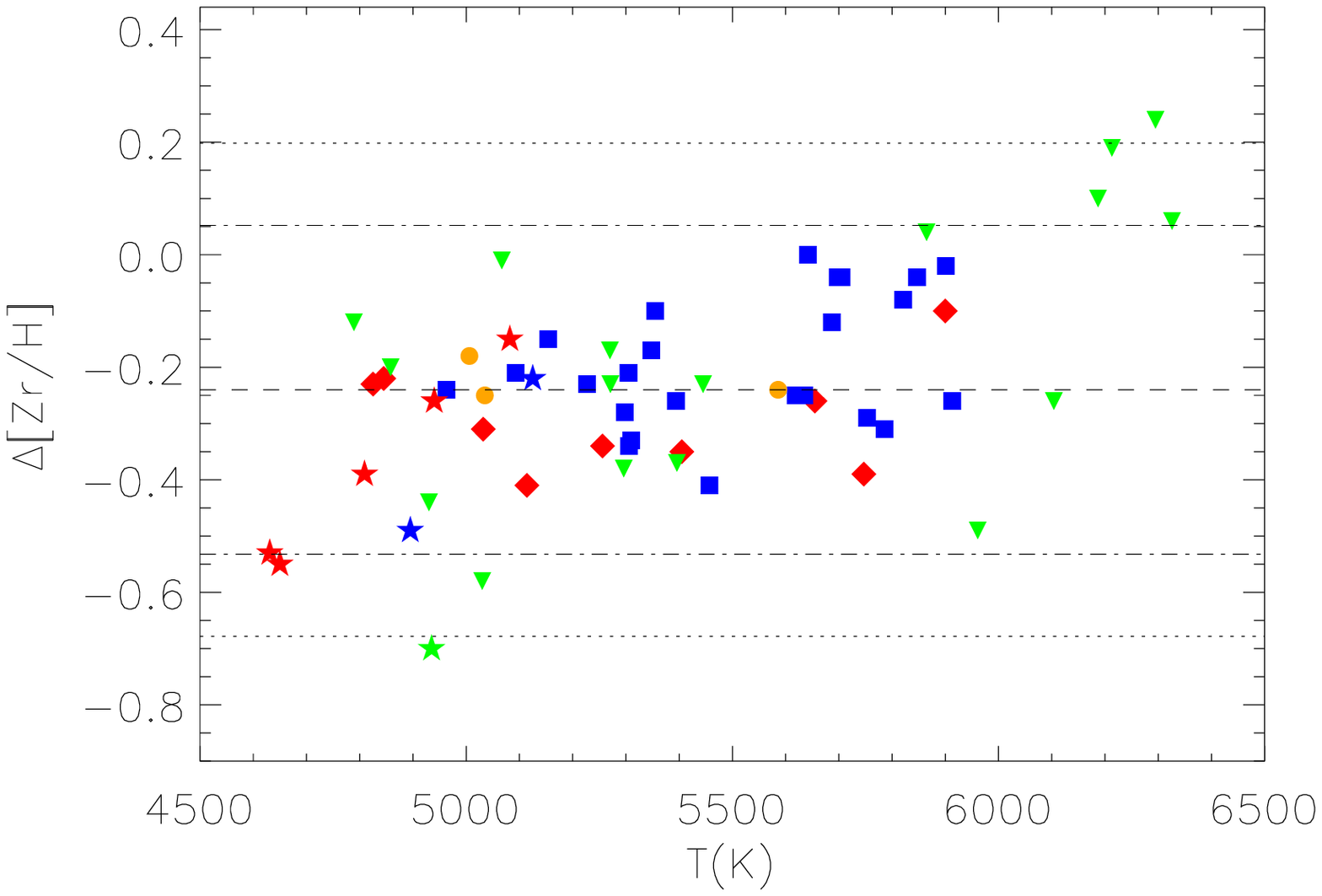}}
\caption{Same as Fig.~\ref{hyadif1} but for the s-process elements  Ba, Ce, Y, and Zr.}
\label{hyadif4}
\end{figure*}

\begin{figure*}
\centering
\centerline{ \includegraphics[scale=0.55]{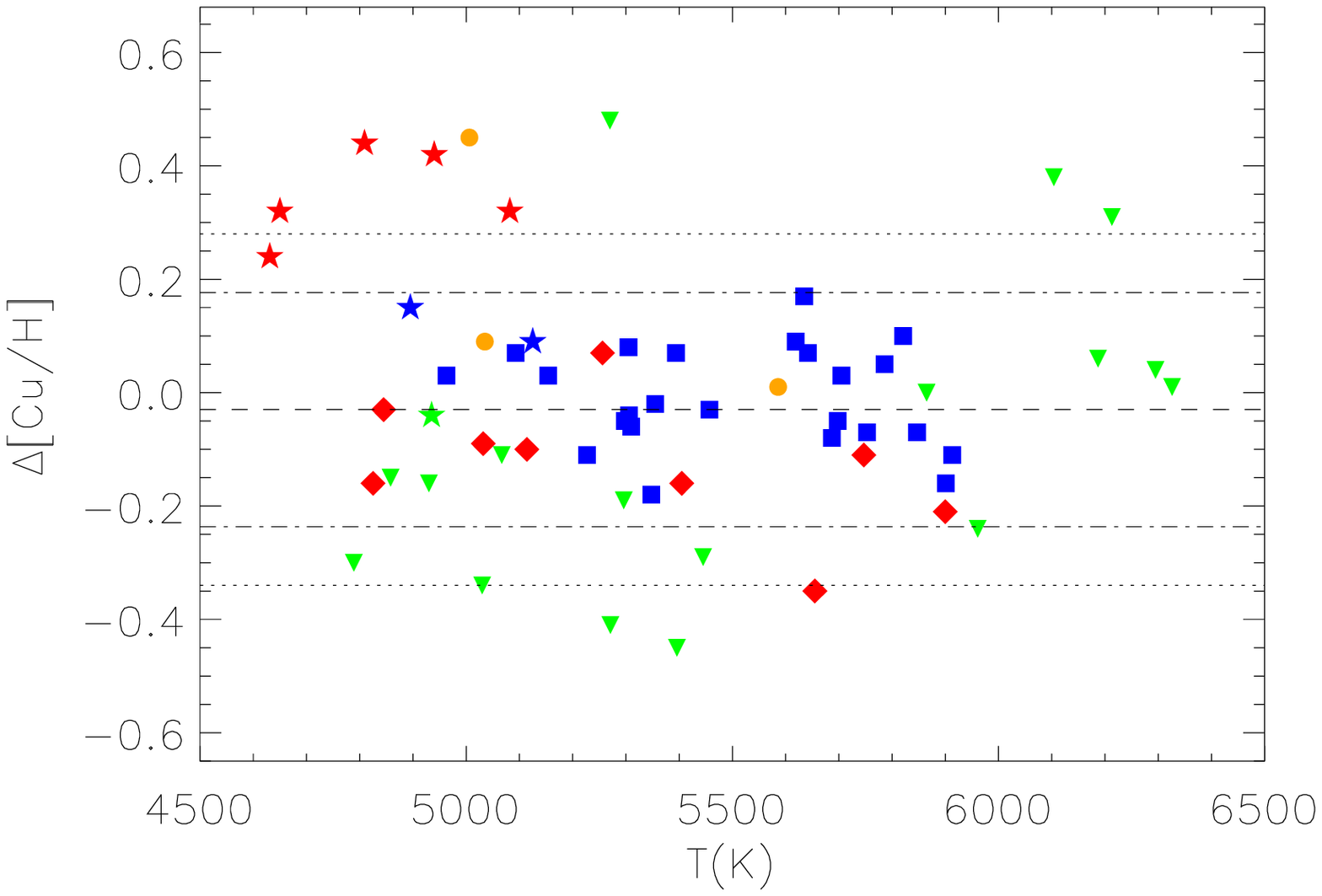}
\includegraphics[scale=0.55]{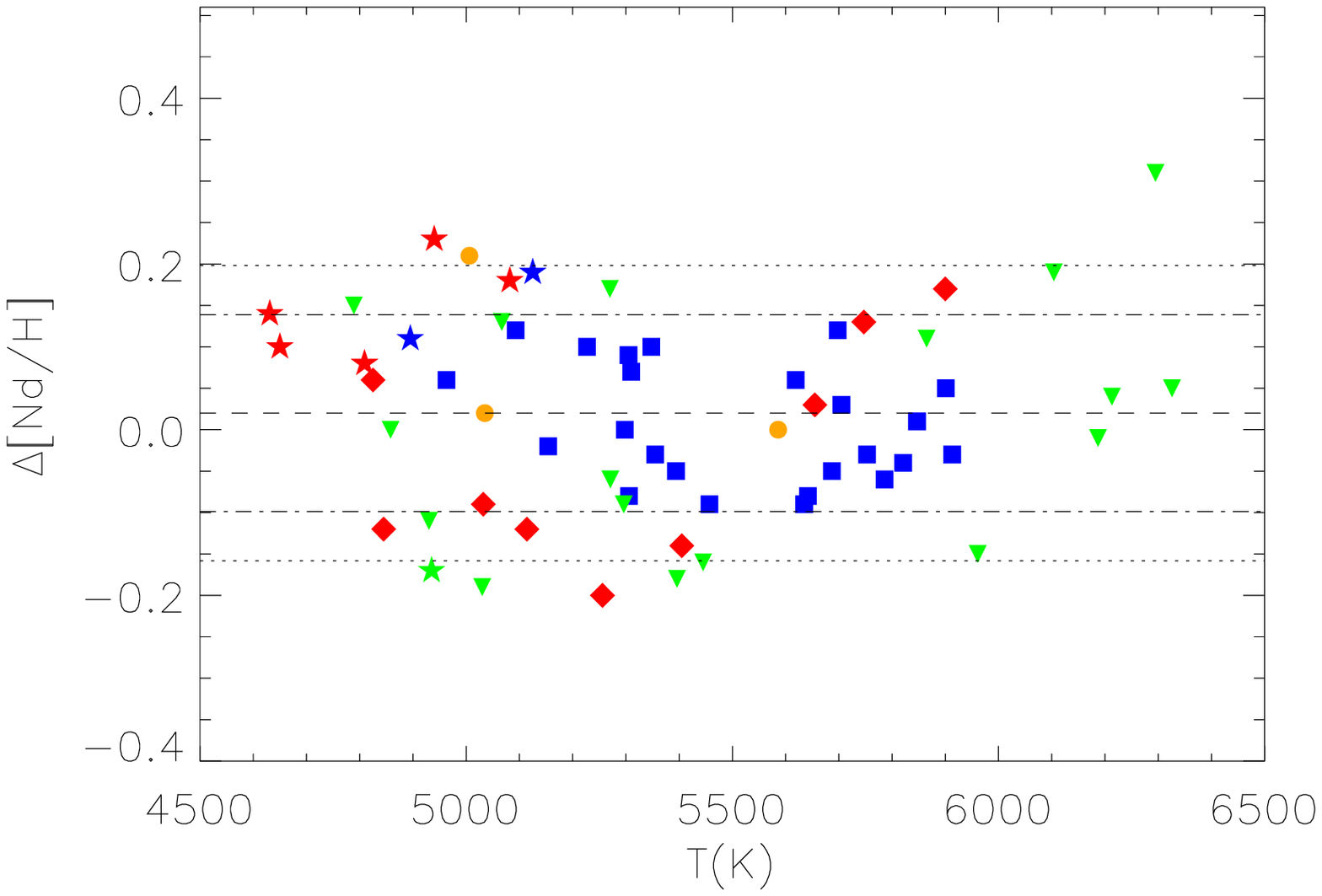}}
\includegraphics[scale=0.55]{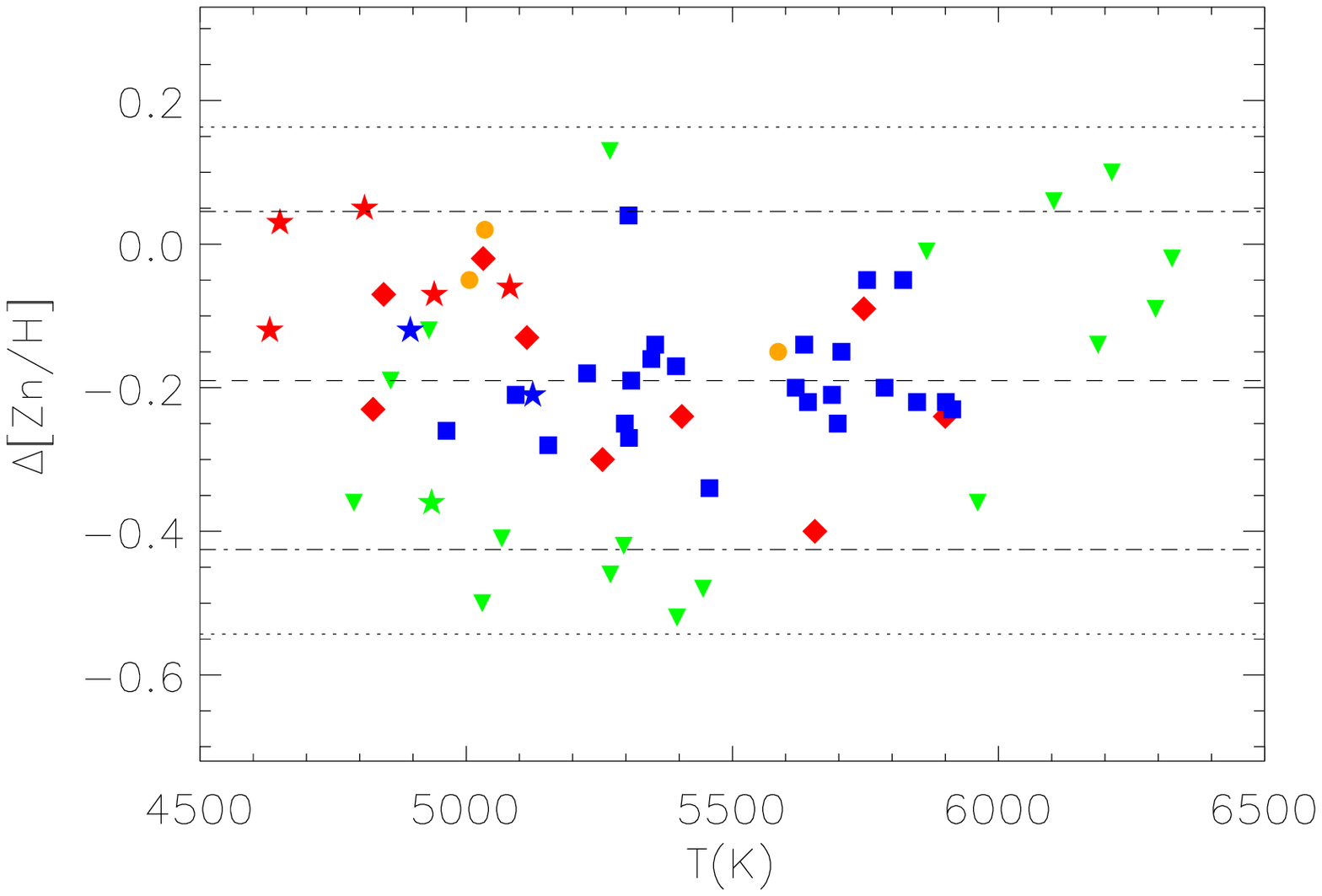}
\caption{Same as Fig.~\ref{hyadif1} but for the s-process elements  Cu, Nd, and Zn.}
\label{hyadif5}
\end{figure*}

%\begin{landscape}
\begin{table*}
\scriptsize
\caption{Properties of the sample, including radial velocities  and derived kinematics.}
\label{tablavel}
\centering
\begin{tabular}{llcccccccccc}
\hline\hline
\noalign{\smallskip}
{\small Name} & {\small HIP } & {\small $m_V$} & {\small SpT}   & {\small $\alpha$} & {\small $\delta$} & {\small $V_r$}  & {\small $U$} & {\small $V$}  & {\small $W$} \\
 & &  (mag) & & (J2000) & (J2000) &  (km s$^{-1}$) &  (km s$^{-1}$) & (km s$^{-1}$) & (km s$^{-1}$) \\ 
\noalign{\smallskip}
\hline
\noalign{\smallskip}
BE Cet & 1803 & 6.38 & G3~V &   00 22 51.55 & -12  12 34.50 & -2.40 $\pm$ 0.02 &  -35.79 $\pm$ 0.67 & -14.60  $\pm$  0.27 &   -0.40 $\pm$ 0.06 \\
HD  5848 & 5372 & 4.24 & K2~III & 01 08 44.16 &  86  15 25.62 & 8.32 $\pm$ 0.02 &  -35.57 $\pm$ 1.52 & -13.00  $\pm$  0.94 &    1.02 $\pm$ 0.27 \\
AZ Ari & 10218 & 7.32 & G5~V & 02 11 22.98 &  21  22 38.56 & 20.04 $\pm$ 0.02 &  -42.59 $\pm$ 0.94 & -19.72  $\pm$  0.92 &   -0.78 $\pm$ 0.39 \\
FT Cet & 12158 & 8.10 & K2.5~V & 02  36 41.57 & -03 09 22.60 & 21.91 $\pm$ 0.02 &  -39.79 $\pm$ 0.84 & -18.27  $\pm$  0.62 &   -0.57 $\pm$ 0.54 \\
BZ Cet & 13976 & 7.95 & K2.5~V &  03 00 02.62 &   07 44 58.93 & 28.86 $\pm$ 0.02 &  -42.74 $\pm$ 0.64 & -18.85  $\pm$  0.67 &   -1.18 $\pm$ 0.54 \\
$\delta$ Ari & 14838 & 4.35 & K2~III &  03 11 37.67 &  19 43 36.11 & 23.08 $\pm$ 0.11 &  -60.17 $\pm$ 5.44 & -48.96  $\pm$  7.19 &   24.71 $\pm$ 4.87 \\
V683 Per & 14976 & 8.15 & G5~V &  03 13  02.66 &  32 53 47.75 & 27.32 $\pm$ 0.02 &  -42.52 $\pm$ 0.98 & -21.27  $\pm$  1.62 &   -0.43 $\pm$ 0.51 \\
V686 Per & 15609 & 7.95 & K0~V &  03 20 59.33 &  33 13 06.96 & 34.84 $\pm$ 0.05 &  -45.63 $\pm$ 0.67 & -20.81  $\pm$  1.45 &  -11.22 $\pm$ 0.17 \\
HD 21663 & 16329 & 8.30 & G5~V &  03 30 30.33 &  20 06 12.30 & 25.71 $\pm$ 0.02 &  -35.55 $\pm$ 0.97 & -27.25  $\pm$  2.26 &   -2.35 $\pm$ 0.77 \\
HD 23356 & 17420 & 7.08 & K2~V &  03 43 55.15 & -19 06 40.61 & 25.16 $\pm$ 0.02 &  -30.46 $\pm$ 0.22 & -14.90  $\pm$  0.10 &   -4.36 $\pm$ 0.19 \\
39 Tau & 19076 & 5.90 & G5~V &  04 05 20.15 &  22 00 33.20 & 23.97 $\pm$ 0.03 &  -25.48 $\pm$ 0.06 & -13.31  $\pm$  0.24 &   -6.53 $\pm$ 0.07 \\
HD 25893 & 19255 & 7.08 & G5~V &  04 07 34.22 &  38 04 30.31 & 27.51 $\pm$ 0.09 &  -34.03 $\pm$ 0.23 & -16.89  $\pm$  0.64 &   -9.46 $\pm$ 0.14 \\
HD 27282 & 20146 & 8.43 & G8~V &  04 19 07.94 &  17 31 29.36 & 38.15 $\pm$ 0.03 &  -41.84 $\pm$ 0.43 & -19.60  $\pm$  1.35 & -1.07 $\pm$ 0.89 \\
HD 27685 & 20441 & 7.82 & G4~V &  04 22 44.68 &  16 47 27.72 & 38.75 $\pm$ 0.03 &  -45.43 $\pm$ 0.90 & -17.84  $\pm$  2.13 &    7.30 $\pm$ 2.13 \\
HD 27989 & 20686 & 7.54 & G5~V &  04 25 51.66 &  18 51 50.93 & 39.69 $\pm$ 0.05 &  -42.59 $\pm$ 0.32 & -18.11  $\pm$  1.10 &   -1.81 $\pm$ 0.72 \\
$\epsilon$ Tau & 20889 & 3.54 & G9.5~III & 04 28 36.93 &  19 10 49.88 & 38.50 $\pm$ 0.12 &  -41.61 $\pm$ 0.25 & -20.13  $\pm$  0.88 &   -0.62 $\pm$ 0.53 \\
111 Tau B & 25220 & 7.92 & K4~V &  05 23 38.23 &  17 19 26.87 & 38.31 $\pm$ 0.02 &  -38.76 $\pm$ 0.04 & -14.56  $\pm$  0.22 &    7.09 $\pm$ 0.31 \\
HD 40979 & 28767 & 6.73 & F8~V &   06 04 29.87 &  44 15 38.93 & 32.63 $\pm$ 0.05 &  -36.63 $\pm$ 0.15 & -21.39  $\pm$  0.77 &    8.34 $\pm$ 0.14 \\
HD 45609 & 30733 & 8.43 & K0~V &  06 27 24.11 & -25 44 05.22 & 26.44 $\pm$ 0.06 &  -28.20 $\pm$ 1.45 & -12.73  $\pm$  0.87 &   -2.26 $\pm$ 0.67 \\
HD 52265 & 33719 & 6.30 & G0~V &  07 00 18.10 & -05 22 02.49 & 53.93 $\pm$ 0.03 &  -52.38 $\pm$ 0.25 & -20.76  $\pm$  0.31 &   -9.26 $\pm$ 0.23 \\
HD 53532 & 34271 & 8.27 & G0~V &  07 06 16.86 &  22 41 01.24 & 42.95 $\pm$ 0.03 &  -44.08 $\pm$ 0.19 & -18.38  $\pm$  0.42 &  -13.15 $\pm$ 1.08 \\
HD 65523 & 39068 & 8.35 & G5~V &  07 59 35.78 &  12 58 59.42 & 33.05 $\pm$ 0.02 &  -43.44 $\pm$ 0.85 & -13.03  $\pm$  0.18 &  -23.57 $\pm$ 1.84 \\
HD 70088 & 40942 & 8.51 & G5~V &  08 21 20.82 &  34 18 36.85 & 33.23 $\pm$ 0.02 &  -40.53 $\pm$ 0.67 & -22.50  $\pm$  0.97 &   -5.93 $\pm$ 1.24 \\
HD 72760 & 42074 & 7.33 & G5~V &  08 34 31.76 &  00 43 34.04 & 35.32 $\pm$ 0.02 &  -35.66 $\pm$ 0.30 & -19.54  $\pm$  0.11 &   -1.50 $\pm$ 0.34 \\
V401 Hya & 42333 & 6.72 & G5~V & 08 37 50.47 & -06 48 25.16 & 35.47 $\pm$ 0.03 &  -43.17 $\pm$ 0.57 & -19.77  $\pm$  0.17 &  -12.07 $\pm$ 0.61 \\
HD 73171 & 42452 & 5.92 & K1~III & 08 39 17.65 &  52  42 42.12 & 26.70 $\pm$ 0.02 &  -26.86 $\pm$ 0.68 &  24.53  $\pm$  1.91 &    2.01 $\pm$ 1.44 \\
CT Pyx & 42281 & 8.74 & K1~V &   08 37 15.61 & -17  29 41.25 & 28.14 $\pm$ 0.02 &  -31.60 $\pm$ 0.82 & -16.78  $\pm$  0.32 &   -3.60 $\pm$ 0.49 \\
$\rho$ Cnc A & 43587 & 5.95 & G8~V &   08 52 36.13 &  28  19 53.00 & 27.53 $\pm$ 0.02 &  -37.27 $\pm$ 0.18 & -18.20  $\pm$  0.13 &   -7.95 $\pm$ 0.27 \\
HD 75898 & 43674 & 8.04 & G0~V &   08 53 50.87 &  33  03 24.77 & 21.85 $\pm$ 0.03 &  -39.23 $\pm$ 1.98 & -11.97  $\pm$  0.84 &  -15.72 $\pm$ 2.59 \\
HD 76151 & 43726 & 6.00 & G3~V &   08 54 18.19 & -05  26 04.32 & 32.10 $\pm$ 0.02 &  -40.35 $\pm$ 0.35 & -20.05  $\pm$  0.06 &  -11.25 $\pm$ 0.37 \\
IK Cnc & 43751 & 8.32 & G5~V &  08 54 41.60 &  16  36 40.60 & 15.89 $\pm$ 0.05 &  -28.55 $\pm$ 2.67 & -10.61  $\pm$  0.94 &  -15.49 $\pm$ 3.76 \\
HD 82106 & 46580 & 7.20 & K3~V &   09 29 55.12 &  05  39 17.52 & 29.98 $\pm$ 0.15 &  -40.89 $\pm$ 0.34 & -13.79  $\pm$  0.11 &    0.12 $\pm$ 0.26 \\
GT Leo & 47587 & 8.88 & K0~V &   09 42 10.04 &  07  35 24.49 & 30.96 $\pm$ 0.05 &  -44.84 $\pm$ 1.58 & -19.70  $\pm$  0.21 &   -4.23 $\pm$ 1.34 \\
HD 85301 & 48423 & 7.73 & G5~V &   09 52 16.96 &  49  11 27.47 & 15.15 $\pm$ 0.07 &  -34.83 $\pm$ 0.80 & -12.31  $\pm$  0.46 &   -7.08 $\pm$ 0.59 \\
HD 86322 & 49163 & 6.89 & K1~III & 10 01 59.43 &  74  45 33.05 & 1.25 $\pm$ 0.09 &  -33.23 $\pm$ 3.04 & -24.67  $\pm$  2.38 &   -6.58 $\pm$ 0.76 \\
HD 89307 & 50473 & 7.03 & G0~V &  10  18 21.45 &  12  37 16.33 & 23.06 $\pm$ 0.03 &  -39.98 $\pm$ 0.93 & -23.15  $\pm$  0.40 &   -5.73 $\pm$ 0.73 \\
HD 89376 & 50524 & 9.01 & K5~V &  10  19 10.74 &  20  33 48.25 & 19.94 $\pm$ 0.06 &  -44.89 $\pm$ 4.53 & -18.38  $\pm$  1.53 &   -9.03 $\pm$ 3.23 \\
$\pi^1$ Leo & 53273 & 5.45 & G8~III &  10  53 43.76 & -02  07 45.27 & 20.10 $\pm$ 0.08 &  -39.32 $\pm$ 2.81 & -18.51  $\pm$  0.53 &    1.94 $\pm$ 1.06 \\
HD 98356 & 55235 & 8.73 & K0~V &  11  18 39.87 & -10 07 35.00 & 7.41 $\pm$ 0.08 &  -55.59 $\pm$ 2.94 & -15.14  $\pm$  0.55 &   -5.38 $\pm$ 0.59 \\
HD 101112 & 56756 & 6.19 & K1~III &  11  38  09.87 &  08  53  01.60 & 11.58 $\pm$ 0.12 &  -30.52 $\pm$ 2.80 & -14.67  $\pm$  0.99 & 2.64 $\pm$ 0.77 \\
HD 106696 & 56583 & 5.19 & K0~III &  11  36  02.62 &  69  19 23.73 & -3.53 $\pm$ 0.02 &   41.53 $\pm$ 1.52 & -20.99  $\pm$  0.74 &   36.59 $\pm$ 1.49 \\
MY UMa & 57859 & 9.57 & K0~V &  11  51 58.07 &  48 05 18.85 & -0.65 $\pm$ 0.12 &  -45.62 $\pm$ 6.96 & -13.06  $\pm$  2.03 &  -14.80 $\pm$ 2.16 \\
GR Leo & 58314 & 8.07 & G5~V &  11  57 29.17 &  19 59 01.58 & 5.09 $\pm$ 0.02 &  -50.60 $\pm$ 1.70 & -18.02  $\pm$  0.59 &   -5.13 $\pm$ 0.34 \\
HD 106023 & 59486 & 8.70 & K0~V &  12  11 59.57 &  23 05 35.70 & 6.33 $\pm$ 0.04 &  -43.06 $\pm$ 2.19 & -30.76  $\pm$  1.54 &   -2.18 $\pm$ 0.44 \\
HD 117860 & 66115 & 7.21 & G0~V &  13  33 11.45 &  -8 26 36.47 & -9.53 $\pm$ 0.03 &  -36.13 $\pm$ 0.93 & -13.30  $\pm$  0.50 &    2.59 $\pm$ 0.30 \\
HD 117936 & 66147 & 7.96 & K3~V &  13  33 32.70 &   8 35 11.52 & -6.21 $\pm$ 0.10 &  -41.05 $\pm$ 0.75 & -19.35  $\pm$  0.39 &    4.07 $\pm$ 0.21 \\
HD 124642 & 69526 & 8.06 & K5~V &  14  13 57.35 &  30 13 00.34 & -16.10 $\pm$ 0.10 &  -36.10 $\pm$ 0.60 & -14.27  $\pm$  0.21 & -5.39 $\pm$ 0.21 \\
HD 126535 & 70608 & 8.86 & K1~V &  14  26 34.82 & -18 49 12.21 & -19.50 $\pm$ 0.03 &  -40.34 $\pm$ 1.78 & -18.86  $\pm$  1.73 &    2.40 $\pm$ 0.98 \\
HD 131023 & 72634 & 7.40 & K0~V &  14  51 02.44 &  09 43 24.72 & -42.70 $\pm$ 0.04 &  -44.39 $\pm$ 0.74 & -15.74  $\pm$  0.47 &  -20.99 $\pm$ 0.51 \\
HD 149026 & 80838 & 8.15 & G0~IV &  16  30 29.68 &  38  20 49.85 & -17.90 $\pm$ 0.04 & -32.36 $\pm$ 1.66 & -20.62  $\pm$  0.61 &    9.60 $\pm$ 1.38 \\
HD 170657 & 90790 & 6.81 & K2~V &  18  31 19.05 & -18  54 30.02 & -43.00 $\pm$ 0.11 &  -37.93 $\pm$ 0.12 & -24.64  $\pm$  0.18 &    5.39 $\pm$ 0.06 \\
MV Dra & 94346 & 7.04 & G8~V &  19  12 11.12 &  57  40 15.56 & -27.00 $\pm$ 0.08 &  -43.99 $\pm$ 0.47 & -21.89  $\pm$  0.09 &  -15.46 $\pm$ 0.09 \\
HD 189087 & 98192 & 7.89 & K1~V &  19  57 13.35 &  29  49 24.39 & -30.30 $\pm$ 0.05 &  -40.64 $\pm$ 0.71 & -15.47  $\pm$  0.31 &    3.52 $\pm$ 0.12 \\
V2425 Cyg & 101227 & 7.45 & K0~V &  20  31  06.89 &  40 51 28.52 & -23.50 $\pm$ 0.04 &  -30.27 $\pm$ 1.26 & -15.57  $\pm$  0.37 &    1.67 $\pm$ 0.28 \\
HD 196885 & 101966 & 6.39 & F8~IV &  20  39 51.85 &  11  14 58.01 & -30.60 $\pm$ 0.04 &  -28.39 $\pm$ 0.34 & -15.68  $\pm$  0.24 &   10.41 $\pm$ 0.11 \\
V2436 Cyg & 103859 & 7.68 & K3~V &  21  02 40.42 &  45  53 03.94 & -14.10 $\pm$ 0.16 &  -34.94 $\pm$ 0.48 & -12.52  $\pm$  0.16 &  -18.89 $\pm$ 0.27 \\
HD 200968 & 104239 & 7.09 & K1~IV &  21  07 10.15 & -13  55 22.14 & -32.80 $\pm$ 0.02 &  -41.86 $\pm$ 0.43 & -18.27  $\pm$  0.08 &   -5.49 $\pm$ 0.53 \\
NS Aqr & 105066 & 8.08 & K0~V &  21  17  02.04 &  -01  04 38.53 & -22.60 $\pm$ 0.02 &  -32.56 $\pm$ 1.18 &  -18.87  $\pm$  0.30 &  -13.38 $\pm$ 1.47 \\
V454 And & 116613 & 6.59 & G3/4~V &  23  37 58.19 &  46  11 58.07 & -0.40 $\pm$ 0.03 &  -33.73 $\pm$ 0.63 & -16.02  $\pm$  0.29 &  -12.02 $\pm$ 0.23 \\
HD 222422 & 116819 & 8.03 & G5~V &  23  40 37.67 & -18  59 19.65 & 10.20 $\pm$ 0.02 &  -28.82 $\pm$ 0.81 & -17.15  $\pm$  0.53 &  -20.72 $\pm$ 0.29 \\
20 Psc & 117375 & 5.51 & G8~III &  23  47 56.49 &  -02  45 41.80 & -15.30 $\pm$ 0.02 &  -37.39 $\pm$ 2.88 & -22.75  $\pm$  1.22 &    4.37 $\pm$ 0.71 \\
\noalign{\smallskip}
\hline
\noalign{\smallskip}
\end{tabular}
\end{table*}
%\end{landscape}

%\begin{landscape}
\begin{table*}
\scriptsize
\caption{Stellar parameters for the whole sample, Fe differential abundance ($\Delta$[Fe/H]), and the resulting membership based on the abundance homogeneity criterion ({\it chemical tagging}).  }
\label{tablapar}
%\centering
\begin{tabular}{lcccccccccc}
\hline
\hline
\noalign{\smallskip}
{\small Name} & {\small $T_{\rm eff}$} & {\small $\log{g}$} & {\small $\xi$ } & {\small $\log{\epsilon ({\ion{Fe}{i}})}$ } & {\small $\log{\epsilon ({\ion{Fe}{ii}}})$ } &{\small [Fe/H]} & {\small$\Delta$[Fe/H]} & {\small Member (Fe)\tablefootmark{a}} & {\small Total\tablefootmark{b}} & {\small Member\tablefootmark{c}} \\
 & (K) & (dex) & (km s$^{-1}$) & & & & & & (1-rms) &  \\
\hline
\hline
\noalign{\smallskip}
vB 153 & 5235 $\pm$ 36  & 4.45 $\pm$ 0.11 & 1.14 $\pm$ 0.06 & 7.53 $\pm$ 0.02 & 7.53 $\pm$ 0.06 &  0.073 $\pm$   0.005  &   ---  &  Y   & 20   &  Y  \\          
\noalign{\smallskip}
\hline
\noalign{\smallskip}
BE Cet & 5865 $\pm$  20 & 4.58 $\pm$ 0.05 & 1.26 $\pm$ 0.03 & 7.67 $\pm$ 0.02 & 7.67 $\pm$ 0.02 &  0.211 $\pm$  0.003  &  0.138 $\pm$  0.005 & N &  14   &   N \\
HD 5848 & 4650 $\pm$  44 & 2.48 $\pm$ 0.17 & 1.89 $\pm$ 0.05 & 7.53 $\pm$ 0.04 & 7.53 $\pm$ 0.12 &  0.068 $\pm$  0.010  & -0.006 $\pm$  0.010 & Y &  14   & N \\
AZ Ari & 5821 $\pm$  10 & 4.54 $\pm$ 0.03 & 1.10 $\pm$ 0.02 & 7.63 $\pm$ 0.01 & 7.63 $\pm$ 0.01 &  0.169 $\pm$  0.002  &  0.097 $\pm$  0.005 & Y &   19   & Y* \\
FT Cet & 5093 $\pm$  45 & 4.49 $\pm$ 0.12 & 1.10 $\pm$ 0.10 & 7.52 $\pm$ 0.03 & 7.52 $\pm$ 0.07 &  0.058 $\pm$  0.006  & -0.015 $\pm$  0.006 & Y &   18   & Y* \\
BZ Cet & 5035 $\pm$  37 & 4.38 $\pm$ 0.11 & 0.98 $\pm$ 0.08 & 7.56 $\pm$ 0.02 & 7.56 $\pm$ 0.07 &  0.106 $\pm$  0.005  &  0.034 $\pm$  0.005 & Y &  19    & Y* \\
$\delta$ Ari & 4940 $\pm$  37 & 3.05 $\pm$ 0.11 & 1.52 $\pm$ 0.04 & 7.62 $\pm$ 0.03 & 7.62 $\pm$ 0.07 &  0.163 $\pm$  0.008 & 0.093 $\pm$  0.008 & Y &  13   & N \\
V683 Per & 5586 $\pm$  20 & 4.50 $\pm$ 0.06 & 1.04 $\pm$ 0.04 & 7.58 $\pm$ 0.02 & 7.58 $\pm$ 0.03 &  0.129 $\pm$  0.004  & 0.053 $\pm$  0.005 & Y &     20   & Y \\
V686 Per & 5705 $\pm$  37 & 4.54 $\pm$ 0.10 & 1.28 $\pm$ 0.05 & 7.60 $\pm$ 0.03 & 7.61 $\pm$ 0.05 &  0.141 $\pm$  0.005  & 0.073 $\pm$  0.006 & Y &     20   & Y \\
HD 21663 & 5457 $\pm$  18 & 4.50 $\pm$ 0.05 & 0.84 $\pm$ 0.04 & 7.43 $\pm$ 0.01 & 7.43 $\pm$ 0.03 & -0.031 $\pm$  0.003  & -0.104 $\pm$  0.004 & Y &  	18   & Y*\\
HD 23356 & 4930 $\pm$  36 & 4.41 $\pm$ 0.09 & 0.63 $\pm$ 0.12 & 7.29 $\pm$ 0.02 & 7.29 $\pm$ 0.06 & -0.167 $\pm$  0.005  & -0.241 $\pm$  0.005 & N &  	 6   & N \\
39 Tau & 5901 $\pm$  13 & 4.55 $\pm$ 0.04 & 1.13 $\pm$ 0.02 & 7.52 $\pm$ 0.01 & 7.52 $\pm$ 0.02 &  0.061 $\pm$  0.002  & -0.012 $\pm$  0.004 & Y &  20    & Y \\
HD 25893 & 5355 $\pm$  43 & 4.50 $\pm$ 0.11 & 1.18 $\pm$ 0.08 & 7.54 $\pm$ 0.03 & 7.54 $\pm$ 0.06 &  0.080 $\pm$  0.006  & 0.010 $\pm$  0.006 & Y &   20  & Y \\
HD 27282 & 5642 $\pm$  19 & 4.49 $\pm$ 0.06 & 1.09 $\pm$ 0.03 & 7.61 $\pm$ 0.01 & 7.61 $\pm$ 0.03 &  0.150 $\pm$  0.003  & 0.076 $\pm$  0.004 & Y &  20    & Y \\
HD 27685 & 5753 $\pm$  18 & 4.47 $\pm$ 0.05 & 0.99 $\pm$ 0.03 & 7.58 $\pm$ 0.01 & 7.57 $\pm$ 0.02 &  0.119 $\pm$  0.005  & 0.050 $\pm$  0.006 & Y &  19    & Y*\\
HD 27989 & 5747 $\pm$  50 & 4.57 $\pm$ 0.12 & 1.37 $\pm$ 0.07 & 7.64 $\pm$ 0.04 & 7.64 $\pm$ 0.05 &  0.179 $\pm$  0.007  & 0.107 $\pm$  0.007 & Y &   14   & N \\
$\epsilon$ Tau & 5006 $\pm$  40 & 2.94 $\pm$ 0.13 & 1.64 $\pm$ 0.04 & 7.66 $\pm$ 0.03 & 7.67 $\pm$ 0.08 &  0.210 $\pm$  0.009  & 0.142 $\pm$  0.008 & N &   9    & N \\
111 Tau B & 4789 $\pm$  64 & 4.53 $\pm$ 0.20 & 1.14 $\pm$ 0.18 & 7.27 $\pm$ 0.03 & 7.27 $\pm$ 0.14 & -0.195 $\pm$  0.009 & -0.260 $\pm$  0.008 & N &  6   & N \\
HD 40979 & 6295 $\pm$  43 & 4.61 $\pm$ 0.10 & 1.42 $\pm$ 0.05 & 7.73 $\pm$ 0.03 & 7.73 $\pm$ 0.05 &  0.267 $\pm$  0.006  & 0.204 $\pm$  0.007 & N &   7   & N \\
HD 45609 & 5271 $\pm$  39 & 4.49 $\pm$ 0.12 & 1.05 $\pm$ 0.08 & 7.36 $\pm$ 0.03 & 7.36 $\pm$ 0.06 & -0.102 $\pm$  0.007  & -0.172 $\pm$  0.007 & N &  10  & N \\
HD 52265 & 6187 $\pm$  16 & 4.42 $\pm$ 0.04 & 1.37 $\pm$ 0.02 & 7.69 $\pm$ 0.01 & 7.69 $\pm$ 0.02 &  0.231 $\pm$  0.003  & 0.162 $\pm$  0.005 & N &   14  & N \\
HD 53532 & 5698 $\pm$  17 & 4.56 $\pm$ 0.05 & 1.10 $\pm$ 0.03 & 7.58 $\pm$ 0.01 & 7.58 $\pm$ 0.02 &  0.119 $\pm$  0.003  & 0.045 $\pm$  0.004 & Y &   20   & Y \\
HD 65523 & 5306 $\pm$  20 & 4.46 $\pm$ 0.05 & 0.70 $\pm$ 0.04 & 7.41 $\pm$ 0.01 & 7.41 $\pm$ 0.03 & -0.050 $\pm$  0.003  & -0.118 $\pm$  0.005 & Y &  18   & Y* \\
HD 70088 & 5655 $\pm$  20 & 4.58 $\pm$ 0.06 & 1.06 $\pm$ 0.03 & 7.41 $\pm$ 0.02 & 7.41 $\pm$ 0.03 & -0.052 $\pm$  0.003  & -0.120 $\pm$  0.005 & Y &  13   & N \\
HD 72760 & 5298 $\pm$  21 & 4.45 $\pm$ 0.06 & 0.84 $\pm$ 0.04 & 7.48 $\pm$ 0.01 & 7.48 $\pm$ 0.03 &  0.018 $\pm$  0.003  & -0.054 $\pm$  0.004 & Y &  20   & Y \\
V401 Hya & 5847 $\pm$  13 & 4.52 $\pm$ 0.03 & 1.10 $\pm$ 0.02 & 7.57 $\pm$ 0.01 & 7.58 $\pm$ 0.02 &  0.115 $\pm$  0.002  & 0.044 $\pm$  0.005 & Y &  20  & Y \\
HD 73171 & 4631 $\pm$  35 & 2.50 $\pm$ 0.13 & 1.63 $\pm$ 0.04 & 7.42 $\pm$ 0.03 & 7.42 $\pm$ 0.09 & -0.039 $\pm$  0.008  & -0.108 $\pm$  0.008 & Y &  15  & N \\
CT Pyx & 5154 $\pm$  27 & 4.41 $\pm$ 0.08 & 1.00 $\pm$ 0.06 & 7.51 $\pm$ 0.02 & 7.51 $\pm$ 0.05 &  0.049 $\pm$  0.004  & -0.025 $\pm$  0.004 & Y &    20  & Y \\
$\rho$ Cnc A & 5270 $\pm$  39 & 4.31 $\pm$ 0.11 & 0.79 $\pm$ 0.08 & 7.77 $\pm$ 0.03 & 7.77 $\pm$ 0.06 &  0.314 $\pm$  0.005  & 0.242 $\pm$  0.004 & N &   2    & N \\
HD 75898 & 6104 $\pm$  14 & 4.36 $\pm$ 0.05 & 1.32 $\pm$ 0.02 & 7.75 $\pm$ 0.01 & 7.75 $\pm$ 0.02 &  0.291 $\pm$  0.003  & 0.218 $\pm$  0.005 & N &   5   & N \\
HD 76151 & 5786 $\pm$  12 & 4.51 $\pm$ 0.04 & 0.98 $\pm$ 0.01 & 7.56 $\pm$ 0.01 & 7.56 $\pm$ 0.02 &  0.101 $\pm$  0.002  & 0.026 $\pm$  0.005 & Y &   19  & Y* \\
IK Cnc & 5396 $\pm$  23 & 4.63 $\pm$ 0.06 & 0.83 $\pm$ 0.06 & 7.20 $\pm$ 0.02 & 7.20 $\pm$ 0.03 & -0.259 $\pm$  0.004  & -0.331 $\pm$  0.005 & N &  2  & N \\
HD 82106 & 4858 $\pm$  45 & 4.50 $\pm$ 0.13 & 0.72 $\pm$ 0.14 & 7.38 $\pm$ 0.02 & 7.38 $\pm$ 0.09 & -0.080 $\pm$  0.005  & -0.161 $\pm$  0.005 & N &  13  &  N \\
GT Leo & 5256 $\pm$  36 & 4.46 $\pm$ 0.09 & 0.85 $\pm$ 0.08 & 7.49 $\pm$ 0.02 & 7.49 $\pm$ 0.05 &  0.034 $\pm$  0.005  & -0.040 $\pm$  0.005 & Y &   18  &   N \\
HD 85301 & 5687 $\pm$  20 & 4.56 $\pm$ 0.06 & 1.18 $\pm$ 0.03 & 7.57 $\pm$ 0.02 & 7.57 $\pm$ 0.03 &  0.112 $\pm$  0.003  & 0.042 $\pm$  0.004 & Y &   20   & Y \\
HD 86322 & 4895 $\pm$  28 & 2.86 $\pm$ 0.11 & 1.58 $\pm$ 0.03 & 7.42 $\pm$ 0.02 & 7.42 $\pm$ 0.06 & -0.043 $\pm$  0.006  & -0.115 $\pm$  0.007 & Y &  18 &   Y* \\
HD 89307 & 5961 $\pm$  20 & 4.52 $\pm$ 0.05 & 1.12 $\pm$ 0.03 & 7.34 $\pm$ 0.02 & 7.34 $\pm$ 0.02 & -0.117 $\pm$  0.003  & -0.189 $\pm$  0.006 & N &  	5 &  N \\
HD 89376 & 5032 $\pm$  44 & 4.46 $\pm$ 0.12 & 1.01 $\pm$ 0.10 & 7.43 $\pm$ 0.02 & 7.43 $\pm$ 0.08 & -0.029 $\pm$  0.006  & -0.099 $\pm$  0.006 & Y &   19 &   N \\
$\pi^1$ Leo & 5082 $\pm$  23 & 2.95 $\pm$ 0.08 & 1.58 $\pm$ 0.02 & 7.56 $\pm$ 0.02 & 7.56 $\pm$ 0.04 &  0.104 $\pm$  0.005  &  0.032 $\pm$  0.006 & Y &   16 &  N \\
HD 98356 & 5310 $\pm$  32 & 4.41 $\pm$ 0.09 & 0.92 $\pm$ 0.07 & 7.54 $\pm$ 0.02 & 7.54 $\pm$ 0.05 &  0.081 $\pm$  0.006  & 0.007 $\pm$  0.006 & Y &   19 &  Y* \\
HD 101112 & 4809 $\pm$  32 & 2.76 $\pm$ 0.11 & 1.53 $\pm$ 0.04 & 7.58 $\pm$ 0.03 & 7.58 $\pm$ 0.07 &  0.122 $\pm$  0.007  & 0.051 $\pm$  0.008 & Y &   12 &   N \\
HD 106696 & 4935 $\pm$  17 & 2.74 $\pm$ 0.05 & 1.50 $\pm$ 0.02 & 7.20 $\pm$ 0.01 & 7.20 $\pm$ 0.03 & -0.256 $\pm$  0.004  &-0.328 $\pm$  0.005 & N &   	3 &  N \\
MY UMa & 5067 $\pm$  57 & 4.52 $\pm$ 0.15 & 1.41 $\pm$ 0.11 & 7.34 $\pm$ 0.03 & 7.34 $\pm$ 0.10 & -0.121 $\pm$  0.009  & -0.189 $\pm$  0.008 & N &   14 &   N \\
GR Leo & 5227 $\pm$  29 & 4.46 $\pm$ 0.08 & 0.75 $\pm$ 0.07 & 7.46 $\pm$ 0.02 & 7.46 $\pm$ 0.04 &  0.003 $\pm$  0.004  & -0.065 $\pm$  0.004 & Y &   20  & Y \\
HD 106023 & 5405 $\pm$  17 & 4.48 $\pm$ 0.05 & 0.93 $\pm$ 0.04 & 7.43 $\pm$ 0.01 & 7.43 $\pm$ 0.03 & -0.028 $\pm$  0.003  & -0.099 $\pm$  0.004 & Y &   17  & N \\
HD 117860 & 5913 $\pm$  12 & 4.54 $\pm$ 0.03 & 1.17 $\pm$ 0.02 & 7.53 $\pm$ 0.01 & 7.53 $\pm$ 0.01 &  0.068 $\pm$  0.002  & -0.002 $\pm$  0.005 & Y &   19 &   Y* \\
HD 117936 & 4845 $\pm$  49 & 4.42 $\pm$ 0.15 & 0.82 $\pm$ 0.17 & 7.43 $\pm$ 0.03 & 7.43 $\pm$ 0.10 & -0.030 $\pm$  0.006  & -0.102 $\pm$  0.006 & Y &   15 & N \\
HD 124642 & 4825 $\pm$  55 & 4.49 $\pm$ 0.16 & 1.00 $\pm$ 0.18 & 7.40 $\pm$ 0.03 & 7.40 $\pm$ 0.11 & -0.060 $\pm$  0.007  & -0.130 $\pm$  0.006 & Y &   15 &  N \\
HD 126535 & 5305 $\pm$  29 & 4.46 $\pm$ 0.08 & 1.01 $\pm$ 0.06 & 7.60 $\pm$ 0.02 & 7.59 $\pm$ 0.04 &  0.141 $\pm$  0.004  & 0.067 $\pm$  0.004 & Y &     19 & Y* \\
HD 131023 & 5619 $\pm$  18 & 4.50 $\pm$ 0.05 & 1.06 $\pm$ 0.03 & 7.61 $\pm$ 0.01 & 7.61 $\pm$ 0.02 &  0.150 $\pm$  0.003  &  0.076 $\pm$  0.004 & Y &    20 & Y \\
HD 149026 & 6213 $\pm$  22 & 4.45 $\pm$ 0.05 & 1.44 $\pm$ 0.03 & 7.80 $\pm$ 0.02 & 7.80 $\pm$ 0.02 &  0.338 $\pm$  0.003  & 0.270 $\pm$  0.006 & N &  	 3  &  N \\
HD 170657 & 5030 $\pm$  22 & 4.46 $\pm$ 0.07 & 0.63 $\pm$ 0.07 & 7.20 $\pm$ 0.01 & 7.20 $\pm$ 0.04 & -0.259 $\pm$  0.003  & -0.332 $\pm$  0.004 & N &  	0 &   N \\
MV Dra & 5394 $\pm$  22 & 4.43 $\pm$ 0.05 & 1.01 $\pm$ 0.03 & 7.56 $\pm$ 0.01 & 7.56 $\pm$ 0.03 &  0.098 $\pm$  0.003  & 0.025 $\pm$  0.004 & Y &  20  & Y \\
HD 189087 & 5296 $\pm$  18 & 4.50 $\pm$ 0.05 & 0.80 $\pm$ 0.04 & 7.36 $\pm$ 0.01 & 7.36 $\pm$ 0.03 & -0.096 $\pm$  0.003  & -0.166 $\pm$  0.004 & N &   12  & N \\
V2425 Cyg & 5348 $\pm$ 43 & 4.69  $\pm$ 0.11 & 1.35 $\pm$  0.09 &  7.50 $\pm$  0.03 &   7.50 $\pm$ 0.06 & 0.044 $\pm$ 0.006 & -0.027 $\pm$  0.006 & Y &  20  & Y \\
HD 196885 & 6326 $\pm$  24 & 4.40 $\pm$ 0.06 & 1.50 $\pm$ 0.03 & 7.72 $\pm$ 0.02 & 7.72 $\pm$ 0.03 &  0.254 $\pm$  0.007  & 0.197 $\pm$  0.009 & N &   	9  & N \\
V2436 Cyg & 4963 $\pm$  40 & 4.43 $\pm$ 0.11 & 0.90 $\pm$ 0.11 & 7.54 $\pm$ 0.02 & 7.54 $\pm$ 0.07 &  0.080 $\pm$  0.005  & 0.008 $\pm$  0.005 & Y &   19  & Y* \\
HD 200968 & 5114 $\pm$  24 & 4.40 $\pm$ 0.07 & 0.90 $\pm$ 0.05 & 7.42 $\pm$ 0.01 & 7.42 $\pm$ 0.04 & -0.044 $\pm$  0.004  & -0.116 $\pm$  0.004 & Y &   13 & N \\
NS Aqr & 5635 $\pm$  15 & 4.45 $\pm$ 0.04 & 1.04 $\pm$ 0.02 & 7.62 $\pm$ 0.01 & 7.62 $\pm$ 0.02 &  0.163 $\pm$  0.003  & 0.091 $\pm$  0.004 & Y &     20  & Y \\
V454 And  & 5900 $\pm$  20 & 4.57 $\pm$ 0.06 & 1.11 $\pm$ 0.02 & 7.61 $\pm$ 0.02 & 7.61 $\pm$ 0.03 &  0.151 $\pm$  0.003  & 0.081 $\pm$  0.005 & Y &   17  & N \\
HD 222422 & 5445 $\pm$  16 & 4.49 $\pm$ 0.05 & 0.76 $\pm$ 0.04 & 7.32 $\pm$ 0.01 & 7.32 $\pm$ 0.02 & -0.137 $\pm$  0.003  & -0.209 $\pm$  0.005 & N & 	 3 & N \\
20 Psc & 5125 $\pm$  19 & 3.08 $\pm$ 0.06 & 1.49 $\pm$ 0.02 & 7.50 $\pm$ 0.02 & 7.50 $\pm$ 0.04 &  0.041 $\pm$  0.004  & -0.031 $\pm$  0.006 & Y &   18  & Y* \\
\noalign{\smallskip}
\hline
\noalign{\smallskip}
\end{tabular}
%\noalign{\smallskip}
\tablefoot{ 
\tablefoottext{a}{If the star satisfies or not the Fe homogeneity criterium at 1-rms.}\\
%\noalign{\smallskip}
\tablefoottext{b}{The number of elements satisfying homogeneity at 1-rms.}\\
%\noalign{\smallskip}
\tablefoottext{c}{If the star satisfies or not the homogeneity criterium in all the analysed elements.  Y* means that the remaing 1 o 2 elements satisfy 1.5-rms criterium but not the 1-rms. }
}
\end{table*}
%\end{landscape}

\begin{landscape}
\begin{table}
\scriptsize
\caption{[X/Fe] ratios for the  $\alpha$-, Fe-peak, and odd-Z elements: Na, Mg, Al, Si, Ca, Sc, Ti, V, Cr, Mn, Co, and Ni. }
\label{galtab2}
\centering
\begin{tabular}{lcccccccccccc}     % 15 columns
\hline
\hline
\noalign{\smallskip}
{\small Name} & {\small [Na/Fe]} & {\small  [Mg/Fe]} & {\small  [Al/Fe]} &{\small  [Si/Fe]} & {\small  [Ca/Fe]} &{\small  [Sc/Fe]} & {\small  [Ti/Fe]} & {\small  [V/Fe]} & {\small [Cr/Fe]}  & {\small  [Mn/Fe]} & {\small  [Co/Fe]} & {\small [Ni/Fe]} \\
\noalign{\smallskip}
\hline
\noalign{\smallskip}
 vB~153  & -0.11 $\pm$ 0.04 & -0.11 $\pm$ 0.08 & -0.07 $\pm$ 0.02 & 0.04 $\pm$ 0.02  & 0.05 $\pm$ 0.02 & 0.01 $\pm$ 0.06 &  0.04 $\pm$ 0.02 & 0.24 $\pm$ 0.05 &  0.04 $\pm$ 0.02 & 0.13 $\pm$ 0.10  & 0.02 $\pm$ 0.02 & -0.01 $\pm$ 0.01 \\
\noalign{\smallskip}
\hline
\noalign{\smallskip}
BE Cet &  0.04 $\pm$  0.03 & -0.01 $\pm$  0.04 &  0.05 $\pm$  0.01 & 
 0.08 $\pm$  0.01 &  0.08 $\pm$  0.01 &  0.02 $\pm$  0.01 &  0.05 $\pm$  0.01 & 
 0.15 $\pm$  0.05 &  0.11 $\pm$  0.02
 &  0.15 $\pm$  0.02 &  0.03 $\pm$  0.02 &  0.04$\pm$   0.01\\
HD 5848 &  0.42 $\pm$  --- & -0.02 $\pm$  0.01 &  0.27 $\pm$  0.03 & 
 0.22 $\pm$  0.06 &  0.07 $\pm$  0.04 &  0.19 $\pm$  0.16 &  0.12 $\pm$  0.03 & 
 0.42 $\pm$  0.08 &  0.06 $\pm$  0.03
 &  0.23 $\pm$  0.14 &  0.16 $\pm$  0.06 &  0.06$\pm$   0.02\\
AZ Ari &  0.02 $\pm$  0.01 & -0.05 $\pm$  0.10 &  0.01 $\pm$  0.02 & 
 0.06 $\pm$  0.01 &  0.08 $\pm$  0.00 & -0.04 $\pm$  0.04 &  0.07 $\pm$  0.01 & 
 0.11 $\pm$  0.04 &  0.09 $\pm$  0.01
 &  0.15 $\pm$  0.06 &  0.03 $\pm$  0.01 &  0.05$\pm$   0.00\\
FT Cet &  0.06 $\pm$  0.03 & -0.03 $\pm$  0.06 &  0.15 $\pm$  0.05 & 
 0.12 $\pm$  0.03 &  0.04 $\pm$  0.05 &  0.14 $\pm$  0.12 &  0.18 $\pm$  0.02 & 
 0.44 $\pm$  0.06 &  0.10 $\pm$  0.03
 &  0.10 $\pm$  0.11 &  0.13 $\pm$  0.03 &  0.06$\pm$   0.01\\
BZ Cet & -0.02 $\pm$  0.09 & -0.02 $\pm$  0.04 &  0.11 $\pm$  0.02 & 
 0.14 $\pm$  0.02 &  0.04 $\pm$  0.03 &  0.09 $\pm$  0.10 &  0.14 $\pm$  0.01 & 
 0.39 $\pm$  0.06 &  0.08 $\pm$  0.03
 &  0.03 $\pm$  0.04 &  0.07 $\pm$  0.02 &  0.08$\pm$   0.01\\
$\delta$ Ari &  0.01 $\pm$  0.23 &  0.06 $\pm$  0.07 &  0.23 $\pm$  0.03 & 
 0.19 $\pm$  0.03 &  0.10 $\pm$  0.03 &  0.01 $\pm$  0.11 &  0.09 $\pm$  0.03 & 
 0.20 $\pm$  0.02 &  0.03 $\pm$  0.03
 &  0.39 $\pm$  0.08 &  0.18 $\pm$  0.06 &  0.05$\pm$   0.01\\
V683 Per &  0.05 $\pm$  0.01 &  -0.07 $\pm$  0.06 &  0.08 $\pm$  0.00 & 
 0.06 $\pm$  0.01 &  0.05 $\pm$  0.05 &  0.05 $\pm$  0.08 &  0.07 $\pm$  0.01 & 
 0.16 $\pm$  0.04 &  0.09 $\pm$  0.01
 &  0.21 $\pm$  0.05 &  0.04 $\pm$  0.02 &  0.06$\pm$   0.01\\
V686 Per & -0.01 $\pm$  0.05 & -0.02 $\pm$  0.03 & -0.03 $\pm$  0.01 & 
 0.05 $\pm$  0.01 &  0.20 $\pm$  0.05 &  0.02 $\pm$  0.02 &  0.12 $\pm$  0.02 & 
 0.14 $\pm$  0.04 &  0.10 $\pm$  0.02
 &  0.05 $\pm$  0.05 &  0.03 $\pm$  0.02 &  0.06$\pm$   0.01\\
HD 21663 &  0.03 $\pm$  0.04 &  0.05 $\pm$  0.03 &  0.08 $\pm$  0.01 & 
 0.07 $\pm$  0.01 &  0.03 $\pm$  0.03 &  0.01 $\pm$  0.02 &  0.09 $\pm$  0.01 & 
 0.17 $\pm$  0.03 &  0.09 $\pm$  0.01
 &  0.11 $\pm$  0.03 &  0.08 $\pm$  0.01 &  0.06$\pm$   0.01\\
HD 23356 & -0.08 $\pm$  0.04 &  0.11 $\pm$  0.07 &  0.11 $\pm$  0.04 & 
 0.13 $\pm$  0.02 &  0.08 $\pm$  0.04 &  0.14 $\pm$  0.11 &  0.21 $\pm$  0.02 & 
 0.48 $\pm$  0.07 &  0.11 $\pm$  0.02
 &  0.04 $\pm$  0.05 &  0.12 $\pm$  0.01 &  0.05$\pm$   0.01\\
39 Tau & -0.02 $\pm$  0.01 &  0.07 $\pm$  0.08 & -0.01 $\pm$  0.01 & 
 0.06 $\pm$  0.01 &  0.09 $\pm$  0.01 & -0.04 $\pm$  0.05 &  0.07 $\pm$  0.01 & 
 0.09 $\pm$  0.04 &  0.08 $\pm$  0.01
 &  0.07 $\pm$  0.04 &  0.00 $\pm$  0.02 &  0.02$\pm$   0.00\\
HD 25893 &  0.02 $\pm$  0.01 & -0.07 $\pm$  0.01 &  0.17 $\pm$  0.01 & 
 0.09 $\pm$  0.02 &  0.14 $\pm$  0.02 & -0.02 $\pm$  0.09 &  0.11 $\pm$  0.02 & 
 0.22 $\pm$  0.07 &  0.13 $\pm$  0.02
 &  0.17 $\pm$  0.08 &  0.06 $\pm$  0.04 &  0.05$\pm$   0.01\\
HD 27282 & -0.03 $\pm$  0.06 & -0.02 $\pm$  0.04 &  0.01 $\pm$  0.02 & 
 0.08 $\pm$  0.01 &  0.11 $\pm$  0.01 & -0.01 $\pm$  0.04 &  0.06 $\pm$  0.01 & 
 0.15 $\pm$  0.04 &  0.08 $\pm$  0.02
 &  0.20 $\pm$  0.08 &  0.02 $\pm$  0.02 &  0.06$\pm$   0.01\\
HD 27685 & -0.01 $\pm$  0.01 &  0.00 $\pm$  0.05 & -0.08 $\pm$  0.01 & 
 0.06 $\pm$  0.01 &  0.08 $\pm$  0.02 &  0.01 $\pm$  0.05 &  0.04 $\pm$  0.01 & 
 0.07 $\pm$  0.03 &  0.09 $\pm$  0.01
 &  0.12 $\pm$  0.02 &  0.01 $\pm$  0.02 &  0.05$\pm$   0.01\\
HD 27989 &  0.21 $\pm$  0.10 & -0.18 $\pm$  0.09 &  0.18 $\pm$  0.03 & 
 0.04 $\pm$  0.02 &  0.12 $\pm$  0.04 &  0.03 $\pm$  0.04 &  0.09 $\pm$  0.03 & 
 0.23 $\pm$  0.07 &  0.12 $\pm$  0.03
 &  0.14 $\pm$  0.05 & -0.01 $\pm$  0.10 &  0.03$\pm$   0.02\\
$\epsilon$ Tau &  0.34 $\pm$  --- &  0.03 $\pm$  0.11 &  0.22 $\pm$  0.04 & 
 0.22 $\pm$  0.05 &  0.07 $\pm$  0.03 &  0.03 $\pm$  0.07 &  0.10 $\pm$  0.03 & 
 0.12 $\pm$  0.03 &  0.05 $\pm$  0.03
 &  0.30 $\pm$  0.07 &  0.24 $\pm$  0.07 &  0.03$\pm$   0.02\\
111 Tau B &  0.02 $\pm$  0.13 & -0.08 $\pm$  0.07 &  0.23 $\pm$  0.03 & 
 0.04 $\pm$  0.04 &  0.21 $\pm$  0.09 &  0.46 $\pm$  0.12 &  0.35 $\pm$  0.02 & 
 0.59 $\pm$  0.07 &  0.19 $\pm$  0.05
 & -0.14 $\pm$  0.09 &  0.10 $\pm$  0.03 &  0.02$\pm$   0.02\\
HD 40979 &  0.02 $\pm$  0.06 & -0.00 $\pm$  0.15 & -0.12 $\pm$  0.37 & 
 0.08 $\pm$  0.02 &  0.07 $\pm$  0.04 &  0.09 $\pm$  0.10 &  0.07 $\pm$  0.03 & 
-0.01 $\pm$  0.11 &  0.12 $\pm$  0.03
 &  0.12 $\pm$  0.09 &  0.06 $\pm$  0.03 &  0.08$\pm$   0.01\\
HD 45609 & -0.06 $\pm$  0.05 &  0.06 $\pm$  0.01 &  0.13 $\pm$  0.04 & 
 0.04 $\pm$  0.02 &  0.04 $\pm$  0.05 &  0.08 $\pm$  0.07 &  0.18 $\pm$  0.02 & 
 0.27 $\pm$  0.05 &  0.15 $\pm$  0.03
 &  0.07 $\pm$  0.05 &  0.11 $\pm$  0.03 &  0.00$\pm$   0.01\\
HD 52265 &  0.08 $\pm$  0.01 &  0.11 $\pm$  0.05 & -0.02 $\pm$  0.02 & 
 0.08 $\pm$  0.01 &  0.09 $\pm$  0.03 & -0.01 $\pm$  0.04 &  0.07 $\pm$  0.01 & 
 0.07 $\pm$  0.04 &  0.06 $\pm$  0.01
 &  0.07 $\pm$  0.02 &  0.03 $\pm$  0.01 &  0.07$\pm$   0.00\\
HD 53532 & -0.04 $\pm$  0.00 &  0.00 $\pm$  --- &  0.02 $\pm$  0.02 & 
 0.07 $\pm$  0.01 &  0.10 $\pm$  0.01 & -0.06 $\pm$  0.07 &  0.08 $\pm$  0.01 & 
 0.11 $\pm$  0.05 &  0.09 $\pm$  0.01
 &  0.07 $\pm$  0.09 &  0.01 $\pm$  0.01 & -0.00$\pm$   0.01\\
HD 65523 & -0.01 $\pm$  0.02 & -0.04 $\pm$  0.04 &  0.09 $\pm$  0.01 & 
 0.07 $\pm$  0.01 &  0.10 $\pm$  0.02 & -0.02 $\pm$  0.05 &  0.14 $\pm$  0.01 & 
 0.23 $\pm$  0.05 &  0.09 $\pm$  0.01
 &  0.13 $\pm$  0.05 &  0.06 $\pm$  0.02 &  0.04$\pm$   0.01\\
HD 70088 & -0.05 $\pm$  0.02 &  0.01 $\pm$  0.02 &  0.06 $\pm$  0.02 & 
 0.04 $\pm$  0.01 &  0.09 $\pm$  0.01 &  0.01 $\pm$  0.04 &  0.09 $\pm$  0.01 & 
 0.11 $\pm$  0.05 &  0.10 $\pm$  0.01
 & -0.04 $\pm$  0.02 & -0.06 $\pm$  0.04 & -0.00$\pm$   0.01\\
HD 72760 & -0.02 $\pm$  0.03 & -0.01 $\pm$  0.06 &  0.06 $\pm$  0.02 & 
 0.07 $\pm$  0.01 &  0.07 $\pm$  0.04 & -0.03 $\pm$  0.06 &  0.12 $\pm$  0.01 & 
 0.23 $\pm$  0.05 &  0.10 $\pm$  0.01
 &  0.15 $\pm$  0.07 &  0.05 $\pm$  0.02 &  0.04$\pm$   0.01\\
V401 Hya & -0.01 $\pm$  0.01 & -0.14 $\pm$  0.15 &  0.02 $\pm$  0.01 &
 0.04 $\pm$  0.01 &  0.07 $\pm$  0.01 &  0.00 $\pm$  0.02 &  0.05 $\pm$  0.01 &
 0.08 $\pm$  0.04 &  0.08 $\pm$  0.01
 &  0.06 $\pm$  0.01 &  0.01 $\pm$  0.02 &  0.03$\pm$   0.01\\
HD 73171 &  0.22 $\pm$  0.03 &  0.04 $\pm$  0.05 &  0.20 $\pm$  0.02 & 
 0.20 $\pm$  0.04 &  0.08 $\pm$  0.02 &  0.10 $\pm$  0.15 &  0.13 $\pm$  0.02 & 
 0.31 $\pm$  0.09 &  0.06 $\pm$  0.02
 &  0.38 $\pm$  0.09 &  0.17 $\pm$  0.04 &  0.03$\pm$   0.02\\
CT Pyx &  0.07 $\pm$  0.00 & -0.01 $\pm$  0.01 &  0.11 $\pm$  0.01 & 
 0.10 $\pm$  0.02 &  0.11 $\pm$  0.02 &  0.03 $\pm$  0.08 &  0.16 $\pm$  0.02 & 
 0.32 $\pm$  0.06 &  0.13 $\pm$  0.02
 &  0.04 $\pm$  0.06 &  0.10 $\pm$  0.02 &  0.07$\pm$   0.01\\
$\rho$ Cnc A &  0.07 $\pm$  0.09 &  0.02 $\pm$  0.09 &  0.18 $\pm$  0.01 & 
 0.13 $\pm$  0.02 &  0.03 $\pm$  0.05 &  0.17 $\pm$  0.10 &  0.15 $\pm$  0.02 & 
 0.47 $\pm$  0.09 &  0.09 $\pm$  0.02
 &  0.12 $\pm$  0.05 &  0.32 $\pm$  0.04 &  0.17$\pm$   0.01\\
HD 75898 &  0.17 $\pm$  0.03 & -0.06 $\pm$  0.00 & -0.03 $\pm$  0.00 & 
 0.08 $\pm$  0.01 &  0.08 $\pm$  0.03 &  0.09 $\pm$  0.07 &  0.07 $\pm$  0.01 & 
 0.13 $\pm$  0.04 &  0.07 $\pm$  0.01
 &  0.16 $\pm$  0.01 &  0.12 $\pm$  0.02 &  0.12$\pm$   0.01\\
HD 76151 &  0.05 $\pm$  0.04 & -0.02 $\pm$  0.08 &  0.02 $\pm$  0.02 & 
 0.06 $\pm$  0.01 &  0.07 $\pm$  0.01 & -0.02 $\pm$  0.01 &  0.07 $\pm$  0.01 & 
 0.12 $\pm$  0.04 &  0.08 $\pm$  0.00
 &  0.11 $\pm$  0.02 &  0.08 $\pm$  0.00 &  0.07$\pm$   0.00\\
IK Cnc &  0.04 $\pm$  0.06 & -0.03 $\pm$  0.03 &  0.07 $\pm$  0.01 & 
 0.10 $\pm$  0.02 &  0.14 $\pm$  0.02 &  0.00 $\pm$  0.09 &  0.15 $\pm$  0.02 & 
 0.21 $\pm$  0.05 &  0.12 $\pm$  0.02
 & -0.02 $\pm$  0.02 &  0.01 $\pm$  0.04 & -0.02$\pm$   0.01\\
HD 82106 & -0.02 $\pm$  0.04 & -0.03 $\pm$  0.04 &  0.10 $\pm$  0.01 & 
 0.11 $\pm$  0.02 &  0.06 $\pm$  0.07 &  0.21 $\pm$  0.12 &  0.24 $\pm$  0.02 & 
 0.53 $\pm$  0.07 &  0.08 $\pm$  0.03
 & -0.08 $\pm$  0.03 &  0.09 $\pm$  0.01 &  0.03$\pm$   0.01\\
GT Leo & -0.03 $\pm$  0.02 &  0.04 $\pm$  0.05 &  0.15 $\pm$  0.02 & 
 0.10 $\pm$  0.02 &  0.05 $\pm$  0.06 &  0.03 $\pm$  0.05 &  0.15 $\pm$  0.01 & 
 0.36 $\pm$  0.06 &  0.11 $\pm$  0.03
 &  0.18 $\pm$  0.11 &  0.14 $\pm$  0.04 &  0.10$\pm$   0.01\\
HD 85301 &  0.00 $\pm$  0.01 & -0.03 $\pm$  0.01 & -0.02 $\pm$  0.00 & 
 0.04 $\pm$  0.01 &  0.12 $\pm$  0.01 & -0.08 $\pm$  0.03 &  0.08 $\pm$  0.01 & 
 0.15 $\pm$  0.05 &  0.12 $\pm$  0.02
 &  0.15 $\pm$  0.07 &  0.02 $\pm$  0.03 &  0.02$\pm$   0.01\\
HD 86322 &  0.09 $\pm$  0.17 &  0.25 $\pm$  0.18 &  0.15 $\pm$  0.01 & 
 0.22 $\pm$  0.04 &  0.07 $\pm$  0.02 &  0.02 $\pm$  0.15 &  0.15 $\pm$  0.02 & 
 0.25 $\pm$  0.06 &  0.07 $\pm$  0.03
 &  0.24 $\pm$  0.02 &  0.18 $\pm$  0.05 &  0.09$\pm$   0.01\\
HD 89307 &  0.02 $\pm$  0.02 &  0.05 $\pm$  0.02 &  0.01 $\pm$  0.01 & 
 0.08 $\pm$  0.01 &  0.07 $\pm$  0.01 &  0.09 $\pm$  0.03 &  0.08 $\pm$  0.01 & 
 0.06 $\pm$  0.07 &  0.05 $\pm$  0.01
 & -0.04 $\pm$  0.06 & -0.00 $\pm$  0.03 &  0.02$\pm$   0.01\\
HD 89376 &  0.12 $\pm$  0.01 & -0.02 $\pm$  0.03 &  0.13 $\pm$  0.03 & 
 0.12 $\pm$  0.02 &  0.13 $\pm$  0.06 &  0.19 $\pm$  0.08 &  0.23 $\pm$  0.03 & 
 0.40 $\pm$  0.06 &  0.14 $\pm$  0.03
 &  0.12 $\pm$  0.17 &  0.09 $\pm$  0.01 &  0.06$\pm$   0.01\\
$\pi^1$ Leo &  0.31 $\pm$  --- &  0.07 $\pm$  0.17 &  0.10 $\pm$  0.01 & 
 0.17 $\pm$  0.03 &  0.11 $\pm$  0.05 & -0.00 $\pm$  0.07 &  0.10 $\pm$  0.02 & 
 0.18 $\pm$  0.06 &  0.08 $\pm$  0.02
 &  0.17 $\pm$  0.07 &  0.18 $\pm$  0.05 &  0.04$\pm$   0.01\\
HD 98356 &  0.11 $\pm$  0.02 &  0.05 $\pm$  0.01 &  0.05 $\pm$  0.02 & 
 0.09 $\pm$  0.02 &  0.06 $\pm$  0.03 &  0.04 $\pm$  0.12 &  0.18 $\pm$  0.02 & 
 0.39 $\pm$  0.06 &  0.09 $\pm$  0.02
 &  0.13 $\pm$  0.05 &  0.20 $\pm$  0.04 &  0.10$\pm$   0.01\\
HD 106696 &  0.15 $\pm$  0.02 &  0.17 $\pm$  0.06 &  0.18 $\pm$  0.01 &
 0.18 $\pm$  0.01 &  0.13 $\pm$  0.02 & -0.02 $\pm$  0.06 &  0.12 $\pm$  0.01 &
 0.20 $\pm$  0.05 &  0.04 $\pm$  0.02
 &  0.15 $\pm$  0.02 &  0.14 $\pm$  0.04 &  0.05$\pm$   0.01\\
HD 101112 &  0.27 $\pm$  --- &  0.04 $\pm$  0.05 &  0.19 $\pm$  0.03 & 
 0.22 $\pm$  0.04 &  0.12 $\pm$  0.03 &  0.09 $\pm$  0.13 &  0.11 $\pm$  0.02 & 
 0.36 $\pm$  0.07 &  0.06 $\pm$  0.02
 &  0.38 $\pm$  0.07 &  0.26 $\pm$  0.12 &  0.12$\pm$   0.02\\
MY UMa &  0.00 $\pm$  0.12 &  0.03 $\pm$  0.09 &  0.12 $\pm$  0.04 & 
 0.06 $\pm$  0.04 &  0.21 $\pm$  0.06 &  0.22 $\pm$  0.11 &  0.29 $\pm$  0.03 & 
 0.51 $\pm$  0.07 &  0.18 $\pm$  0.04
 & -0.27 $\pm$  0.21 &  0.06 $\pm$  0.04 &  0.01$\pm$   0.02\\
GR Leo & -0.07 $\pm$  0.01 &  0.04 $\pm$  0.01 &  0.05 $\pm$  0.02 & 
 0.08 $\pm$  0.01 &  0.09 $\pm$  0.02 & -0.01 $\pm$  0.07 &  0.17 $\pm$  0.01 & 
 0.30 $\pm$  0.06 &  0.11 $\pm$  0.02
 &  0.03 $\pm$  0.05 &  0.07 $\pm$  0.02 &  0.03$\pm$   0.01\\
HD 106023 &  0.02 $\pm$  0.01 &  0.03 $\pm$  0.03 &  0.03 $\pm$  0.02 & 
 0.07 $\pm$  0.01 &  0.10 $\pm$  0.01 & -0.02 $\pm$  0.06 &  0.12 $\pm$  0.01 & 
 0.20 $\pm$  0.04 &  0.10 $\pm$  0.01
 &  0.09 $\pm$  0.08 &  0.04 $\pm$  0.01 &  0.04$\pm$   0.01\\
HD 117860 & -0.01 $\pm$  0.00 & -0.01 $\pm$  0.02 & -0.00 $\pm$  0.02 & 
 0.05 $\pm$  0.01 &  0.08 $\pm$  0.01 &  0.05 $\pm$  0.02 &  0.04 $\pm$  0.01 & 
 0.05 $\pm$  0.04 &  0.09 $\pm$  0.01
 &  0.02 $\pm$  0.09 & -0.02 $\pm$  0.01 &  0.02$\pm$   0.00\\
HD 117936 & -0.12 $\pm$  --- &  0.01 $\pm$  0.08 &  0.17 $\pm$  0.01 & 
 0.07 $\pm$  0.03 &  0.16 $\pm$  0.04 &  0.36 $\pm$  0.15 &  0.28 $\pm$  0.02 & 
 0.61 $\pm$  0.08 &  0.16 $\pm$  0.03
 &  0.05 $\pm$  0.04 &  0.18 $\pm$  0.01 &  0.09$\pm$   0.01\\
HD 124642 &  0.08 $\pm$  0.01 & -0.04 $\pm$  0.10 &  0.14 $\pm$  0.04 & 
 0.07 $\pm$  0.03 &  0.10 $\pm$  0.08 &  0.33 $\pm$  0.14 &  0.30 $\pm$  0.02 & 
 0.59 $\pm$  0.07 &  0.17 $\pm$  0.04
 & -0.08 $\pm$  0.07 &  0.09 $\pm$  0.01 &  0.06$\pm$   0.01\\
HD 126535 &  0.07 $\pm$  0.01 &  0.07 $\pm$  0.02 &  0.07 $\pm$  0.01 & 
 0.09 $\pm$  0.01 &  0.08 $\pm$  0.02 &  0.06 $\pm$  0.08 &  0.14 $\pm$  0.01 & 
 0.29 $\pm$  0.05 &  0.09 $\pm$  0.02
 &  0.18 $\pm$  0.08 &  0.13 $\pm$  0.02 &  0.08$\pm$   0.01\\
HD 131023 &  0.05 $\pm$  0.01 &  0.02 $\pm$  0.04 &  0.05 $\pm$  0.03 & 
 0.09 $\pm$  0.01 &  0.09 $\pm$  0.01 & -0.01 $\pm$  0.04 &  0.09 $\pm$  0.01 & 
 0.18 $\pm$  0.04 &  0.11 $\pm$  0.01
 &  0.19 $\pm$  0.07 &  0.09 $\pm$  0.02 &  0.06$\pm$   0.01\\
HD 149026 &  0.14 $\pm$  0.03 &  0.01 $\pm$  0.02 &  0.05 $\pm$  0.05 & 
 0.10 $\pm$  0.01 &  0.03 $\pm$  0.02 &  0.02 $\pm$  0.02 &  0.07 $\pm$  0.01 & 
 0.15 $\pm$  0.06 &  0.07 $\pm$  0.02
 &  0.16 $\pm$  0.03 &  0.09 $\pm$  0.02 &  0.10$\pm$   0.01\\
HD 170657 &  0.03 $\pm$  0.01 &  0.03 $\pm$  0.02 &  0.11 $\pm$  0.00 & 
 0.11 $\pm$  0.01 &  0.11 $\pm$  0.02 &  0.06 $\pm$  0.08 &  0.21 $\pm$  0.02 & 
 0.37 $\pm$  0.06 &  0.10 $\pm$  0.02
 &  0.07 $\pm$  0.06 &  0.09 $\pm$  0.01 &  0.06$\pm$   0.01\\
MV Dra &  0.02 $\pm$  0.01 &  0.03 $\pm$  0.02 &  0.04 $\pm$  0.02 & 
 0.08 $\pm$  0.01 &  0.10 $\pm$  0.02 & -0.02 $\pm$  0.04 &  0.09 $\pm$  0.01 & 
 0.19 $\pm$  0.04 &  0.12 $\pm$  0.01
 &  0.19 $\pm$  0.06 &  0.05 $\pm$  0.03 &  0.04$\pm$   0.01\\
HD 189087 &  0.00 $\pm$  0.01 &  0.02 $\pm$  0.03 &  0.08 $\pm$  0.02 & 
 0.09 $\pm$  0.01 &  0.08 $\pm$  0.03 & -0.02 $\pm$  0.06 &  0.16 $\pm$  0.01 & 
 0.24 $\pm$  0.05 &  0.10 $\pm$  0.01
 &  0.09 $\pm$  0.05 &  0.04 $\pm$  0.02 &  0.03$\pm$   0.01\\
V2425 Cyg &  0.08 $\pm$  0.04 & -0.12 $\pm$  0.15 & -0.01 $\pm$  0.06 &
 0.12 $\pm$  0.03 &  0.09 $\pm$  0.01 &  0.08 $\pm$  0.11 &  0.16 $\pm$  0.02 &
 0.36 $\pm$  0.08 &  0.12 $\pm$  0.03
 &  0.11 $\pm$  0.06 &  0.01 $\pm$  0.05 &  0.04$\pm$   0.01\\
HD 196885 &  0.08 $\pm$  0.02 & -0.03 $\pm$  0.10 & -0.00 $\pm$  0.02 & 
 0.08 $\pm$  0.02 &  0.07 $\pm$  0.02 &  0.03 $\pm$  0.05 &  0.07 $\pm$  0.01 & 
 0.09 $\pm$  0.08 &  0.05 $\pm$  0.03
 &  0.09 $\pm$  0.07 & -0.00 $\pm$  0.04 &  0.06$\pm$   0.01\\
V2436 Cyg & -0.02 $\pm$  0.04 & -0.06 $\pm$  0.01 &  0.08 $\pm$  0.01 & 
 0.12 $\pm$  0.02 &  0.10 $\pm$  0.04 &  0.05 $\pm$  0.11 &  0.12 $\pm$  0.01 & 
 0.43 $\pm$  0.07 &  0.07 $\pm$  0.03
 &  0.12 $\pm$  0.10 &  0.07 $\pm$  0.01 &  0.07$\pm$   0.01\\
HD 200968 & -0.06 $\pm$  0.09 & -0.06 $\pm$  0.02 &  0.07 $\pm$  0.02 & 
 0.10 $\pm$  0.02 &  0.12 $\pm$  0.03 & -0.01 $\pm$  0.08 &  0.12 $\pm$  0.01 & 
 0.30 $\pm$  0.06 &  0.12 $\pm$  0.02
 &  0.11 $\pm$  0.09 &  0.06 $\pm$  0.02 &  0.05$\pm$   0.01\\
NS Aqr &  0.09 $\pm$  0.01 & -0.03 $\pm$  0.04 &  0.05 $\pm$  0.02 & 
 0.08 $\pm$  0.01 &  0.09 $\pm$  0.01 &  0.02 $\pm$  0.03 &  0.08 $\pm$  0.01 & 
 0.15 $\pm$  0.04 &  0.11 $\pm$  0.01
 &  0.08 $\pm$  0.04 &  0.07 $\pm$  0.01 &  0.07$\pm$   0.01\\
V454 And & -0.11 $\pm$  0.04 &  0.04 $\pm$  0.07 &  0.04 $\pm$  0.01 & 
-0.01 $\pm$  0.01 &  0.02 $\pm$  0.03 &  0.01 $\pm$  0.04 &  0.07 $\pm$  0.01 & 
 0.07 $\pm$  0.03 &  0.08 $\pm$  0.01
 &  0.02 $\pm$  0.12 & -0.01 $\pm$  0.01 &  0.03$\pm$   0.01\\
HD 222422 & -0.02 $\pm$  0.00 &  0.10 $\pm$  0.03 &  0.11 $\pm$  0.03 & 
 0.09 $\pm$  0.01 &  0.15 $\pm$  0.02 & -0.03 $\pm$  0.08 &  0.17 $\pm$  0.01 & 
 0.18 $\pm$  0.04 &  0.09 $\pm$  0.01
 &  0.02 $\pm$  0.03 &  0.05 $\pm$  0.01 &  0.03$\pm$   0.01\\
20 Psc &  0.16 $\pm$  0.06 &  0.10 $\pm$  0.06 &  0.09 $\pm$  0.01 & 
 0.15 $\pm$  0.02 &  0.12 $\pm$  0.03 & -0.06 $\pm$  0.04 &  0.09 $\pm$  0.01 & 
 0.19 $\pm$  0.05 &  0.06 $\pm$  0.02
 &  0.22 $\pm$  0.04 &  0.06 $\pm$  0.03 &  0.01$\pm$   0.01\\
\noalign{\smallskip}
\hline
\noalign{\smallskip}
\end{tabular}
\end{table}
\end{landscape}

%\begin{landscape}
\begin{table*}
\scriptsize
\caption{[X/Fe] ratios for the s-process elements: Cu, Zn, Y, Zr, Ba, Ce, and Nd.}
\label{galtab3}
\centering
\begin{tabular}{lccccccc}     % 15 columns
\hline
\hline
\noalign{\smallskip}
{\small Name} & {\small [Cu/Fe]} & {\small [Zn/Fe]} & {\small [Y/Fe]} & {\small [Zr/Fe]} & {\small [Ba/Fe]} & {\small [Ce/Fe]} & {\small [Nd/Fe]}\\
\noalign{\smallskip}
\hline
\noalign{\smallskip}
vB 153 & 0.00 $\pm$ 0.04 &  0.21 $\pm$   0.23 &  0.08  $\pm$  0.04 & 0.25 $\pm$ 0.34 & -0.01 $\pm$ 0.03 & 0.07 $\pm$ 0.03 & 0.12 $\pm$ 0.10 \\
\noalign{\smallskip}
\hline
\noalign{\smallskip}
BE Cet & -0.09 $\pm$  0.02 &  0.06 $\pm$  0.28 &  0.08 $\pm$  0.14 &  0.16 $\pm$  0.28 &  0.21 $\pm$  0.22 &  0.03 $\pm$  0.14 &  0.04 $\pm$  0.17 \\
HD 5848 &  0.23 $\pm$  0.02 &  0.23 $\pm$  0.64 &  0.03 $\pm$  0.39 & -0.30 $\pm$  0.46 & -0.09 $\pm$  0.17 &  0.04 $\pm$  0.10 &  0.17 $\pm$  0.28 \\
AZ Ari & -0.03 $\pm$  0.02 &  0.06 $\pm$  0.28 &  0.07 $\pm$  0.20 &  0.08 $\pm$  0.49 &  0.12 $\pm$  0.17 &  0.07 $\pm$  0.00 & -0.06 $\pm$  0.32 \\
FT Cet &  0.00 $\pm$  0.02 &  0.00 $\pm$  0.36 & -0.07 $\pm$  0.28 &  0.05 $\pm$  0.22 &  0.11 $\pm$  0.17 &  0.19 $\pm$  0.14 &  0.20 $\pm$  0.37 \\
BZ Cet & -0.05 $\pm$  0.04 & -0.05 $\pm$  0.26 &  0.03 $\pm$  0.32 & -0.03 $\pm$  0.24 &  0.02 $\pm$  0.17 &  0.11 $\pm$  0.20 &  0.06 $\pm$  0.14 \\
$\delta$  Ari &  0.20 $\pm$  0.03 &  0.05 $\pm$  0.48 &  0.13 $\pm$  0.20 & -0.10 $\pm$  0.10 &  0.01 $\pm$  0.22 &  0.14 $\pm$  0.14 &  0.22 $\pm$  0.35\\
V683 Per & -0.10 $\pm$  0.05 &  0.00 $\pm$  0.20 &  0.07 $\pm$  0.14 & -0.04 $\pm$  0.22 &  0.11 $\pm$  0.10 &  0.04 $\pm$  0.17 &  0.02 $\pm$  0.26 \\
V686 Per & -0.07 $\pm$  0.01 & -0.02 $\pm$  0.26 & -0.02 $\pm$  0.01 &  0.14 $\pm$  0.10 &  0.07 $\pm$  0.10 &  0.06 $\pm$  0.14 &  0.04 $\pm$  0.32\\
HD 21663 &  0.05 $\pm$  0.02 & -0.03 $\pm$  0.17 &  0.10 $\pm$  0.17 & -0.06 $\pm$  0.26 &  0.16 $\pm$  0.20 &  0.19 $\pm$  0.24 &  0.09 $\pm$  0.10 \\
HD 23356 &  0.02 $\pm$  0.01 &  0.10 $\pm$  0.37 &  0.21 $\pm$  0.17 &  0.05 $\pm$  0.30 &  0.06 $\pm$  0.24 &  0.09 $\pm$  0.14 &  0.21 $\pm$  0.35 \\
39 Tau & -0.09 $\pm$  0.03 & -0.01 $\pm$  0.28 &  0.07 $\pm$  0.17 &  0.24 $\pm$  0.30 &  0.18 $\pm$  0.17 &  0.16 $\pm$  0.10 &  0.14 $\pm$  ---\\
HD 25893 &  0.02 $\pm$  0.04 &  0.06 $\pm$  0.32 &  0.14 $\pm$  0.14 &  0.14 $\pm$  0.32 &  0.13 $\pm$  0.22 &  0.03 $\pm$  0.10 &  0.03 $\pm$  0.45\\
HD 27282 & -0.04 $\pm$  0.02 & -0.09 $\pm$  0.17 &  0.02 $\pm$  0.00 &  0.17 $\pm$  0.41 &  0.10 $\pm$  0.10 & -0.05 $\pm$  0.14 & -0.09 $\pm$  0.24 \\
HD 27685 & -0.06 $\pm$  0.05 &  0.11 $\pm$  0.36 &  0.02 $\pm$  0.17 & -0.09 $\pm$  0.24 &  0.06 $\pm$  0.14 &  0.02 $\pm$  0.17 &  0.00 $\pm$  0.24\\
HD 27989 & -0.15 $\pm$  0.13 &  0.01 $\pm$  0.28 &  0.11 $\pm$  0.14 &  0.09 $\pm$  0.01 &  0.19 $\pm$  0.14 & -0.12 $\pm$  0.20 &  0.10 $\pm$  0.22\\
$\epsilon$ Tau &  0.28 $\pm$  0.11 &  0.02 $\pm$  0.50 & -0.01 $\pm$  0.17 & -0.07 $\pm$  0.20 & -0.14 $\pm$  0.20 &  0.03 $\pm$  0.17 &  0.15 $\pm$  0.20\\
111 Tau B &  0.00 $\pm$  0.03 & -0.12 $\pm$  0.20 &  0.57 $\pm$  0.58 &  0.39 $\pm$  0.35 &  0.13 $\pm$  0.25 &  0.40 $\pm$  0.20 &  0.49 $\pm$  0.45\\
HD 40979 & -0.12 $\pm$  0.07 & -0.08 $\pm$  0.17 &  0.00 $\pm$  0.14 & -0.05 $\pm$  0.01 &  0.12 $\pm$  0.14 & -0.10 $\pm$  0.26 &  0.14 $\pm$  0.01\\
HD 45609 & -0.21 $\pm$  0.06 & -0.08 $\pm$  0.14 &  0.08 $\pm$  0.32 &  0.19 $\pm$  0.30 &  0.07 $\pm$  0.28 &  0.39 $\pm$  0.30 &  0.19 $\pm$  0.33\\
HD 52265 & -0.02 $\pm$  0.03 & -0.09 $\pm$  0.41 & -0.04 $\pm$  0.10 &  0.19 $\pm$  0.26 &  0.05 $\pm$  0.10 & -0.03 $\pm$  0.10 & -0.10 $\pm$  0.35 \\
HD 53532 & -0.13 $\pm$  0.02 & -0.09 $\pm$  0.00 &  0.12 $\pm$  --- &  0.16 $\pm$  0.45 &  0.16 $\pm$  0.17 &  0.16 $\pm$  0.14 &  0.14 $\pm$  0.10 \\
HD 65523 &  0.01 $\pm$  0.01 &  0.06 $\pm$  0.28 &  0.12 $\pm$  0.26 &  0.03 $\pm$  0.28 &  0.10 $\pm$  0.17 &  0.24 $\pm$  0.00 &  0.12 $\pm$  0.10 \\
HD 70088 & -0.16 $\pm$  0.02 & -0.14 $\pm$  0.00 &  0.16 $\pm$  0.10 &  0.11 $\pm$  0.14 &  0.20 $\pm$  0.17 &  0.23 $\pm$  0.14 &  0.23 $\pm$  --- \\
HD 72760 & -0.03 $\pm$  0.03 &  0.00 $\pm$  0.22 &  0.15 $\pm$  0.10 &  0.02 $\pm$  0.24 &  0.19 $\pm$  0.30 &  0.24 $\pm$  0.17 &  0.12 $\pm$  0.24 \\
V401 Hya & -0.05 $\pm$  0.01 & -0.08 $\pm$  0.00 &  0.09 $\pm$  0.10 &  0.17 $\pm$  0.37 &  0.14 $\pm$  0.10 &  0.16 $\pm$  0.20 &  0.05 $\pm$  0.24 \\
HD 73171 &  0.25 $\pm$  0.04 &  0.19 $\pm$  0.57 &  0.13 $\pm$  0.39 & -0.17 $\pm$  0.40 &  0.22 $\pm$  0.17 &  0.23 $\pm$  0.25 &  0.33 $\pm$  0.32 \\
CT Pyx &  0.02 $\pm$  0.04 & -0.05 $\pm$  0.26 &  0.05 $\pm$  0.00 &  0.12 $\pm$  0.49 &  0.02 $\pm$  0.20 &  0.12 $\pm$  0.22 &  0.07 $\pm$  0.24 \\
$\rho$ Cnc A &  0.17 $\pm$  0.04 &  0.16 $\pm$  0.00 & -0.02 $\pm$  0.00 & -0.16 $\pm$  0.33 & -0.08 $\pm$  0.24 &  0.07 $\pm$  0.14 &  0.01 $\pm$  0.30 \\
HD 75898 &  0.09 $\pm$  0.02 &  0.05 $\pm$  0.20 & -0.02 $\pm$  0.17 & -0.23 $\pm$  0.22 & -0.02 $\pm$  0.10 & -0.14 $\pm$  0.17 &  0.05 $\pm$  0.26 \\
IK Cnc & -0.09 $\pm$  0.04 &  0.02 $\pm$  0.17 &  0.24 $\pm$  0.10 &  0.21 $\pm$  0.35 &  0.19 $\pm$  0.17 &  0.23 $\pm$  0.10 &  0.22 $\pm$  0.22\\
HD 76151 & -0.01 $\pm$  0.05 & -0.03 $\pm$  0.00 & -0.01 $\pm$  0.10 & -0.09 $\pm$  0.17 & -0.01 $\pm$  0.10 & -0.04 $\pm$  0.10 & -0.01 $\pm$  0.14\\
HD 82106 & 0.00 $\pm$  0.01 & -0.07 $\pm$  0.14 &  0.34 $\pm$  0.36 &  0.20 $\pm$  0.28 &  0.10 $\pm$  0.20 &  0.29 $\pm$  0.26 &  0.22 $\pm$  0.44\\
HD 83983 &  0.02 $\pm$  0.07 & -0.06 $\pm$  0.14 &  0.07 $\pm$  0.10 & -0.04 $\pm$  0.01 & -0.02 $\pm$  0.22 &  0.03 $\pm$  0.35 & -0.08 $\pm$  0.46\\
HD 85301 & -0.08 $\pm$  0.04 & -0.05 $\pm$  0.14 &  0.06 $\pm$  0.00 &  0.09 $\pm$  0.33 &  0.11 $\pm$  0.14 &  0.09 $\pm$  0.17 & -0.01 $\pm$  0.24\\
HD 86322 &  0.20 $\pm$  0.02 &  0.20 $\pm$  0.52 &  0.14 $\pm$  0.22 & -0.12 $\pm$  0.32 &  0.05 $\pm$  0.22 &  0.26 $\pm$  0.22 &  0.30 $\pm$  0.25\\
HD 89307 & -0.06 $\pm$  0.03 &  0.04 $\pm$  0.32 &  0.03 $\pm$  0.00 & -0.04 $\pm$  0.26 &  0.11 $\pm$  0.20 &  0.05 $\pm$  0.14 &  0.12 $\pm$  0.00\\
HD 89376 &  0.02 $\pm$  0.01 &  0.05 $\pm$  0.25 &  0.22 $\pm$  0.17 &  0.04 $\pm$  0.22 & -0.05 $\pm$  0.10 &  0.06 $\pm$  0.10 &  0.09 $\pm$  0.37\\
$\pi^1$ Leo &  0.18 $\pm$  0.02 &  0.12 $\pm$  0.42 &  0.08 $\pm$  0.25 &  0.07 $\pm$  0.10 &  0.07 $\pm$  0.10 &  0.15 $\pm$  0.26 &  0.22 $\pm$  0.26\\
HD 98356 & -0.08 $\pm$  0.05 & 0.00 $\pm$  0.39 &  0.10 $\pm$  0.14 & -0.09 $\pm$  0.36 & -0.04 $\pm$  0.17 &  0.03 $\pm$  0.25 &  0.14 $\pm$  0.17\\
HD 101112 &  0.36 $\pm$  0.11 &  0.21 $\pm$  0.56 &  0.13 $\pm$  0.22 & -0.19 $\pm$  0.33 & -0.09 $\pm$  0.25 &  0.07 $\pm$  0.17 &  0.11 $\pm$  0.32\\
HD 106696 &  0.14 $\pm$  0.07 &  0.18 $\pm$  0.22 & -0.03 $\pm$  0.20 & -0.12 $\pm$  0.14 &  0.11 $\pm$  0.26 &  0.12 $\pm$  0.17 &  0.24 $\pm$  0.17 \\
MY UMa & -0.01 $\pm$  0.03 & -0.01 $\pm$  0.44 &  0.24 $\pm$  0.52 &  0.43 $\pm$  0.45 &  0.14 $\pm$  0.22 &  0.31 $\pm$  0.28 &  0.40 $\pm$  0.39\\
GR Leo &  0.03 $\pm$  0.01 &  0.10 $\pm$  0.42 &  0.16 $\pm$  0.00 &  0.10 $\pm$  0.22 &  0.19 $\pm$  0.20 &  0.27 $\pm$  0.00 &  0.25 $\pm$  0.17\\
HD 106023 & -0.12 $\pm$  0.06 &  0.07 $\pm$  0.33 &  0.10 $\pm$  0.10 & 0.00 $\pm$  0.28 &  0.10 $\pm$  0.10 &  0.19 $\pm$  0.14 &  0.04 $\pm$  0.14\\
HD 117860 & -0.15 $\pm$  0.07 & -0.03 $\pm$  0.26 &  0.07 $\pm$  0.14 & -0.01 $\pm$  0.17 &  0.19 $\pm$  0.17 &  0.12 $\pm$  0.10 &  0.05 $\pm$  0.24\\
HD 117936 &  0.04 $\pm$  0.02 & 0.00 $\pm$  0.20 &  0.31 $\pm$  0.60 &  0.14 $\pm$  0.30 & -0.03 $\pm$  0.28 &  0.27 $\pm$  0.33 &  0.06 $\pm$  0.56\\
HD 124642 & -0.03 $\pm$  0.01 & -0.13 $\pm$  0.20 &  0.46 $\pm$  0.59 &  0.16 $\pm$  0.35 &  0.08 $\pm$  0.28 &  0.29 $\pm$  0.26 &  0.26 $\pm$  0.49\\
HD 126535 &  0.01 $\pm$  0.02 &  0.18 $\pm$  0.44 &  0.04 $\pm$  0.00 & -0.02 $\pm$  0.22 & -0.01 $\pm$  0.22 &  0.11 $\pm$  0.17 &  0.10 $\pm$  0.28\\
HD 131023 & -0.03 $\pm$  0.09 & -0.07 $\pm$  0.20 &  0.03 $\pm$  0.14 & -0.07 $\pm$  0.26 &  0.07 $\pm$  0.14 &  0.04 $\pm$  0.17 &  0.05 $\pm$  0.20\\
HD 149026 &  0.04 $\pm$  0.01 &  0.04 $\pm$  0.24 & -0.03 $\pm$  0.14 & -0.17 $\pm$  0.00 & -0.04 $\pm$  0.10 & -0.12 $\pm$  0.14 & -0.15 $\pm$  0.10\\
HD 170657 & -0.02 $\pm$  0.02 &  0.04 $\pm$  0.22 &  0.07 $\pm$  0.28 &  0.01 $\pm$  0.24 &  0.12 $\pm$  0.26 &  0.12 $\pm$  0.24 &  0.21 $\pm$  0.24\\
MV Dra & -0.03 $\pm$  0.01 & -0.04 $\pm$  0.00 &  0.07 $\pm$  0.10 & -0.04 $\pm$  0.20 &  0.00 $\pm$  0.17 &  0.11 $\pm$  0.22 &  0.00 $\pm$  0.14\\
HD 189087 & -0.03 $\pm$  0.03 & -0.04 $\pm$  0.10 &  0.16 $\pm$  0.14 &  0.04 $\pm$  0.26 &  0.19 $\pm$  0.00 &  0.25 $\pm$  0.17 &  0.15 $\pm$  0.10\\
V2425 Cyg & -0.11 $\pm$  0.04 &  0.08 $\pm$  0.26 & -0.05 $\pm$  0.39 &  0.11 $\pm$  0.33 &  0.24 $\pm$  0.20 &  0.05 $\pm$  0.01 &  0.21 $\pm$  0.30 \\
HD 196885 & -0.05 $\pm$  0.06 &  0.01 $\pm$  0.26 & -0.11 $\pm$  0.10 &  0.13 $\pm$  0.30 &  0.03 $\pm$  0.17 & -0.06 $\pm$  0.14 & -0.06 $\pm$  0.17 \\
V2436 Cyg & -0.01 $\pm$  0.03 & -0.06 $\pm$  0.22 &  0.24 $\pm$  0.25 & 0.00 $\pm$  0.28 &  0.08 $\pm$  0.26 &  0.13 $\pm$  0.01 &  0.13 $\pm$  0.17\\
HD 200968 & -0.04 $\pm$  0.01 &  0.18 $\pm$  0.35 &  0.10 $\pm$  0.00 & -0.05 $\pm$  0.26 &  0.07 $\pm$  0.20 &  0.10 $\pm$  0.30 &  0.07 $\pm$  0.20\\
NS Aqr &  0.04 $\pm$  0.05 & -0.03 $\pm$  0.14 & -0.04 $\pm$  0.10 & -0.09 $\pm$  0.28 & -0.03 $\pm$  0.10 &  0.02 $\pm$  0.17 & -0.10 $\pm$  0.26 \\
V454 And & -0.25 $\pm$  0.00 & -0.11 $\pm$  0.24 &  0.12 $\pm$  0.10 &  0.07 $\pm$  0.20 &  0.16 $\pm$  0.17 &  0.09 $\pm$  0.14 &  0.16 $\pm$  0.10\\
HD 222422 & -0.04 $\pm$  0.01 & -0.06 $\pm$  0.14 &  0.12 $\pm$  0.14 &  0.24 $\pm$  0.33 &  0.30 $\pm$  0.10 &  0.17 $\pm$  0.14 &  0.13 $\pm$  0.28 \\
20 Psc &  0.04 $\pm$  0.04 &  0.03 $\pm$  0.35 &  0.09 $\pm$  0.14 &  0.06 $\pm$  0.17 &  0.28 $\pm$  0.24 &  0.22 $\pm$  0.20 &  0.29 $\pm$  0.26 \\
\noalign{\smallskip}
\hline
\noalign{\smallskip}
\end{tabular}
\end{table*}
%\end{landscape}

\begin{landscape}
\begin{table}
\scriptsize
\caption{Differential abundances $\Delta$[X/H] for the  $\alpha$-, Fe-peak, and odd-Z elements: Na, Mg, Al, Si, Ca, Sc, Ti, V, Cr, Mn, Co, and Ni.}
\label{diftable2}
\centering
\begin{tabular}{lcccccccccccc}     % 15 columns
\hline
\hline
\noalign{\smallskip}
{\small Name} & {\small $\Delta$[Na/H]} & {\small $\Delta$[Mg/H]} & {\small $\Delta$[Al/H]} & {\small $\Delta$[Si/H]} & {\small $\Delta$[Ca/H]} & {\small $\Delta$[Sc/H]} & {\small $\Delta$[Ti/H]} & {\small $\Delta$[V/H]} & {\small $\Delta$[Cr/H]} & {\small $\Delta$[Mn/H]} & {\small $\Delta$[Co/H]} & {\small  $\Delta$[Ni/H]} \\
\noalign{\smallskip}
\hline
\noalign{\smallskip}
BE Cet &  0.22 $\pm$  0.01 &  0.17 $\pm$  0.04 &  0.18 $\pm$  0.01 & 
 0.10 $\pm$  0.03 &  0.13 $\pm$  0.02 &  0.09 $\pm$  0.06 &  0.08 $\pm$  0.02 & 
-0.02 $\pm$  0.05 &  0.16 $\pm$  0.02 & 
 0.10 $\pm$  0.07 &  0.06 $\pm$  0.03 &  0.12 $\pm$  0.01\\
HD 5848 &  0.49 $\pm$  --- &  0.01 $\pm$  0.07 &  0.26 $\pm$  0.04 & 
 0.11 $\pm$  0.04 & -0.01 $\pm$  0.04 &  0.10 $\pm$  0.16 &  0.00 $\pm$  0.03 & 
 0.10 $\pm$  0.05 & -0.09 $\pm$  0.04 & 
 0.03 $\pm$  0.17 &  0.13 $\pm$  0.05 & -0.03 $\pm$  0.02\\
AZ Ari &  0.16 $\pm$  0.05 &  0.09 $\pm$  0.01 &  0.10 $\pm$  0.00 & 
 0.04 $\pm$  0.02 &  0.07 $\pm$  0.01 & -0.02 $\pm$  0.01 &  0.07 $\pm$  0.02 & 
-0.10 $\pm$  0.05 &  0.10 $\pm$  0.02 & 
 0.04 $\pm$  0.05 &  0.04 $\pm$  0.02 &  0.07 $\pm$  0.01\\
FT Cet &  0.09 $\pm$  0.07 & -0.01 $\pm$  0.02 &  0.12 $\pm$  0.07 & 
-0.06 $\pm$  0.03 & -0.07 $\pm$  0.04 &  0.05 $\pm$  0.07 &  0.05 $\pm$  0.02 & 
 0.12 $\pm$  0.04 &  0.02 $\pm$  0.02 & 
-0.22 $\pm$  0.11 &  0.05 $\pm$  0.02 & -0.02 $\pm$  0.01\\
BZ Cet &  0.05 $\pm$  0.13 &  0.05 $\pm$  0.05 &  0.14 $\pm$  0.04 & 
 0.04 $\pm$  0.03 & -0.03 $\pm$  0.04 &  0.04 $\pm$  0.08 &  0.06 $\pm$  0.01 & 
 0.11 $\pm$  0.04 & -0.01 $\pm$  0.03 & 
-0.12 $\pm$  0.08 &  0.03 $\pm$  0.03 &  0.04 $\pm$  0.01\\
$\delta$ Ari &  0.14 $\pm$  0.26 &  0.19 $\pm$  0.02 &  0.31 $\pm$  0.05 & 
 0.16 $\pm$  0.03 &  0.10 $\pm$  0.04 &  0.03 $\pm$  0.09 &  0.06 $\pm$  0.03 & 
 0.02 $\pm$  0.07 &  0.02 $\pm$  0.02 & 
 0.24 $\pm$  0.08 &  0.25 $\pm$  0.09 &  0.10 $\pm$  0.01\\
V683 Per &  0.14 $\pm$  0.03 &  0.10 $\pm$  0.02 &  0.12 $\pm$  0.02 & 
-0.01 $\pm$  0.03 & -0.03 $\pm$  0.04 &  0.02 $\pm$  0.04 & -0.01 $\pm$  0.02 & 
-0.09 $\pm$  0.04 &  0.05 $\pm$  0.02 & 
 0.05 $\pm$  0.06 &  0.00 $\pm$  0.03 &  0.05 $\pm$  0.01\\
V686 Per &  0.10 $\pm$  0.09 &  0.09 $\pm$  0.05 &  0.04 $\pm$  0.02 & 
 0.02 $\pm$  0.02 &  0.06 $\pm$  0.02 &  0.01 $\pm$  0.04 &  0.11 $\pm$  0.02 & 
-0.10 $\pm$  0.04 &  0.06 $\pm$  0.02 & 
-0.05 $\pm$  0.11 &  0.03 $\pm$  0.01 &  0.05 $\pm$  0.01\\
HD 21663 & -0.03 $\pm$  0.08 & -0.01 $\pm$  0.11 & -0.03 $\pm$  0.01 & 
-0.14 $\pm$  0.03 & -0.14 $\pm$  0.03 & -0.17 $\pm$  0.03 & -0.12 $\pm$  0.02 & 
-0.25 $\pm$  0.03 & -0.12 $\pm$  0.03 & 
-0.18 $\pm$  0.06 & -0.11 $\pm$  0.02 & -0.12 $\pm$  0.01\\
HD 23356 & -0.28 $\pm$  0.08 & -0.09 $\pm$  0.15 & -0.14 $\pm$  0.05 & 
-0.22 $\pm$  0.03 & -0.23 $\pm$  0.04 & -0.18 $\pm$  0.07 & -0.13 $\pm$  0.02 & 
-0.07 $\pm$  0.05 & -0.25 $\pm$  0.02 & 
-0.39 $\pm$  0.04 & -0.21 $\pm$  0.03 & -0.23 $\pm$  0.01\\
39 Tau &  0.01 $\pm$  0.05 &  0.10 $\pm$  0.16 & -0.03 $\pm$  0.02 & 
-0.04 $\pm$  0.02 & -0.03 $\pm$  0.02 & -0.12 $\pm$  0.02 & -0.04 $\pm$  0.01 & 
-0.23 $\pm$  0.05 & -0.02 $\pm$  0.02 & 
-0.12 $\pm$  0.07 & -0.09 $\pm$  0.02 & -0.07 $\pm$  0.01\\
HD 25893 &  0.07 $\pm$  0.05 & -0.02 $\pm$  0.07 &  0.17 $\pm$  0.02 & 
-0.00 $\pm$  0.02 &  0.03 $\pm$  0.03 & -0.08 $\pm$  0.04 & -0.00 $\pm$  0.02 & 
-0.08 $\pm$  0.04 &  0.02 $\pm$  0.01 & 
 0.09 $\pm$  0.12 & -0.02 $\pm$  0.05 & -0.02 $\pm$  0.01\\
HD 27282 &  0.09 $\pm$  0.01 &  0.10 $\pm$  0.04 &  0.08 $\pm$  0.04 & 
 0.06 $\pm$  0.02 &  0.07 $\pm$  0.02 & -0.01 $\pm$  0.02 &  0.04 $\pm$  0.02 & 
-0.08 $\pm$  0.04 &  0.05 $\pm$  0.02 & 
 0.05 $\pm$  0.05 &  0.03 $\pm$  0.01 &  0.07 $\pm$  0.01\\
HD 27685 &  0.08 $\pm$  0.04 &  0.09 $\pm$  0.03 & -0.04 $\pm$  0.03 & 
 0.01 $\pm$  0.02 &  0.03 $\pm$  0.02 & -0.02 $\pm$  0.05 & -0.02 $\pm$  0.02 & 
-0.19 $\pm$  0.04 &  0.05 $\pm$  0.02 & 
-0.03 $\pm$  0.08 & -0.03 $\pm$  0.03 &  0.03 $\pm$  0.01\\
HD 27989 &  0.35 $\pm$  0.14 & -0.04 $\pm$  0.00 &  0.28 $\pm$  0.04 & 
 0.04 $\pm$  0.03 &  0.13 $\pm$  0.05 &  0.06 $\pm$  0.03 &  0.08 $\pm$  0.03 & 
 0.10 $\pm$  0.05 &  0.10 $\pm$  0.02 & 
 0.05 $\pm$  0.08 &  0.01 $\pm$  0.10 &  0.06 $\pm$  0.02\\
$\epsilon$ Tau &  0.56 $\pm$  --- &  0.22 $\pm$  0.02 &  0.36 $\pm$  0.05 & 
 0.23 $\pm$  0.03 &  0.13 $\pm$  0.04 &  0.10 $\pm$  0.05 &  0.15 $\pm$  0.03 & 
 0.01 $\pm$  0.07 &  0.09 $\pm$  0.02 & 
 0.24 $\pm$  0.12 &  0.29 $\pm$  0.07 &  0.10 $\pm$  0.01\\
111 Tau B & -0.20 $\pm$  0.09 & -0.30 $\pm$  0.15 & -0.03 $\pm$  0.04 & 
-0.36 $\pm$  0.05 & -0.20 $\pm$  0.05 &  0.13 $\pm$  0.08 & -0.01 $\pm$  0.02 & 
 0.02 $\pm$  0.06 & -0.17 $\pm$  0.02 & 
-0.58 $\pm$  0.08 & -0.22 $\pm$  0.04 & -0.27 $\pm$  0.02\\
HD 40979 &  0.26 $\pm$  0.10 &  0.24 $\pm$  0.23 &  0.08 $\pm$  0.38 & 
 0.19 $\pm$  0.02 &  0.19 $\pm$  0.04 &  0.21 $\pm$  0.08 &  0.16 $\pm$  0.04 & 
-0.16 $\pm$  0.10 &  0.21 $\pm$  0.03 & 
 0.15 $\pm$  0.06 &  0.18 $\pm$  0.03 &  0.22 $\pm$  0.02\\
HD 45609 & -0.19 $\pm$  0.01 & -0.08 $\pm$  0.09 & -0.05 $\pm$  0.05 & 
-0.23 $\pm$  0.02 & -0.23 $\pm$  0.03 & -0.17 $\pm$  0.05 & -0.13 $\pm$  0.02 & 
-0.21 $\pm$  0.03 & -0.12 $\pm$  0.03 & 
-0.29 $\pm$  0.05 & -0.15 $\pm$  0.04 & -0.23 $\pm$  0.01\\
HD 52265 &  0.28 $\pm$  0.04 &  0.31 $\pm$  0.13 &  0.14 $\pm$  0.04 & 
 0.12 $\pm$  0.02 &  0.17 $\pm$  0.04 &  0.08 $\pm$  0.09 &  0.12 $\pm$  0.02 & 
-0.08 $\pm$  0.06 &  0.12 $\pm$  0.03 & 
 0.05 $\pm$  0.10 &  0.10 $\pm$  0.02 &  0.17 $\pm$  0.01\\
HD 53532 &  0.05 $\pm$  0.04 &  0.01 $\pm$  --- &  0.06 $\pm$  0.04 & 
 0.00 $\pm$  0.02 &  0.05 $\pm$  0.02 & -0.09 $\pm$  0.02 &  0.02 $\pm$  0.02 & 
-0.15 $\pm$  0.04 &  0.05 $\pm$  0.02 & 
-0.08 $\pm$  0.03 & -0.03 $\pm$  0.02 & -0.02 $\pm$  0.01\\
HD 65523 & -0.09 $\pm$  0.06 & -0.12 $\pm$  0.05 & -0.04 $\pm$  0.01 & 
-0.18 $\pm$  0.03 & -0.12 $\pm$  0.03 & -0.22 $\pm$  0.02 & -0.11 $\pm$  0.02 & 
-0.20 $\pm$  0.02 & -0.12 $\pm$  0.02 & 
-0.19 $\pm$  0.05 & -0.12 $\pm$  0.02 & -0.17 $\pm$  0.01\\
HD 70088 & -0.13 $\pm$  0.06 & -0.07 $\pm$  0.10 & -0.06 $\pm$  0.03 & 
-0.19 $\pm$  0.02 & -0.14 $\pm$  0.02 & -0.18 $\pm$  0.02 & -0.11 $\pm$  0.01 & 
-0.32 $\pm$  0.06 & -0.09 $\pm$  0.02 & 
-0.33 $\pm$  0.09 & -0.27 $\pm$  0.04 & -0.20 $\pm$  0.01\\
HD 72760 & -0.03 $\pm$  0.07 & -0.03 $\pm$  0.03 &  0.00 $\pm$  0.00 & 
-0.12 $\pm$  0.03 & -0.08 $\pm$  0.02 & -0.16 $\pm$  0.03 & -0.05 $\pm$  0.02 & 
-0.11 $\pm$  0.01 & -0.03 $\pm$  0.02 & 
-0.12 $\pm$  0.04 & -0.09 $\pm$  0.02 & -0.08 $\pm$  0.01\\
V401 Hya &  0.07 $\pm$  0.05 & -0.06 $\pm$  0.06 &  0.06 $\pm$  0.02 &
-0.03 $\pm$  0.03 &  0.01 $\pm$  0.02 & -0.03 $\pm$  0.04 &  0.02 $\pm$  0.02 &
-0.19 $\pm$  0.05 &  0.05 $\pm$  0.02 &
-0.08 $\pm$  0.09 & -0.03 $\pm$  0.02 &  0.00 $\pm$  0.01\\
HD 73171 &  0.16 $\pm$  0.02 & -0.03 $\pm$  0.03 &  0.08 $\pm$  0.00 & 
-0.04 $\pm$  0.04 & -0.14 $\pm$  0.01 & -0.08 $\pm$  0.13 & -0.13 $\pm$  0.03 & 
-0.11 $\pm$  0.07 & -0.17 $\pm$  0.02 & 
 0.03 $\pm$  0.08 &  0.03 $\pm$  0.07 & -0.13 $\pm$  0.01\\
CT Pyx &  0.09 $\pm$  0.04 &  0.01 $\pm$  0.09 &  0.08 $\pm$  0.01 & 
-0.06 $\pm$  0.02 & -0.01 $\pm$  0.02 & -0.07 $\pm$  0.04 &  0.01 $\pm$  0.01 & 
-0.01 $\pm$  0.03 & -0.01 $\pm$  0.01 & 
-0.16 $\pm$  0.09 &  0.02 $\pm$  0.02 & -0.03 $\pm$  0.01\\
$\rho$ Cnc A &  0.35 $\pm$  0.12 &  0.31 $\pm$  0.00 &  0.42 $\pm$  0.01 & 
 0.27 $\pm$  0.02 &  0.17 $\pm$  0.03 &  0.33 $\pm$  0.07 &  0.33 $\pm$  0.02 & 
 0.40 $\pm$  0.06 &  0.23 $\pm$  0.01 & 
 0.19 $\pm$  0.09 &  0.52 $\pm$  0.04 &  0.33 $\pm$  0.01\\
HD 75898 &  0.43 $\pm$  0.02 &  0.20 $\pm$  0.08 &  0.19 $\pm$  0.02 & 
 0.19 $\pm$  0.02 &  0.18 $\pm$  0.02 &  0.23 $\pm$  0.10 &  0.20 $\pm$  0.02 & 
 0.04 $\pm$  0.04 &  0.19 $\pm$  0.02 & 
 0.20 $\pm$  0.09 &  0.25 $\pm$  0.02 &  0.26 $\pm$  0.01\\
HD 76151 &  0.12 $\pm$  0.01 &  0.05 $\pm$  0.01 &  0.04 $\pm$  0.00 & 
-0.03 $\pm$  0.03 & -0.01 $\pm$  0.01 & -0.07 $\pm$  0.05 & -0.01 $\pm$  0.02 & 
-0.16 $\pm$  0.04 &  0.04 $\pm$  0.02 & 
-0.03 $\pm$  0.09 &  0.03 $\pm$  0.02 &  0.03 $\pm$  0.01\\
IK Cnc & -0.25 $\pm$  0.10 & -0.32 $\pm$  0.06 & -0.27 $\pm$  0.02 & 
-0.38 $\pm$  0.03 & -0.27 $\pm$  0.05 & -0.41 $\pm$  0.05 & -0.23 $\pm$  0.03 & 
-0.43 $\pm$  0.05 & -0.30 $\pm$  0.02 & 
-0.52 $\pm$  0.10 & -0.41 $\pm$  0.03 & -0.41 $\pm$  0.01\\
HD 82106 & -0.14 $\pm$  0.00 & -0.15 $\pm$  0.12 & -0.07 $\pm$  0.02 & 
-0.21 $\pm$  0.03 & -0.16 $\pm$  0.04 & -0.03 $\pm$  0.09 & -0.01 $\pm$  0.02 & 
 0.06 $\pm$  0.06 & -0.22 $\pm$  0.01 & 
-0.37 $\pm$  0.09 & -0.15 $\pm$  0.03 & -0.18 $\pm$  0.01\\
GT Leo & -0.04 $\pm$  0.06 &  0.04 $\pm$  0.03 &  0.10 $\pm$  0.00 & 
-0.07 $\pm$  0.02 & -0.05 $\pm$  0.02 & -0.09 $\pm$  0.05 & -0.00 $\pm$  0.02 & 
 0.01 $\pm$  0.03 & -0.06 $\pm$  0.02 & 
-0.06 $\pm$  0.08 &  0.02 $\pm$  0.05 & -0.02 $\pm$  0.01\\
HD 85301 &  0.08 $\pm$  0.04 &  0.05 $\pm$  0.09 &  0.02 $\pm$  0.02 & 
-0.02 $\pm$  0.02 &  0.04 $\pm$  0.03 & -0.11 $\pm$  0.03 &  0.03 $\pm$  0.02 & 
-0.12 $\pm$  0.05 &  0.02 $\pm$  0.02 & 
 0.02 $\pm$  0.06 & -0.03 $\pm$  0.03 &  0.00 $\pm$  0.01\\
HD 86322 &  0.02 $\pm$  0.20 &  0.18 $\pm$  0.10 &  0.04 $\pm$  0.02 & 
-0.03 $\pm$  0.03 & -0.11 $\pm$  0.03 & -0.17 $\pm$  0.12 & -0.10 $\pm$  0.03 & 
-0.11 $\pm$  0.06 & -0.20 $\pm$  0.02 & 
-0.06 $\pm$  0.09 & -0.01 $\pm$  0.06 & -0.10 $\pm$  0.01\\
HD 89307 & -0.13 $\pm$  0.06 & -0.09 $\pm$  0.07 & -0.18 $\pm$  0.02 & 
-0.23 $\pm$  0.02 & -0.21 $\pm$  0.01 & -0.17 $\pm$  0.04 & -0.19 $\pm$  0.02 & 
-0.41 $\pm$  0.07 & -0.22 $\pm$  0.03 & 
-0.38 $\pm$  0.11 & -0.28 $\pm$  0.03 & -0.23 $\pm$  0.01\\
HD 89376 &  0.06 $\pm$  0.03 & -0.08 $\pm$  0.05 &  0.02 $\pm$  0.05 & 
-0.09 $\pm$  0.02 & -0.07 $\pm$  0.06 &  0.01 $\pm$  0.04 &  0.01 $\pm$  0.02 & 
-0.01 $\pm$  0.05 & -0.11 $\pm$  0.02 & 
-0.14 $\pm$  0.07 & -0.08 $\pm$  0.03 & -0.11 $\pm$  0.01\\
$\pi^1$ Leo &  0.42 $\pm$  --- &  0.14 $\pm$  0.09 &  0.13 $\pm$  0.02 & 
 0.07 $\pm$  0.02 &  0.03 $\pm$  0.02 & -0.05 $\pm$  0.04 & -0.01 $\pm$  0.03 & 
-0.06 $\pm$  0.03 &  0.00 $\pm$  0.02 & 
 0.01 $\pm$  0.16 &  0.14 $\pm$  0.05 & -0.00 $\pm$  0.01\\
HD 98356 &  0.16 $\pm$  0.02 &  0.10 $\pm$  0.07 &  0.05 $\pm$  0.03 & 
-0.02 $\pm$  0.03 & -0.05 $\pm$  0.02 & -0.03 $\pm$  0.08 &  0.09 $\pm$  0.03 & 
 0.06 $\pm$  0.03 & -0.01 $\pm$  0.01 & 
-0.06 $\pm$  0.10 &  0.12 $\pm$  0.03 &  0.03 $\pm$  0.01\\
HD 101112 &  0.40 $\pm$  --- &  0.13 $\pm$  0.03 &  0.24 $\pm$  0.04 & 
 0.15 $\pm$  0.03 &  0.10 $\pm$  0.04 &  0.07 $\pm$  0.12 &  0.06 $\pm$  0.03 & 
 0.10 $\pm$  0.06 & -0.04 $\pm$  0.03 & 
 0.23 $\pm$  0.14 &  0.31 $\pm$  0.11 &  0.09 $\pm$  0.01\\
HD 106696 & -0.14 $\pm$  0.06 & -0.12 $\pm$  0.02 & -0.16 $\pm$  0.03 &
-0.24 $\pm$  0.02 & -0.29 $\pm$  0.03 & -0.43 $\pm$  0.02 & -0.30 $\pm$  0.02 &
-0.41 $\pm$  0.02 & -0.42 $\pm$  0.03 &
-0.37 $\pm$  0.07 & -0.27 $\pm$  0.05 & -0.35 $\pm$  0.01\\
MY UMa & -0.14 $\pm$  0.07 & -0.12 $\pm$  0.17 & -0.08 $\pm$  0.06 & 
-0.26 $\pm$  0.05 & -0.08 $\pm$  0.06 & -0.05 $\pm$  0.05 & -0.02 $\pm$  0.04 & 
 0.04 $\pm$  0.02 & -0.13 $\pm$  0.03 & 
-0.66 $\pm$  0.11 & -0.22 $\pm$  0.03 & -0.23 $\pm$  0.02\\
GR Leo & -0.09 $\pm$  0.04 &  0.01 $\pm$  0.09 & -0.02 $\pm$  0.00 & 
-0.05 $\pm$  0.02 & -0.07 $\pm$  0.03 & -0.15 $\pm$  0.03 & -0.01 $\pm$  0.02 & 
-0.08 $\pm$  0.02 & -0.08 $\pm$  0.02 & 
-0.21 $\pm$  0.10 & -0.07 $\pm$  0.02 & -0.11 $\pm$  0.01\\
HD 106023 & -0.03 $\pm$  0.05 & -0.02 $\pm$  0.06 & -0.07 $\pm$  0.00 & 
-0.15 $\pm$  0.03 & -0.11 $\pm$  0.01 & -0.20 $\pm$  0.01 & -0.12 $\pm$  0.02 & 
-0.21 $\pm$  0.02 & -0.08 $\pm$  0.01 & 
-0.20 $\pm$  0.03 & -0.13 $\pm$  0.02 & -0.12 $\pm$  0.01\\
HD 117860 &  0.03 $\pm$  0.04 &  0.03 $\pm$  0.07 & -0.01 $\pm$  0.03 & 
-0.06 $\pm$  0.02 & -0.02 $\pm$  0.01 & -0.03 $\pm$  0.06 & -0.06 $\pm$  0.02 & 
-0.27 $\pm$  0.06 & -0.02 $\pm$  0.02 & 
-0.18 $\pm$  0.04 & -0.11 $\pm$  0.02 & -0.06 $\pm$  0.01\\
HD 117936 & -0.14 $\pm$  --- & -0.06 $\pm$  0.16 &  0.07 $\pm$  0.01 & 
-0.17 $\pm$  0.04 & -0.10 $\pm$  0.03 &  0.18 $\pm$  0.12 &  0.10 $\pm$  0.02 & 
 0.20 $\pm$  0.06 & -0.07 $\pm$  0.02 & 
-0.25 $\pm$  0.08 & -0.01 $\pm$  0.04 & -0.06 $\pm$  0.02\\
HD 124642 & -0.00 $\pm$  0.04 & -0.13 $\pm$  0.18 &  0.00 $\pm$  0.06 & 
-0.21 $\pm$  0.03 & -0.13 $\pm$  0.05 &  0.12 $\pm$  0.11 &  0.07 $\pm$  0.02 & 
 0.15 $\pm$  0.06 & -0.07 $\pm$  0.02 & 
-0.37 $\pm$  0.11 & -0.12 $\pm$  0.03 & -0.12 $\pm$  0.02\\
HD 126535 &  0.18 $\pm$  0.03 &  0.17 $\pm$  0.09 &  0.13 $\pm$  0.02 & 
 0.01 $\pm$  0.02 &  0.05 $\pm$  0.01 &  0.05 $\pm$  0.03 &  0.09 $\pm$  0.02 & 
 0.05 $\pm$  0.03 &  0.07 $\pm$  0.01 & 
-0.05 $\pm$  0.14 &  0.12 $\pm$  0.02 &  0.08 $\pm$  0.01\\
HD 131023 &  0.17 $\pm$  0.03 &  0.14 $\pm$  0.04 &  0.12 $\pm$  0.01 & 
 0.05 $\pm$  0.02 &  0.07 $\pm$  0.01 & -0.01 $\pm$  0.02 &  0.05 $\pm$  0.02 & 
-0.05 $\pm$  0.03 &  0.10 $\pm$  0.01 & 
 0.07 $\pm$  0.07 &  0.08 $\pm$  0.02 &  0.06 $\pm$  0.01\\
HD 149026 &  0.44 $\pm$  0.02 &  0.32 $\pm$  0.06 &  0.31 $\pm$  0.03 & 
 0.23 $\pm$  0.03 &  0.21 $\pm$  0.02 &  0.22 $\pm$  0.07 &  0.24 $\pm$  0.02 & 
 0.11 $\pm$  0.07 &  0.26 $\pm$  0.02 & 
 0.23 $\pm$  0.12 &  0.28 $\pm$  0.03 &  0.30 $\pm$  0.01\\
HD 170657 & -0.26 $\pm$  0.05 & -0.26 $\pm$  0.07 & -0.22 $\pm$  0.02 & 
-0.34 $\pm$  0.03 & -0.32 $\pm$  0.02 & -0.35 $\pm$  0.03 & -0.25 $\pm$  0.02 & 
-0.28 $\pm$  0.03 & -0.34 $\pm$  0.02 & 
-0.46 $\pm$  0.04 & -0.33 $\pm$  0.02 & -0.35 $\pm$  0.01\\
MV Dra &  0.09 $\pm$  0.05 &  0.09 $\pm$  0.07 &  0.06 $\pm$  0.00 & 
-0.03 $\pm$  0.03 &  0.02 $\pm$  0.02 & -0.07 $\pm$  0.01 &  0.01 $\pm$  0.02 & 
-0.10 $\pm$  0.03 &  0.04 $\pm$  0.01 & 
 0.01 $\pm$  0.05 &  0.01 $\pm$  0.02 & -0.01 $\pm$  0.01\\
HD 189087 & -0.12 $\pm$  0.05 & -0.11 $\pm$  0.05 & -0.09 $\pm$  0.00 & 
-0.21 $\pm$  0.03 & -0.18 $\pm$  0.03 & -0.26 $\pm$  0.01 & -0.12 $\pm$  0.02 & 
-0.23 $\pm$  0.02 & -0.15 $\pm$  0.01 & 
-0.27 $\pm$  0.04 & -0.17 $\pm$  0.01 & -0.22 $\pm$  0.01\\
V2425 Cyg &  0.09 $\pm$  0.01 & -0.10 $\pm$  0.22 & -0.05 $\pm$  0.08 &
-0.06 $\pm$  0.04 &  0.00 $\pm$  0.04 & -0.03 $\pm$  0.07 &  0.02 $\pm$  0.03 &
 0.03 $\pm$  0.05 & -0.05 $\pm$  0.03 &
-0.10 $\pm$  0.05 & -0.10 $\pm$  0.05 & -0.06 $\pm$  0.01\\
HD 196885 &  0.31 $\pm$  0.03 &  0.21 $\pm$  0.18 &  0.19 $\pm$  0.00 & 
 0.14 $\pm$  0.03 &  0.22 $\pm$  0.03 &  0.15 $\pm$  0.04 &  0.17 $\pm$  0.02 & 
-0.02 $\pm$  0.08 &  0.15 $\pm$  0.03 & 
 0.12 $\pm$  0.09 &  0.11 $\pm$  0.04 &  0.18 $\pm$  0.01\\
V2436 Cyg &  0.03 $\pm$  0.00 & -0.01 $\pm$  0.08 &  0.08 $\pm$  0.01 & 
-0.01 $\pm$  0.03 & -0.03 $\pm$  0.03 & -0.02 $\pm$  0.08 &  0.04 $\pm$  0.01 & 
 0.13 $\pm$  0.05 & -0.01 $\pm$  0.02 & 
-0.07 $\pm$  0.03 &  0.01 $\pm$  0.03 & -0.01 $\pm$  0.01\\
HD 200968 & -0.14 $\pm$  0.05 & -0.14 $\pm$  0.07 & -0.05 $\pm$  0.00 & 
-0.17 $\pm$  0.03 & -0.10 $\pm$  0.02 & -0.20 $\pm$  0.02 & -0.11 $\pm$  0.02 & 
-0.13 $\pm$  0.03 & -0.11 $\pm$  0.01 & 
-0.20 $\pm$  0.03 & -0.12 $\pm$  0.02 & -0.15 $\pm$  0.01\\
NS Aqr &  0.22 $\pm$  0.05 &  0.10 $\pm$  0.05 &  0.14 $\pm$  0.00 & 
 0.03 $\pm$  0.03 &  0.09 $\pm$  0.02 &  0.03 $\pm$  0.03 &  0.06 $\pm$  0.02 & 
-0.07 $\pm$  0.04 &  0.10 $\pm$  0.02 & 
-0.00 $\pm$  0.08 &  0.11 $\pm$  0.01 &  0.09 $\pm$  0.01\\
V454 And &  0.01 $\pm$  0.01 &  0.16 $\pm$  0.15 &  0.12 $\pm$  0.03 & 
 0.02 $\pm$  0.02 &  0.03 $\pm$  0.04 &  0.01 $\pm$  0.04 &  0.07 $\pm$  0.02 & 
-0.16 $\pm$  0.06 &  0.09 $\pm$  0.03 & 
-0.08 $\pm$  0.03 &  0.00 $\pm$  0.03 &  0.03 $\pm$  0.01\\
HD 222422 & -0.19 $\pm$  0.04 & -0.07 $\pm$  0.11 & -0.10 $\pm$  0.01 & 
-0.26 $\pm$  0.03 & -0.15 $\pm$  0.02 & -0.31 $\pm$  0.04 & -0.16 $\pm$  0.02 & 
-0.34 $\pm$  0.02 & -0.21 $\pm$  0.02 & 
-0.38 $\pm$  0.06 & -0.25 $\pm$  0.02 & -0.25 $\pm$  0.01\\
20 Psc &  0.17 $\pm$  0.01 &  0.11 $\pm$  0.02 &  0.05 $\pm$  0.03 & 
-0.00 $\pm$  0.02 & -0.01 $\pm$  0.03 & -0.17 $\pm$  0.06 & -0.07 $\pm$  0.02 & 
-0.11 $\pm$  0.03 & -0.08 $\pm$  0.02 & 
 0.01 $\pm$  0.10 & -0.03 $\pm$  0.05 & -0.09 $\pm$  0.01\\
\noalign{\smallskip}
\hline
\noalign{\smallskip}
\end{tabular}
\end{table}
\end{landscape}

%\begin{landscape}
\begin{table*}
\scriptsize
\caption{Differential abundances $\Delta$[X/H] for the s-process elements: Cu, Zn, Y, Zr, Ba, Ce, and Nd.}
\label{diftable3}
\centering
\begin{tabular}{lccccccc}     % 15 columns
\hline
\hline
\noalign{\smallskip}
{\small Name} & {\small $\Delta$[Cu/H]} & {\small $\Delta$[Zn/H]} & {\small $\Delta$[Y/H]} & {\small $\Delta$[Zr/H]} & {\small $\Delta$[Ba/H]} & {\small $\Delta$[Ce/H]} & {\small $\Delta$[Nd/H]}\\
\noalign{\smallskip}
\hline
\noalign{\smallskip}
BE Cet & 0.00 $\pm$  0.06 & -0.01 $\pm$  0.17 &  0.16 $\pm$  0.06 &  0.04 $\pm$  0.26 &  0.36 $\pm$  0.05 &  0.08 $\pm$  0.01 &  0.11 $\pm$  0.08 \\
HD 5848 &  0.32 $\pm$  0.02 &  0.03 $\pm$  0.20 & -0.24 $\pm$  0.04 & -0.55 $\pm$  0.13 & -0.02 $\pm$  0.05 & -0.07 $\pm$  0.03 &  0.10 $\pm$  0.12 \\
AZ Ari &  0.10 $\pm$  0.06 & -0.05 $\pm$  0.23 &  0.18 $\pm$  0.05 & -0.08 $\pm$  0.11 &  0.22 $\pm$  0.02 &  0.06 $\pm$  0.04 & -0.04 $\pm$  0.15 \\
FT Cet &  0.07 $\pm$  0.01 & -0.21 $\pm$  0.11 & -0.08 $\pm$  0.04 & -0.21 $\pm$  0.29 &  0.13 $\pm$  0.07 &  0.15 $\pm$  0.03 &  0.12 $\pm$  0.10 \\
BZ Cet &  0.09 $\pm$  0.02 &  0.02 $\pm$ --- &  0.07 $\pm$  0.01 & -0.25 $\pm$  0.28 & 0.00 $\pm$  0.02 &  0.11 $\pm$  0.07 &  0.02 $\pm$  0.03 \\
$\delta$ Ari &  0.42 $\pm$  0.08 & -0.07 $\pm$  0.14 &  0.12 $\pm$  0.02 & -0.26 $\pm$  0.33 &  0.15 $\pm$  0.02 &  0.12 $\pm$  0.03 &  0.23 $\pm$  0.17 \\
V683 Per &  0.01 $\pm$  0.07 & -0.15 $\pm$  0.20 &  0.04 $\pm$  0.01 & -0.24 $\pm$  0.29 &  0.14 $\pm$  0.02 &  0.01 $\pm$  0.02 & 0.00 $\pm$  0.12 \\
V686 Per &  0.03 $\pm$  0.03 & -0.15 $\pm$  0.17 & -0.04 $\pm$  0.05 & -0.04 $\pm$  0.35 &  0.15 $\pm$  0.02 &  0.07 $\pm$ --- &  0.03 $\pm$  0.14 \\
HD 21663 & -0.03 $\pm$  0.04 & -0.34 $\pm$  0.21 & -0.12 $\pm$  0.07 & -0.41 $\pm$  0.27 & -0.04 $\pm$  0.01 & -0.10 $\pm$  0.05 & -0.09 $\pm$  0.06 \\
HD 23356 & -0.16 $\pm$  0.02 & -0.12 $\pm$  0.20 & -0.17 $\pm$  0.05 & -0.44 $\pm$  0.25 & -0.22 $\pm$  0.03 & -0.23 $\pm$  0.03 & -0.11 $\pm$  0.08 \\
39 Tau & -0.16 $\pm$  0.09 & -0.22 $\pm$  0.17 &  0.02 $\pm$  0.01 & -0.02 $\pm$  0.25 &  0.16 $\pm$  0.05 & 0.00 $\pm$  0.02 &  0.05 $\pm$  0.05 \\
HD 25893 & -0.02 $\pm$  0.01 & -0.14 $\pm$  0.13 &  0.06 $\pm$  0.02 & -0.10 $\pm$  0.24 &  0.08 $\pm$  0.03 & -0.03 $\pm$  0.01 & -0.03 $\pm$  0.25 \\
HD 27282 &  0.07 $\pm$  0.02 & -0.22 $\pm$  0.26 &  0.04 $\pm$  0.01 & 0.00 $\pm$  0.17 &  0.09 $\pm$  0.04 & -0.01 $\pm$  0.03 & -0.08 $\pm$  0.10 \\
HD 27685 & -0.07 $\pm$  0.03 & -0.05 $\pm$  0.11 & -0.02 $\pm$  0.05 & -0.29 $\pm$  0.28 &  0.11 $\pm$  0.03 & -0.09 $\pm$  0.01 & -0.03 $\pm$  0.11 \\
HD 27989 & -0.11 $\pm$  0.16 & -0.09 $\pm$  0.21 &  0.16 $\pm$  0.01 & -0.39 $\pm$ --- &  0.12 $\pm$  0.02 & -0.08 $\pm$ --- &  0.13 $\pm$ --- \\
$\epsilon$ Tau &  0.45 $\pm$  0.07 & -0.05 $\pm$  0.14 &  0.08 $\pm$  0.04 & -0.18 $\pm$  0.38 &  0.10 $\pm$  0.06 &  0.06 $\pm$  0.02 &  0.21 $\pm$  0.09 \\
111 Tau B & -0.30 $\pm$  0.11 & -0.36 $\pm$  0.03 &  0.24 $\pm$  0.20 & -0.12 $\pm$  0.22 & -0.14 $\pm$  0.04 & -0.08 $\pm$  0.06 &  0.15 $\pm$  0.16 \\
HD 40979 &  0.04 $\pm$  0.10 & -0.09 $\pm$  0.24 &  0.15 $\pm$  0.04 &  0.24 $\pm$ --- &  0.40 $\pm$  0.06 & -0.01 $\pm$  0.07 &  0.31 $\pm$ --- \\
HD 45609 & -0.41 $\pm$  0.01 & -0.46 $\pm$  0.23 & -0.18 $\pm$  0.04 & -0.23 $\pm$  0.25 & -0.16 $\pm$  0.09 &  0.11 $\pm$  0.07 & -0.06 $\pm$  0.15 \\
HD 52265 &  0.06 $\pm$  0.08 & -0.14 $\pm$  0.10 &  0.08 $\pm$  0.02 &  0.10 $\pm$  0.27 &  0.24 $\pm$  0.02 &  0.06 $\pm$  0.01 & -0.01 $\pm$  0.16 \\
HD 53532 & -0.05 $\pm$  0.05 & -0.25 $\pm$  0.23 &  0.09 $\pm$  0.03 & -0.04 $\pm$  0.14 &  0.23 $\pm$  0.02 &  0.13 $\pm$  0.01 &  0.12 $\pm$  0.06 \\
HD 65523 & -0.04 $\pm$  0.02 & -0.27 $\pm$  0.16 & -0.12 $\pm$  0.07 & -0.34 $\pm$  0.26 & -0.01 $\pm$  0.01 &  0.08 $\pm$  0.01 & -0.08 $\pm$  0.06 \\
HD 70088 & -0.35 $\pm$  0.07 & -0.40 $\pm$  0.17 & -0.07 $\pm$  0.04 & -0.26 $\pm$  0.32 &  0.10 $\pm$  0.03 &  0.02 $\pm$  0.05 &  0.03 $\pm$  0.05 \\
HD 72760 & -0.05 $\pm$  0.02 & -0.25 $\pm$  0.28 & -0.01 $\pm$  0.03 & -0.28 $\pm$  0.28 &  0.14 $\pm$  0.07 &  0.01 $\pm$  0.01 & 0.00 $\pm$  0.11 \\
V401 Hya & -0.07 $\pm$  0.04 & -0.22 $\pm$  0.21 &  0.04 $\pm$  0.01 & -0.04 $\pm$  0.20 &  0.17 $\pm$  0.01 &  0.07 $\pm$  0.01 &  0.01 $\pm$  0.11 \\
HD 73171 &  0.24 $\pm$  0.05 & -0.12 $\pm$  0.12 & -0.16 $\pm$  0.04 & -0.53 $\pm$  0.18 &  0.09 $\pm$  0.03 &  0.08 $\pm$  0.05 &  0.14 $\pm$  0.14 \\
CT Pyx &  0.03 $\pm$  0.02 & -0.28 $\pm$  0.22 &  0.01 $\pm$  0.02 & -0.15 $\pm$  0.10 & -0.04 $\pm$  0.01 &  0.02 $\pm$  0.06 & -0.02 $\pm$  0.01 \\
$\rho$ Cnc A &  0.48 $\pm$  0.01 &  0.13 $\pm$  0.20 &  0.17 $\pm$ --- & -0.17 $\pm$  0.24 &  0.07 $\pm$  0.01 &  0.23 $\pm$  0.04 &  0.17 $\pm$  0.04 \\
HD 75898 &  0.38 $\pm$  0.03 &  0.06 $\pm$  0.21 &  0.21 $\pm$  0.05 & -0.26 $\pm$  0.39 &  0.22 $\pm$  0.04 & 0.00 $\pm$  0.01 &  0.19 $\pm$  0.02 \\
HD 76151 &  0.05 $\pm$  0.04 & -0.20 $\pm$  0.24 & -0.07 $\pm$  0.02 & -0.31 $\pm$  0.31 & 0.00 $\pm$  0.02 & -0.10 $\pm$  0.01 & -0.06 $\pm$  0.07 \\ 
IK Cnc & -0.45 $\pm$  0.03 & -0.52 $\pm$  0.21 & -0.18 $\pm$  0.08 & -0.37 $\pm$  0.22 & -0.06 $\pm$  0.05 & -0.20 $\pm$ --- & -0.18 $\pm$ --- \\
HD 82106 & -0.15 $\pm$  0.05 & -0.19 $\pm$  0.05 &  0.03 $\pm$  0.09 & -0.20 $\pm$  0.26 & -0.03 $\pm$  0.02 &  0.07 $\pm$  0.07 & 0.00 $\pm$  0.14 \\
GT Leo &  0.07 $\pm$  0.04 & -0.30 $\pm$  0.24 & -0.07 $\pm$  0.03 & -0.34 $\pm$  0.34 &  0.01 $\pm$  0.07 & -0.17 $\pm$  0.07 & -0.20 $\pm$  0.16 \\
HD 85301 & -0.08 $\pm$  0.08 & -0.21 $\pm$  0.21 &  0.12 $\pm$  0.05 & -0.12 $\pm$  0.24 &  0.15 $\pm$  0.02 &  0.05 $\pm$  0.05 & -0.05 $\pm$  0.10 \\
HD 86322 &  0.15 $\pm$  0.07 & -0.12 $\pm$  0.07 & -0.11 $\pm$  0.02 & -0.49 $\pm$  0.25 & -0.04 $\pm$  0.02 &  0.07 $\pm$  0.07 &  0.11 $\pm$  0.11 \\
HD 89307 & -0.24 $\pm$  0.08 & -0.36 $\pm$  0.15 & -0.20 $\pm$  0.01 & -0.49 $\pm$  0.27 & -0.08 $\pm$  0.04 & -0.19 $\pm$  0.02 & -0.15 $\pm$  0.05 \\
HD 89376 & -0.09 $\pm$  0.02 & -0.02 $\pm$  0.13 & -0.02 $\pm$  0.04 & -0.31 $\pm$  0.29 & -0.16 $\pm$ --- &  0.03 $\pm$  0.14 & -0.09 $\pm$  0.10 \\
$\pi^1$ Leo &  0.32 $\pm$  0.04 & -0.06 $\pm$  0.09 &  0.08 $\pm$  0.01 & -0.15 $\pm$  0.35 &  0.13 $\pm$  0.01 &  0.14 $\pm$  0.06 &  0.18 $\pm$  0.12 \\
HD 98356 & -0.06 $\pm$  0.07 & -0.19 $\pm$  0.13 &  0.02 $\pm$  0.04 & -0.33 $\pm$  0.21 & -0.09 $\pm$  0.04 &  0.03 $\pm$  0.02 &  0.07 $\pm$  0.02 \\
HD 101112 &  0.44 $\pm$  0.07 &  0.05 $\pm$  0.13 & -0.01 $\pm$  0.03 & -0.39 $\pm$  0.24 &  0.01 $\pm$  0.01 &  0.01 $\pm$  0.02 &  0.08 $\pm$  0.14 \\
HD 106696 & -0.04 $\pm$  0.11 & -0.36 $\pm$  0.27 & -0.37 $\pm$  0.04 & -0.70 $\pm$  0.37 & -0.24 $\pm$  0.02 & -0.29 $\pm$  0.01 & -0.17 $\pm$  0.07 \\
MY UMa & -0.11 $\pm$  0.02 & -0.41 $\pm$  0.05 & -0.04 $\pm$  0.13 & -0.01 $\pm$  0.14 & -0.06 $\pm$  0.04 &  0.01 $\pm$  0.05 &  0.13 $\pm$  0.10 \\
GR Leo & -0.11 $\pm$  0.03 & -0.18 $\pm$  0.06 &  0.04 $\pm$ --- & -0.23 $\pm$  0.29 &  0.14 $\pm$  0.02 &  0.09 $\pm$  0.01 &  0.10 $\pm$  0.07 \\
HD 106023 & -0.16 $\pm$  0.08 & -0.24 $\pm$  0.13 & -0.10 $\pm$  0.03 & -0.35 $\pm$  0.26 & -0.04 $\pm$  0.01 & -0.06 $\pm$  0.04 & -0.14 $\pm$  0.06 \\
HD 117860 & -0.11 $\pm$  0.11 & -0.23 $\pm$  0.18 &  0.02 $\pm$  0.06 & -0.26 $\pm$  0.37 &  0.22 $\pm$  0.04 & -0.01 $\pm$  0.05 & -0.03 $\pm$  0.11 \\
HD 117936 & -0.03 $\pm$  0.04 & -0.07 $\pm$  0.03 &  0.13 $\pm$  0.28 & -0.22 $\pm$  0.25 & -0.18 $\pm$ --- &  0.06 $\pm$  0.08 & -0.12 $\pm$  0.27 \\
HD 124642 & -0.16 $\pm$  0.06 & -0.23 $\pm$  0.03 &  0.24 $\pm$  0.21 & -0.23 $\pm$  0.23 & -0.11 $\pm$  0.02 &  0.08 $\pm$  0.08 &  0.06 $\pm$  0.19 \\
HD 126535 &  0.08 $\pm$  0.08 &  0.04 $\pm$  0.04 & 0.00 $\pm$  0.02 & -0.21 $\pm$  0.29 & -0.02 $\pm$  0.04 &  0.11 $\pm$  0.04 &  0.09 $\pm$  0.04 \\
HD 131023 &  0.09 $\pm$  0.09 & -0.20 $\pm$  0.27 &  0.01 $\pm$  0.01 & -0.25 $\pm$  0.27 &  0.14 $\pm$  0.03 &  0.07 $\pm$  0.02 &  0.06 $\pm$  0.08 \\
HD 149026 &  0.31 $\pm$  0.05 &  0.10 $\pm$  0.20 &  0.22 $\pm$  0.03 &  0.19 $\pm$ --- &  0.23 $\pm$ --- &  0.06 $\pm$ --- &  0.04 $\pm$  0.06 \\
HD 170657 & -0.34 $\pm$  0.02 & -0.50 $\pm$  0.19 & -0.27 $\pm$  0.01 & -0.58 $\pm$  0.28 & -0.27 $\pm$  0.02 & -0.24 $\pm$  0.03 & -0.19 $\pm$  0.01 \\
MV Dra &  0.07 $\pm$  0.02 & -0.17 $\pm$  0.26 & 0.00 $\pm$  0.05 & -0.26 $\pm$  0.30 &  0.03 $\pm$  0.02 &  0.03 $\pm$  0.03 & -0.05 $\pm$  0.03 \\
HD 189087 & -0.19 $\pm$ --- & -0.42 $\pm$  0.22 & -0.11 $\pm$  0.02 & -0.38 $\pm$  0.27 & -0.01 $\pm$ --- & -0.05 $\pm$  0.02 & -0.09 $\pm$  0.06 \\
V2425 Cyg & -0.18 $\pm$  0.03 & -0.16 $\pm$  0.20 & -0.05 $\pm$  0.04 & -0.17 $\pm$  0.23 &  0.10 $\pm$  0.07 & -0.06 $\pm$  0.02 &  0.10 $\pm$  0.05 \\
HD 196885 &  0.01 $\pm$  0.11 & -0.02 $\pm$  0.20 &  0.14 $\pm$  0.08 &  0.06 $\pm$  0.25 &  0.30 $\pm$  0.02 &  0.09 $\pm$  0.04 &  0.05 $\pm$  0.07 \\
V2436 Cyg &  0.03 $\pm$  0.01 & -0.26 $\pm$  0.24 &  0.11 $\pm$  0.04 & -0.24 $\pm$  0.26 &  0.09 $\pm$  0.06 &  0.13 $\pm$  0.06 &  0.06 $\pm$  0.02 \\
HD 200968 & -0.10 $\pm$  0.03 & -0.13 $\pm$  0.17 & -0.03 $\pm$  0.02 & -0.41 $\pm$  0.27 & -0.08 $\pm$  0.01 & -0.08 $\pm$  0.05 & -0.12 $\pm$  0.08 \\
NS Aqr &  0.17 $\pm$  0.04 & -0.14 $\pm$  0.25 & -0.02 $\pm$ --- & -0.25 $\pm$  0.26 &  0.09 $\pm$  0.02 &  0.02 $\pm$  0.04 & -0.09 $\pm$  0.12 \\
V454 And & -0.21 $\pm$  0.04 & -0.24 $\pm$  0.18 &  0.11 $\pm$  0.03 & -0.10 $\pm$  0.38 &  0.26 $\pm$  0.03 &  0.06 $\pm$  0.04 &  0.17 $\pm$  0.06 \\
HD 222422 & -0.29 $\pm$  0.04 & -0.48 $\pm$  0.24 & -0.19 $\pm$  0.01 & -0.23 $\pm$  0.24 & -0.01 $\pm$  0.05 & -0.12 $\pm$  0.02 & -0.16 $\pm$  0.13 \\
20 Psc &  0.09 $\pm$  0.10 & -0.21 $\pm$  0.13 & -0.01 $\pm$  0.05 & -0.22 $\pm$  0.38 &  0.25 $\pm$  0.03 &  0.11 $\pm$  0.02 &  0.19 $\pm$  0.12 \\
\noalign{\smallskip}
\hline
\noalign{\smallskip}
\end{tabular}
\end{table*}
%\end{landscape}

\begin{landscape}
\begin{table}
\scriptsize
\caption{ Differential abundance sensitivities for Na, Mg, Al, Si, Ca, Sc, Ti, V, Cr, Mn, Co, and Ni.}
\label{tablasense}
%\centering
\begin{tabular}{lcccccccccccc}
\hline
\hline
\noalign{\smallskip}

\noalign{\smallskip}
{\small Name} & {\small $\Delta$[Na/H]} & {\small $\Delta$[Mg/H]} & {\small $\Delta$[Al/H]} & {\small $\Delta$[Si/H]} & {\small $\Delta$[Ca/H]} & {\small $\Delta$[Sc/H]} & {\small $\Delta$[Ti/H]} & {\small $\Delta$[V/H]} & {\small $\Delta$[Cr/H]} & {\small $\Delta$[Mn/H]} & {\small $\Delta$[Co/H]} & {\small  $\Delta$[Ni/H]} \\
\hline
\noalign{\smallskip}
$\Delta T_{\rm eff}$ = $\pm$ 100 K       &      &      &      &      &      &      &      &      &      &      &      &     \\
\hline
\noalign{\smallskip}
HD 27685 & 0.06 & 0.05 & 0.05 & 0.01 & 0.07 & 0.09 & 0.10 & 0.11 & 0.07 & 0.06 & 0.07 & 0.05\\
HD 98356 & 0.07 & 0.04 & 0.05 & 0.01 & 0.08 & 0.10 & 0.12 & 0.13 & 0.08 & 0.07 & 0.05 & 0.03\\
BZ Cet   & 0.08 & 0.03 & 0.06 & 0.03 & 0.09 & 0.11 & 0.13 & 0.14 & 0.09 & 0.06 & 0.02 & 0.01\\
\hline
\noalign{\smallskip}
$\Delta$$\log{g}$ = $\pm$ 0.30 dex  &      &      &      &      &      &      &      &      &      &      &      &     \\
\hline
\noalign{\smallskip}
HD 27685 & 0.03 & 0.03 & 0.01 & 0.00 & 0.07 & 0.00 & 0.01 & 0.01 & 0.03 & 0.05 & 0.01 & 0.00\\
HD 98356 & 0.06 & 0.04 & 0.02 & 0.02 & 0.09 & 0.01 & 0.03 & 0.02 & 0.05 & 0.07 & 0.02 & 0.02\\
BZ Cet   & 0.07 & 0.04 & 0.04 & 0.04 & 0.11 & 0.02 & 0.04 & 0.03 & 0.06 & 0.07 & 0.04 & 0.04\\
\hline
\noalign{\smallskip}
$\Delta$ $\xi$ = $\pm$ 0.50 km s$^{-1}$ &      &      &      &      &      &      &      &      &      &      &      &     \\
\hline
\noalign{\smallskip}
HD 27685 & 0.02 & 0.03 & 0.01 & 0.02 & 0.08 & 0.01 & 0.06 & 0.04 & 0.07 & 0.11 & 0.04 & 0.06\\
HD 98356 & 0.03 & 0.03 & 0.02 & 0.02 & 0.08 & 0.04 & 0.11 & 0.16 & 0.08 & 0.09 & 0.08 & 0.07\\
BZ Cet   & 0.03 & 0.03 & 0.04 & 0.02 & 0.08 & 0.08 & 0.16 & 0.20 & 0.08 & 0.09 & 0.08 & 0.07\\
\hline
\noalign{\smallskip}
$\Delta$[Fe/H] = $\pm$ 0.30 dex   &      &      &      &      &      &      &      & 	 &      &      &      &     \\
\hline
\noalign{\smallskip}
HD 27685 & 0.01 & 0.01 & 0.00 & 0.04 & 0.02 & 0.00 & 0.00 & 0.00 & 0.01 & 0.02 & 0.02 & 0.03\\
HD 98356 & 0.03 & 0.04 & 0.01 & 0.07 & 0.04 & 0.00 & 0.00 & 0.01 & 0.03 & 0.06 & 0.06 & 0.07\\
BZ Cet   & 0.04 & 0.06 & 0.02 & 0.08 & 0.05 & 0.00 & 0.02 & 0.01 & 0.05 & 0.08 & 0.08 & 0.09\\
\noalign{\smallskip}
\hline
\noalign{\smallskip}

\end{tabular}

\tablefoot{Sensitivities to changes of 100 K in $T_{\rm eff}$, 0.30 dex in $\log{g}$, 0.50 km s$^{-1}$  in $\xi$, and 0.30 dex in [Fe/H]. We verified these sensitivities for the stars HD 27685 (5753 K, 4.47 dex, 0.99 km s$^{-1}$, 0.12 dex), HD 98356 (5310 K, 4.41 dex, 0.92 km s$^{-1}$, 0.08 dex), and BZ Cet (5035 K, 4.38 dex, 0.98 km s$^{-1}$, 0.11 dex).}
\end{table}
\end{landscape}

%\begin{landscape}
\begin{table*}
\scriptsize
\caption{ Differential abundance sensitivities for Cu, Zn, Y, Zr, Ba, and Ce. Differences are computed as in Table~\ref{tablasense}}
\label{tablasense2}
\centering
\begin{tabular}{lccccccc}
\hline
\hline
\noalign{\smallskip}

\noalign{\smallskip}
{\small Name} & {\small $\Delta$[Cu/H]} &  {\small $\Delta$[Zn/H]} & {\small $\Delta$[Y/H]} & {\small $\Delta$[Zr/H]} & {\small $\Delta$[Ba/H]} & {\small $\Delta$[Ce/H]} & {\small  $\Delta$[Nd/H]} \\
\hline
\noalign{\smallskip}
$\Delta T_{\rm eff}$ = $\pm$ 100 K       &      &      &     &    &   &   &   \\
\hline
\noalign{\smallskip}
HD 27685 & 0.07 & 0.00 & 0.01 & 0.11 & 0.02 & 0.02 & 0.02 \\
HD 98356 & 0.05 & 0.02 & 0.01 & 0.13 & 0.02 & 0.01 & 0.02 \\
BZ Cet   & 0.04 & 0.02 & 0.01 & 0.14 & 0.00 & 0.02 & 0.03 \\
\hline
\noalign{\smallskip}
$\Delta$$\log{g}$ = $\pm$ 0.30 dex  &      &      &     &       &     &  &      \\
\hline
\noalign{\smallskip}
HD 27685 & 0.02 & 0.02 & 0.11 & 0.01 & 0.04 & 0.13 & 0.13 \\
HD 98356 & 0.02 & 0.03 & 0.11 & 0.01 & 0.04 & 0.12 & 0.13 \\
BZ Cet   & 0.01 & 0.02 & 0.10 & 0.01 & 0.04 & 0.12 & 0.13 \\
\hline
\noalign{\smallskip}
$\Delta$ $\xi$ = $\pm$ 0.50 km s$^{-1}$ &      &      &     &       &     &  &      \\ 
\hline
\noalign{\smallskip}
HD 27685 & 0.14 & 0.13 & 0.17 & 0.02 & 0.18 & 0.05 & 0.03 \\
HD 98356 & 0.12 & 0.10 & 0.15 & 0.06 & 0.17 & 0.10 & 0.04 \\
BZ Cet   & 0.12 & 0.13 & 0.19 & 0.16 & 0.15 & 0.11 & 0.04 \\
\hline
\noalign{\smallskip}
$\Delta$[Fe/H] = $\pm$ 0.30 dex  &      &      &     &       &     &  &      \\ 
\hline
\noalign{\smallskip}
HD 27685 & 0.04 & 0.08 & 0.09 & 0.01 & 0.15 & 0.11 & 0.11 \\
HD 98356 & 0.09 & 0.10 & 0.11 & 0.01 & 0.17 & 0.11 & 0.12 \\
BZ Cet   & 0.12 & 0.12 & 0.12 & 0.01 & 0.18 & 0.12 & 0.11 \\
\noalign{\smallskip}
\hline
\noalign{\smallskip}

\end{tabular}
\end{table*}
%\end{landscape}

\begin{table}
\caption{Atomic parameters employed for the lines of Cu, Zn, Y, Zr, Ba, Ce, and Nd. }
\label{linerstab}
\centering
\begin{tabular}{lcccc}     
\hline
\hline
\noalign{\smallskip}
Element & $\lambda$ & $\chi_{\rm l}$ & $\log{gf}$  & Ref. \\
\noalign{\smallskip}
 & ($\AA$) & (eV) &  &  \\
\noalign{\smallskip}
\hline
\noalign{\smallskip}
$\ion{Cu}{i}$ & 5105.55   &   1.39 &  -1.520 &    1    \\
      & 5218.21   &   3.82 &   0.480 &    1    \\
      & 5220.09   &   3.82 &  -0.450 &    1    \\
      & 5782.12   &   1.64 &  -1.720 &    1    \\
\hline
\noalign{\smallskip}
$\ion{Zn}{i}$  & 4722.16   &   4.03 &  -0.370 &     1   \\
      & 4810.54   &   4.08 &  -0.170 &     1   \\
      & 6362.35   &   5.79 &   0.140 &     1   \\
\hline
\noalign{\smallskip}
$\ion{Y}{ii}$  & 4900.12   &   1.03 &  -0.090 &     1   \\
      & 5087.43   &   1.08 &  -0.160 &     1   \\     
      & 5200.42   &   0.99 &  -0.570 &     1   \\
      & 5402.78   &   1.84 &  -0.440 &     1   \\
\hline
\noalign{\smallskip}
$\ion{Zr}{i}$  & 4687.81   &   0.73 &   0.550 &   2     \\
      & 4739.48   &   0.65 &   0.230 &   2     \\  
$\ion{Zr}{ii}$ & 5112.28   &   1.66 &  -0.590 &   1    \\
\hline
\noalign{\smallskip}
$\ion{Ba}{ii}$ & 4554.03   &   0.00 &   0.140 &    2    \\
      & 4934.08   &   0.00 &  -0.157 &    2    \\
      & 5853.67   &   0.60 &  -0.909 &    1    \\
      & 6141.71   &   0.70 &  -0.030 &    1    \\
      & 6496.90   &   0.60 &  -0.406 &    1    \\
\hline
\noalign{\smallskip}
$\ion{Ce}{ii}$ & 4523.08  &   0.51 &   0.040 &     1   \\
      & 4562.36  &   0.48 &   0.230 &     2   \\
      & 4628.16  &   0.52 &   0.230 &     1   \\
      & 4773.96  &   0.92 &   0.250 &     1   \\
      & 5274.23  &   1.04 &   0.150 &     1   \\
\hline
\noalign{\smallskip}
$\ion{Nd}{ii}$  & 5092.80  &   0.38 &  -0.650 &    1   \\
      & 5319.82  &   0.55 &  -0.140 &    2   \\
\noalign{\smallskip}
\hline
\noalign{\smallskip}

\end{tabular}

%\noalign{\smallskip}
\tablefoot{
%\noalign{\smallskip}
\tablefoottext{1}{ \citet{gon10} }\\
%\noalign{\smallskip}
\tablefoottext{2}{ \citet{pom11} }\\
%\noalign{\smallskip}
 }

\end{table}
\end{appendix}
\end{document}